\documentclass[11pt]{article}
\usepackage{xcolor}
\usepackage{array}
\usepackage{float}
\usepackage{textcomp}
\usepackage{amsmath}
\usepackage{amsthm}
\usepackage{amssymb}
\usepackage{graphicx}
\usepackage{comment}

\usepackage{mathtools}
\usepackage{authblk}

\makeatletter

\providecommand{\tabularnewline}{\\}

\usepackage{array}
\usepackage{url}

\usepackage[shortlabels]{enumitem}

\providecommand{\tabularnewline}{\\}

\usepackage{array}

\usepackage{subfigure}

\usepackage{amsfonts}
\usepackage{amsthm} % \usepackage{enumerate}
\usepackage{ifthen}
\usepackage{latexsym}\usepackage{syntonly}\usepackage{setspace}
\usepackage{mathabx}
\usepackage{epstopdf}
\usepackage{calc}
\usepackage{xspace}
\usepackage{soul}

\RequirePackage{fancyhdr}

\newtheorem{thm}{Theorem}[section]
\newtheorem{lem}{Lemma}[section]
\newtheorem{defi}{Definition}[section]

\newtheorem{prop}[thm]{Proposition}\newtheorem{rem}{Remark}[section]
\newtheorem{cor}{Corollary}[section]

\def\bt{\begin{thm}}
	\def\et{\end{thm}}
\def\bl{\begin{lem}}
	\def\el{\end{lem}}
\def\bd{\begin{defi}}
	\def\ed{\end{defi}}
\def\bc{\begin{cor}}
	\def\ec{\end{cor}}
\def\bp{\begin{proof}}
	\def\ep{\end{proof}}
\def\br{\begin{rem}}
	\def\er{\end{rem}}
\def\bprop{\begin{prop}}
	\def\eprop{\end{prop}}

\def\be{\begin{equation}}
	\def\ee{\end{equation}}
\def\bes{\begin{equation*}}
	\def\ees{\end{equation*}}
\def\bea{\begin{equation} \begin{aligned}}
		\def\eea{\end{aligned} \end{equation}}
\def\beas{\begin{equation*} \begin{aligned}}
		\def\eeas{\end{aligned} \end{equation*}}

\def\bi{\begin{itemize}}
	\def\ei{\end{itemize}}
\def\ben{\begin{enumerate}}
	\def\een{\end{enumerate}}

\usepackage[style=nature, sorting=none, natbib=true, defernumbers]{biblatex}
\AtEveryCitekey{%
	\ifcsundef{blx@entry@refsegment@\the\c@refsection @\thefield{entrykey}}
	{\csnumgdef{blx@entry@refsegment@\the\c@refsection @\thefield{entrykey}}{\the\c@refsegment}}
	{}}
\defbibcheck{onlynew}{%
	\ifnumless{0\csuse{blx@entry@refsegment@\the\c@refsection @\thefield{entrykey}}}{\the\c@refsegment}
	{\skipentry}
	{}}

\makeatother

	\title{\textbf{A stable hothouse}\\
		\textbf{triggered by a tipping mechanism}}
\author[1,2,*]{Erik Chavez}
\affil[*]{Corresponding author}
\affil[1]{Brevan Howard Centre for Financial Analysis, Imperial College Business School, Imperial College London; London, SW7 2AZ, United Kingdom}
\affil[2]{Laboratoire de M\'et\'eorologie Dynamique (CNRS \& IPSL), Ecole Polytechnique, Paris, France}
\author[3]{Jan Rombouts}
\affil[3]{Unit of Theoretical Chronobiology, Faculty of Sciences, Université Libre de Bruxelles; Brussels, 1050, Belgium}
\author[4,5,6]{Michael Ghil}
\affil[4]{Geosciences Department and Laboratoire de M\'et\'eorologie Dynamique (CNRS \& IPSL), Ecole Normale Sup\'erieure and PSL University; Paris, 75005, France}
\affil[5]{Department of Atmospheric and Oceanic Sciences, University of California, Los Angeles; Los Angeles, 90095, USA}
\affil[6]{Department of Mathematics, Imperial College London; London, SW7 2AZ, United Kingdom}

\date{}

\begin{document}
	
	\begin{refsegment}
		
	\maketitle
	
%	\textbf{\textcolor{blue}{
			{\bf The climate system's nonlinear dynamics is influenced by various external forcings and internal  feedbacks that can give rise to regional and even global tipping points that may lead to significant and potentially irreversible changes. Paleoclimatic records reveal that Earth's climate has shifted between distinct equilibria, including a ``hothouse Earth" state with temperatures about 10~K higher than at present. However, a specific mechanism for a sudden tipping to an alternate stable state, several degrees warmer than the present climate, has yet to be presented. We introduce a temperature–carbon–vegetation (TCV) model comprising an energy balance model of global temperature, coupled with global terrestrial and ocean CO$_{2}$ dynamics, and with vegetation ecosystem change. Our model exhibits a new tipping mechanism that leads to a hothouse Earth under a high-emissions scenario. Its simulations align with both observations and IPCC-class global climate models prior to tipping. The two processes that produce global tipping are: (i) temperature–albedo feedback due to darkening of the terrestrial cryosphere by glacial microalgae; and (ii) limits to vegetation adaptation that lead to reduced carbon absorption.}

\vspace{0.1cm}

\section{Introduction and motivation}	
\label{sec:introduction}

The climate system is characterized by nonlinear and chaotic dynamics on global as well as regional scales \parencite{Lenton2016,Ghil2020}. However, the relationship between the globally averaged temperature increase and cumulative anthropogenic carbon emissions is generally approximated by a linear function \parencite{Matthews2009} also known as the Transient Climate Response to Cumulative Emissions (TCRE). This linear approximation is widely accepted for the present-day and moderate future emission scenarios, as reported in the 6$^{\rm {th}}$ and most recent Assessment Report (AR6) of the Intergovernmental Panel on Climate Change (IPCC) \parencite{Forster2021}. Nonetheless, several studies also cited in the same AR6 report\parencite{Forster2021} (section 7.4.3.1 State-dependence of Feedbacks in Models) stress the limits of this proportionality relationship which no longer holds at higher atmospheric carbon levels and global temperature \parencite{Zhu2019,Caballero2013}. As feedbacks could become more pronounced and climate sensitivity be state-dependent, high emission scenarios may translate into stronger warming than the TCRE predicts. Furthermore, AR6 cites studies \parencite{Steffen.ea.2018,Aswhin2020} that indicate "the possibility of more substantial changes in climate feedbacks, sometimes accompied by hysteresis and/or irreversibility" that "could occur on a global scale and across relatively narrow temperature changes".

The impact of anthropogenic forcing on several of the climate system's processes has led to concern about the triggering of internal feedbacks that may lead to tipping of subsystems and its potentially irreversible consequences \parencite{Lenton2016, Lenton.tip.2008,Ghil2019, Ghil2020}. The consequences of a regime shift to an extreme, Venus-like runaway greenhouse would be most dire \parencite{Goldblatt2012} but are not very plausible. More plausible and long-lasting "hotouse" episodes have been observed in the Earth's past \parencite{Zeebe_2013,Gutjahr_2017,Winguth2015,Sun2012} and could be expected in its future \parencite{Steffen.ea.2018}. One of the major sources of uncertainty is the climate system's complexity and nonlinear response to several external forcings, both natural and anthropogenic, operating at overlapping timescales\parencite{Heydt2021}, which could lead to such type of global-scale changes. However, specific mechanisms for rapid global tipping to a hothouse Earth have not been explored so far to a satisfactory extent \parencite{Ferreira2011,Popp.ea.2016, Wunderling_2024} and remain quite uncertain, as stressed by the AR6 \parencite{Forster2021}.
Representing fundamental uncertainties in future temperature values and other key climate features \parencite{Rising_2022} is essential for shaping mitigation and adaptation policy \parencite{Barnett2021,Lackner.ea.2022}. 

% 5While  has been made,  remain %in ESMs' ability 
Accurately simulating and predicting major nonlinear features of the climate system remains quite challenging, in spite of significant progress \parencite{Palmer.Stevens.2019, Slingo2011}. Complementary modeling approaches to ESMs have proven useful in contributing to understand these nonlinear dynamics. The approach of ``emergent constraints," for instance, integrates observational data and reconstructed paleoclimate proxies \parencite{Hall.ea.2019, Sherwood2020, Eisenman_2024}. An alternative method employs a hierarchy of models to address specific aspects of the climate system \parencite{Schneider.Dick.1974,Ghil.2001}. 

Thus, Held (2005) \parencite{Held.gap.2005} argues that the leaps molecular biology has achieved in understanding complex systems like the human organism are largely due to the systematic and constant study, validation, and integration of hypotheses and learning from the lower rungs of the biological model hierarchy into the higher rungs. Their use is also valuable in the climate sciences to better understand and predict the response of more detailed ESM models \parencite{Ghil2019,Ghil2020,IPCC.2021}. 

In the climate science, the initial rung of the model hierarchy is that of global energy balance models (EBMs) \parencite{Budyko.1969, Sellers.1969, Held.Suarez.1974, Ghil1976, North1981} that can track the evolution of the globally averaged temperature subject to changes in radiative in- and output. Similar simple models with a few scalar variables have been used to characterise nonlinear phenomena such as bistability \parencite{Stommel1961} and chaotic behavior \parencite{Lorenz1963} in the oceans or atmosphere, as observed in the Earth's paleoclimatic records \parencite{Boers_2022}. 

Following the evolution of atmospheric carbon concentration on different timescales also requires the modeling of carbon fluxes through the Earth's subsystems \parencite{Lenton2000,Friedlingstein2015}. The lowest rung in the carbon model hierarchy is that of using a small number of boxes. Typically, these boxes 
% the carbon fluxes of the system, including 
encompass the atmosphere, the land soil and the vegetation, the upper ocean, and the deep ocean. Accordingly, two- \parencite{Svirezhev1997}, three-  \parencite{Anderies2013}, four-\parencite{Lenton2000,Lade_2018,Friedling2015} or even eight-box models \parencite{Hartin2015} have been proposed. Such simplified models do not account for spatially  heterogenous dynamics but can still be successful at reproducing the global dynamics of the underlying processes, including carbon uptake by vegetation. For instance the zero-dimensional (0-D) models developed by Lade et al \parencite{Lade_2018}, Svirezhev and von Bloh \parencite{Svirezhev1997}, or Smith et al \parencite{Smith_2018}, 
% while simplified and without representation of spatial heterogeneity, 
allow one to capture essential feedback mechanisms between global vegetation and climate.

No coupled EBM model on human timescales, though, has --- to the best of our knowledge --- described abrupt tipping mechanism in the climate system from the present state to a stable hothouse, in which the Earth's climate would stabilize at roughly 10~K above the current climate, as conceptually proposed by Steffen et al. \parencite{Steffen.ea.2018}, and in line with paleoclimatic proxies \parencite{Zeebe_2013}. Ghil \parencite{Ghil1976} showed, in fact, that a spatially one-dimensional EBM with temperature dependent solely on latitude and time can yield a pole-to-equator steady-state profile in very good agreement with seasonally averaged mid-20th century temperatures. Furthermore, EBMs such as these have helped study the double-fold bifurcation giving rise to an alternative steady state called, at the time, a ``deep freeze." \parencite{Held.Suarez.1974,North1981,Ghil1976} The theoretical prediction of such a state was subsequently confirmed through geochemical analysis of Neoproterozoic and other paleoclimatic records \parencite{Hoffman1998}, and is called now a ``snowball Earth". Encouraged by this and other successes of such models \cite{Ghil2019,Anderies2013,Lade_2018}, we formulate herein a 0-D model coupling the evolution of global average temperatures with global carbon levels and a global vegetation index.

The purpose of this paper is to study a fundamental source of uncertainty in the future trajectory of the climate system in response to anthropogenic carbon forcing; namely the likelihood of a potential planetary regime shift or tipping that could take the system into an alternate but plausible regime of behavior \cite{Ghil2019, Lenton2019}. In this study, we identify a tipping mechanism that relies on two specific and credible biogeophysical feedback processes.  

% \textcolor{blue}{
	%{ \mg 
	Clearly, confidence in our results needs to be tempered by the simplicity of the model and seen in the perspective of the aforementioned hierarchy of models.
	Global EBMs and idealized coupled models like ours are able to exhibit planetary-scale tipping when their few state variables permit including feedback processes that may lead to saddle-node \parencite{Ghil1976} or Hopf \parencite{Kallen1979} bifurcations. %the linear nature of 
	Spatial averaging in EBMs and box models, however, mutes heterogeneity, which leads to loss of the ability to capture tipping when one or more regions exceed local thresholds, jointly or independently. 
	
	High-resolution ESMs, though, can resolve such heterogenous processes leading to the potential crossing of regional thresholds, such as in ecosystem shifts \parencite{Lenton2016}. Their use thus allows one to study potential tipping of regional subsystems \parencite{Lenton2016} but their high resolution in today's IPCC-class models, of the order of $10^9$ scalar variables for a mesh size of 0.25~deg $\times$ 0.25~deg $\times$ 0.2~km, makes their results much less transparent, as do the tens of parameters \cite{Hourdin.ea.2017} used in fitting their various unresolved processes to the climate of the 150 years of the instrumental era.
	
%	However, the discretization by ESMs of the governing partial differential equations (PDEs) across $N$ grid cells generates $\mathcal{O}(N)$ eigenmodes ($N \approx 10^9$ in a 0.25$^\circ$ resolution ESM), many with real parts $\operatorname{Re}(\sigma_k)\!\approx\!0$. Lucarini and Bodai \parencite{Lucarini2017} show how with a dense near‑neutral eigenmodes spectrum, ESMs large‑scale diagnostics can reorganise abruptly without any single eigenvalue crossing zero (i.e. lack of dynamic bifurcation) through a quasi‑static transition. Furthermore, 

	Moreover, the numerical diffusion of ESMs, as well as their limited run length due to computational cost, can also postpone or damp a genuine global bifurcation \cite{Kim2022}. Therefore, low‑dimensional models such as those used herein can reveal and help characterize dynamic bifurcations, while ESMs with high spatial resolution allow one to 
%	encode processes driven by spatially heterogenous processes and local thresholds and potential 
	explore regional tippings that may, or may not, aggregate into a global regime change. The two ends of the model hierarchy can thus complement each other. Reduced models help explore potential tipping mechanisms of the global climate system, while high-resolution models 
%	 are useful to understand the range of potential bifurcation mechanisms and identify parameters worth probing, while high-resolution gridded model simulations 
	supply the spatially explicit representation of these mechanisms that is needed to calibrate and refine the  reduced-model results \cite{Schneider.Dick.1974, Ghil.2001,Held.gap.2005}.  
%}
		
%		creating a two-way interaction as proposed by \textcite{Held.gap.2005} that closes the scale gap and better understand potential planetary-scale nonlinearities and tipping.
%}

%The purpose of this paper is to study a fundamental source of uncertainty in the future trajectory of the climate system in response to anthropogenic carbon forcing; namely the likelihood of a potential planetary regime shift or tipping that could take the system into an alternate but plausible regime of behavior \supercite{Ghil2019, Lenton2019}. In this study, we identify a tipping mechanism that relies on two specific and credible biogeophysical feedback processes.

\section{A model for coupled globally averaged temperature and carbon dynamics}
	
\label{ssec:climdyn}
	 
Our 0-D EBM is based on the 0-D version \parencite{Craf.Kall.1978} of the original, spatially one-dimensional EBM of refs. \parencite{Ghil1976, Sellers.1969}, coupled to a three-box carbon model. This coupled model is based on physical, chemical and biological processes that operate on the decadal timescales of interest in the adaptation and mitigation problem \parencite{Lackner.ea.2022}. More precisely, the model's energy balance separates ocean from land albedo, following Källen et al. \parencite{Kallen1979}. 

Encouraged by the performance and high tractability of the framework presented by Lade et al.\parencite{Lade_2018}, we adopt its choice of global-level carbon fluxes representation. Our modelling of the terrestrial carbon sink's  global dynamics relies on the Carnegie-Ames-Stanford biosphere model \parencite{Potter_1993}, on the 0-D formulation of vegetation carbon uptake in ref.~\parencite{Svirezhev1997}, and on the global terrestrial respiration's temperature sensitivity as in ref.~\parencite{Mahecha_2010}. In order to represent the 0-D global carbon dynamics of the ocean sink, we rely on detailed studies of the carbon cycle in the ocean \parencite{Eppley1972, Henson2012, Lade_2018}. 

The coupled model is designated as the TCV model, for its key global variables: temperature, carbon content, and vegetation. Its parameter values are based on output of IPCC-class model simulations and on observational data; see the three Tables with model parameters in the Supplementary Material.

This biogeophysical TCV model is governed by a set of four ordinary differential equations (ODEs): Eq.~\eqref{eq:TempEq} for the evolution of the globally averaged atmospheric and near-surface ocean temperature $T$; Eq.~\eqref{eq:CarbAt} for the evolution of atmospheric carbon stock $C_{\mathrm{A}}$; Eq.~\eqref{eq:CarbOc} for the evolution of the carbon stock $C_{\mathrm{M}}$ in the ocean's mixed layer; and Eq.~\eqref{eq:EnvDegr} for the level  $V$ of degradation of the land vegetation. The full four-ODE model is given by:  

\begin{subequations} \label{eq:TC-4box}
	\begin{align}
& c \dot{T} = Q_{0}\left(1-p\alpha_{{\rm{L}}}(T)-(1-p)\alpha_{{\rm {O}}}(T)\right)-\kappa\left(T-T_{\kappa}\right) + a\ln\left(C_{\mathrm{A}}/C_{0}\right),
\label{eq:TempEq}\\
& \dot{C}_{\mathrm{A}} = -F_{\mathrm{A}\rightarrow \mathrm{L}}(T,C_{\mathrm{A}},V)-F_{\mathrm{A}\rightarrow \mathrm{M}}(T,C_{\mathrm{A}},C_{\mathrm{M}}) + e(t),
\label{eq:CarbAt}\\
& \dot{C}_{\mathrm{M}}=F_{\mathrm{A}\rightarrow \mathrm{M}}(T,C_{\mathrm{A}},C_{\mathrm{M}})-F_{\mathrm{M}\rightarrow \mathrm{D}}^P(T,C_{\mathrm{M}})-F_{\mathrm{M} \to \mathrm{D}}^B(T,C_{\mathrm{M}}),
\label{eq:CarbOc}\\
& \dot{V}=-\mu V(1-V)+f(T)(1-V).\label{eq:EnvDegr}
	\end{align}
\end{subequations}

The temperature equation~\eqref{eq:TempEq} is governed by the balance between incoming radiation $R_{{\rm {i}}}$ and outgoing radiation $R_{{\rm {o}}}$. The incoming radiation from the sun $Q_0$ is modified by the albedo $\alpha_{{\rm {L}}}$ of the land and $\alpha_{{\rm {O}}}$ of the ocean, which in turn depend on the temperature $T$; the fraction of land is given by $p$. 
The outgoing radiation $R_{{\rm {out}}}=\kappa\left(T-T_{\kappa}\right)$ is diminished by the greenhouse effect due to the  atmospheric CO$_2$ concentration $\mathcal C$; this effect is given by $a\ln\left(C_{\mathrm{A}}/C_{0}\right)$, with $C_0$ the preindustrial value of $C_{\mathrm{A}}$, and  $a$ the intensity parameter of radiative forcing \parencite{Myhre1998}. Note that $C_{\mathrm{A}}$ and $\mathcal C$ differ by a constant factor, since the volume of the atmosphere can be assumed to be fixed on these timescales. %\cj{I wonder whether it is really necessary to introduce this distinction between concentration and mass with a symbol}.
The heat capacity $c$ in this equation is taken equal to that of the atmosphere and ocean mixed layer combined, since we are interested here in global temperature variations on the time scales of years to decades. 

In equation~\eqref{eq:CarbAt}, $F_{\mathrm{A}\rightarrow \mathrm{{L}}}$ captures the flux of atmospheric carbon into the terrestrial carbon sink, while $F_{\mathrm{A}\rightarrow \mathrm{M}}$ captures the carbon flux from the atmosphere to the ocean mixed layer. In this equation, $e = e(t)$ is the anthropogenic GHG forcing, expressed in GtC/yr of carbon equivalent, with $e(0) = 0$ and $t=0$ in the year 1800. 

In equation~\eqref{eq:CarbOc}, $F_{\mathrm{M}\rightarrow \mathrm{D}}^P$ captures the flux of carbon from the ocean mixed layer to the deeper ocean through the physical effect of the temperature-dependent overturning circulation, while $F_{\mathrm{M}\rightarrow \mathrm{D}}^B$ captures the flux of carbon to the deeper ocean due to the biopump associated with Wallace Broecker's  name \parencite{Broecker.Peng.1982}. All these carbon fluxes depend in the model on the global temperature $T$.

Equation~\eqref{eq:EnvDegr} describes the evolution of $V$, a proxy for ecosystem degradation. Its evolution depends on temperature and the equation is derived from a modified logistic growth equation. The function $f(T)$ models how high temperatures lead to vegetation loss, and $\mu$ dictates the timescale of recovery; see the Methods section for further details.

In order to understand our TCV model's  behavior in a more transparent, graphical way, we first reduce the four-ODE  system~\eqref{eq:TC-4box} to a three-ODE system~\eqref{eq:TC-3box}, by assuming a fast equilibration \parencite{Zeebe2001} across the ocean surface of the carbon concentrations, and hence of $C_{\mathrm{A}}$ and $C_{\mathrm{M}}$; the two ODEs~\eqref{eq:CarbAt} and \eqref{eq:CarbOc} above are thus reduced to a single ODE~\eqref{eq:TotalCarb} and to the algebraic equation~\eqref{eq:AlEq} below. For the details of the model reduction, see the Methods section. In the next section, we will first analyze the model's behavior for fixed $V$. In this case, we have a system of two ODEs and one algebraic equation, whose steady states can be studied graphically in the system's phase plane.
	
The reduced differential-algebraic system of equations (DAE) below couples the evolution of $T$ and of the total atmospheric and mixed layer carbon $C_{\mathrm{S}} = C_{\mathrm{A}} + C_{\mathrm{M}}$:
\begin{subequations}
\label{eq:TC-3box} 
\begin{align}
	& c \dot{T}=Q_{0}\left(1-p\alpha_{{\rm{L}}}(T)-(1-p)\alpha_{{\rm {O}}}(T)\right)-\kappa\left(T-T_{\kappa}\right)+a\ln\left(C_{\mathrm{A}}/C_{0}\right),
	\label{eq:TempEq3}\\
	& \dot{C_{\mathrm{S}}} = -F_{\mathrm{A}\rightarrow \mathrm{L}}(T,C_{\mathrm{A}},V)-F_{\mathrm{M}\rightarrow \mathrm{D}}^P(T,C_{\mathrm{M}})-F_{\mathrm{M}\rightarrow \mathrm{D}}^B(T,C_{\mathrm{M}}) + e(t),
	\label{eq:TotalCarb}\\
	& \dot{V}=-\mu V(1-V)+f(T)(1-V) .
	\label{eq:EnvDegr3}\\
	& 0=F_{\mathrm{A}\rightarrow \mathrm{M}}(T,C_{\mathrm{A}},C_{\mathrm{M}}).
	\label{eq:AlEq}
\end{align}
\end{subequations}

We refer to this simplified DAE version of the TCV model as TCV-DAE for short. The parameter values used in the equations above are listed in Tables~\ref{tab:param-temp}, \ref{tab:param-terrestrial-carbon}, and \refeq{tab:param-ocean-carbon}, and the annual anthropogenic emissions $e(t)$ are expressed in GtC/yr of 
carbon equivalent.

Both the full, four-ODE model \eqref{eq:TC-4box} and the reduced, DAE version \eqref{eq:TC-3box} allow one to reproduce reliably historical globally averaged temperature and carbon fluxes to the ocean and land and are in agreement with IPCC-class ESMs simulations of global averages of these temperature and carbon fields.
Moreover, our model captures two essential, regional mechanisms that drive the tipping from the current, mild climate to a stable hothouse. These two biogeophysical mechanisms are part of the Earth system's response to increased atmospheric GHG concentrations, namely:
\begin{enumerate}[(i)] % \nospace]
	\item the self-sustained rapid darkening of the terrestrial cryosphere due to 
glacial microalgal growth \parencite{Williamson2020,Millar2024,Tedesco2016} that drives the decrease of planetary albedo; and 
    \item the limited ability of land vegetation to adapt to the higher temperatures that will prevail,\parencite{Allen2010,Xu_2020,Hammond_2022}, and that will decrease therewith its carbon absorbing capacity. 
\end{enumerate} 
The way that these mechanisms enter equations ~\eqref{eq:TempEq} or \eqref{eq:TempEq3} for mechanism (i) and ~\eqref{eq:EnvDegr} or \eqref{eq:EnvDegr3} for mechanism (ii), respectively, is described in the Methods section.
	
\section{Results}	
\label{sec:result}

The predictive robustness of the proposed TCV model is first tested by comparing its simulations with historical data, as well as with the simulations of IPCC-class GCMs subject to several anthropogenic emissions scenarios (Fig.~\ref{Fig:IPCC_model_comparison}). The good approximation by TCV model simulations of observations and IPCC-class model simulations allows one to validate its soundness and usefulness in studying how the addition of a feedback process not currently included in IPCC-class models might affect global temperature--carbon dynamics (Fig.~\ref{Fig:Terrestrial-Albedo-Darkening}). Here, we integrate in the TCV model the effect of terrestrial darkening due to mechanism (i) above that is currently not included in ESMs \parencite{IPCC.2021}.

Subsequently, we show that the TCV model, has two distinct stable steady states, and illustrate how this leads to tipping when it is forced with increased anthropogenic carbon emissions such as in Representative Carbon Pathway (RCP) 8.5 scenario.
The coupled DAE model of system~\eqref{eq:TC-3box} above is simple enough to analyze graphically, yet includes the just described physical and biogeochemical processes that operate at the decadal timescales of interest in adaptation to and mitigation of climate change \parencite{IPCC.2021, Lackner.ea.2022}. In the absence of time-dependent forcing, the model exhibits a single stable fixed point $P_1$ that corresponds to the present climate. The presence of anthropogenic forcing can then give rise to an additional fixed point $P_2$, which corresponds to a stable hothouse that lies about 10~K above the present global temperature; the two points $P_1$ and $P_2$ lie at the left- and rightmost intersection of the two nullclines in Fig.~\ref{Fig: Nullclines-model5}. 
	
The transitions between these two stable steady states, the current climate and the hothouse, will be studied next, in connection with Fig.~\ref{Fig: Ca-bifurcation}. In spite of its simplicity, when forced with several  RCPs, our 0-D TCV model produces results that are close to three-dimensional GCMs and ESMs, as shown in section ``Time evolution of the model under anthropogenic forcing” below. Moreover, by including the representation of ecosystem loss in equations~\eqref{eq:EnvDegr} or \eqref{eq:EnvDegr3}, the phenomenon of climate commitment \parencite{Held2010, Pierrehumbert2014} --- the temperature warming locked in without any further anthropogenic forcing ---  is likewise captured, as seen in Fig.~\ref{Fig:TempCarb_Evolution} .
	
%Thus, on the one hand, our model's parsimony and flexibility should allow its use in systematic exercises of quantifying climatic uncertainties to help optimize future economic evolution of our societies \parencite{Barnett2021,Lackner.ea.2022}; on the other, and in spite of this parsimony, the model can mimick instrumental observational data as well as features of IPCC-class models in its simulations of future climate.
	
	\subsection*{TCV model evaluation vs. observations and high-end model simulations} 
	
	The parsimonious formulation of our TCV model relies on functional forms for the processes included that are derived from validated representations of each relevant field, as shown in the Methods section. The model's realism is supported by the choice of model parameters values. All of them are extracted from refereed publications based on detailed observational studies or on high-end model simulations. The corresponding references are listed in Table~\ref{tab:param-temp} for the energy balance equation \eqref{eq:TempEq}, in Table~\ref{tab:param-terrestrial-carbon} for the flux of carbon into the terrestrial carbon sink in Eq.~\eqref{eq:CarbAt} and environmental degradation in Eq.~\eqref{eq:EnvDegr}, and in Table  \ref{tab:param-ocean-carbon} for the flux of carbon into the ocean carbon sink of \eqref{eq:CarbOc}. All the parameter values in this paper lie within the corresponding confidence intervals reported in the references cited in these tables. 
	
\begin{figure}[H] %[!ht]
		
		\centering
		\includegraphics[scale=0.3]{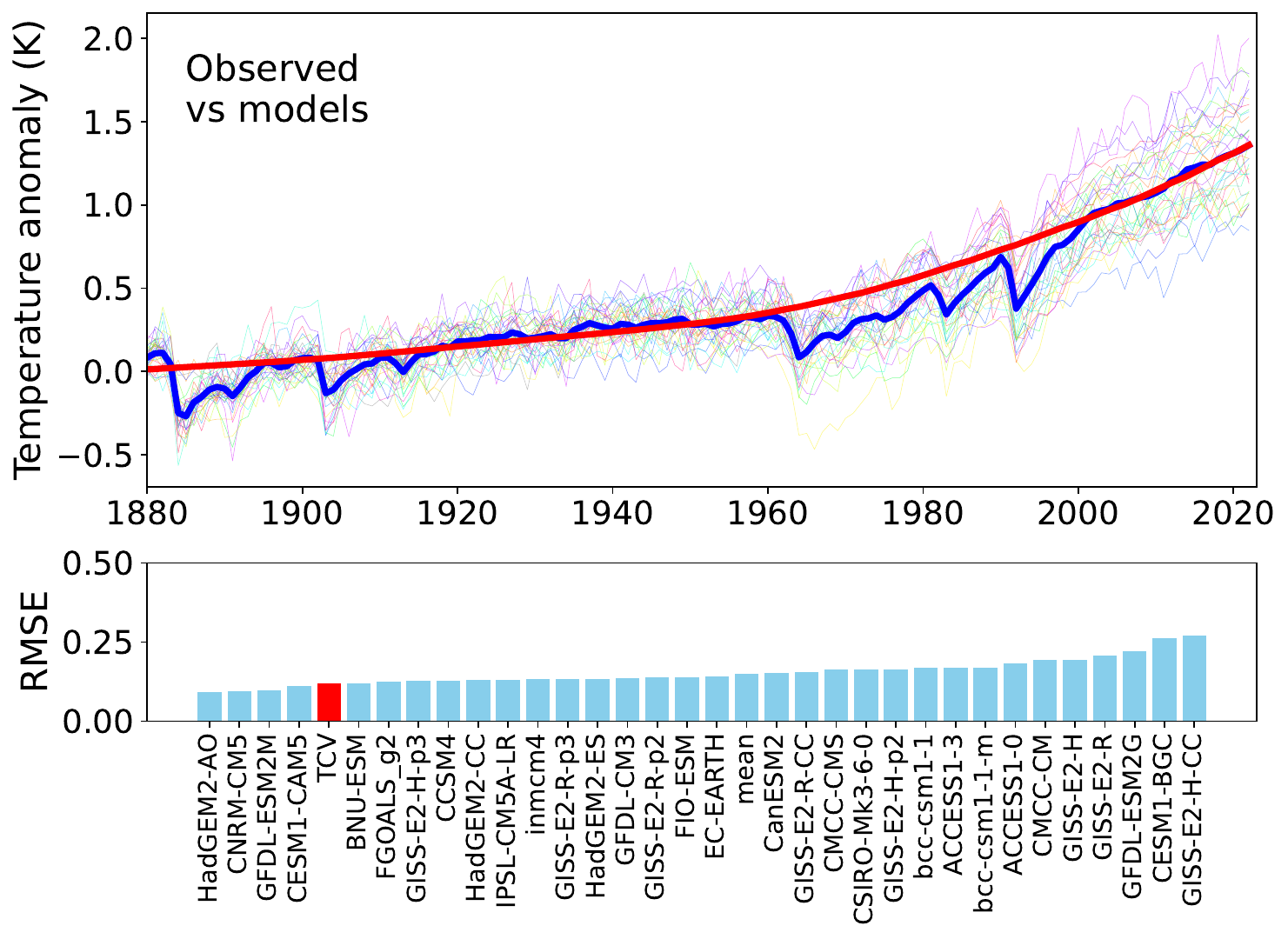} % RCP 26
		
		%%% middle panels %%%
		
		\centering
		\includegraphics[scale=0.3]{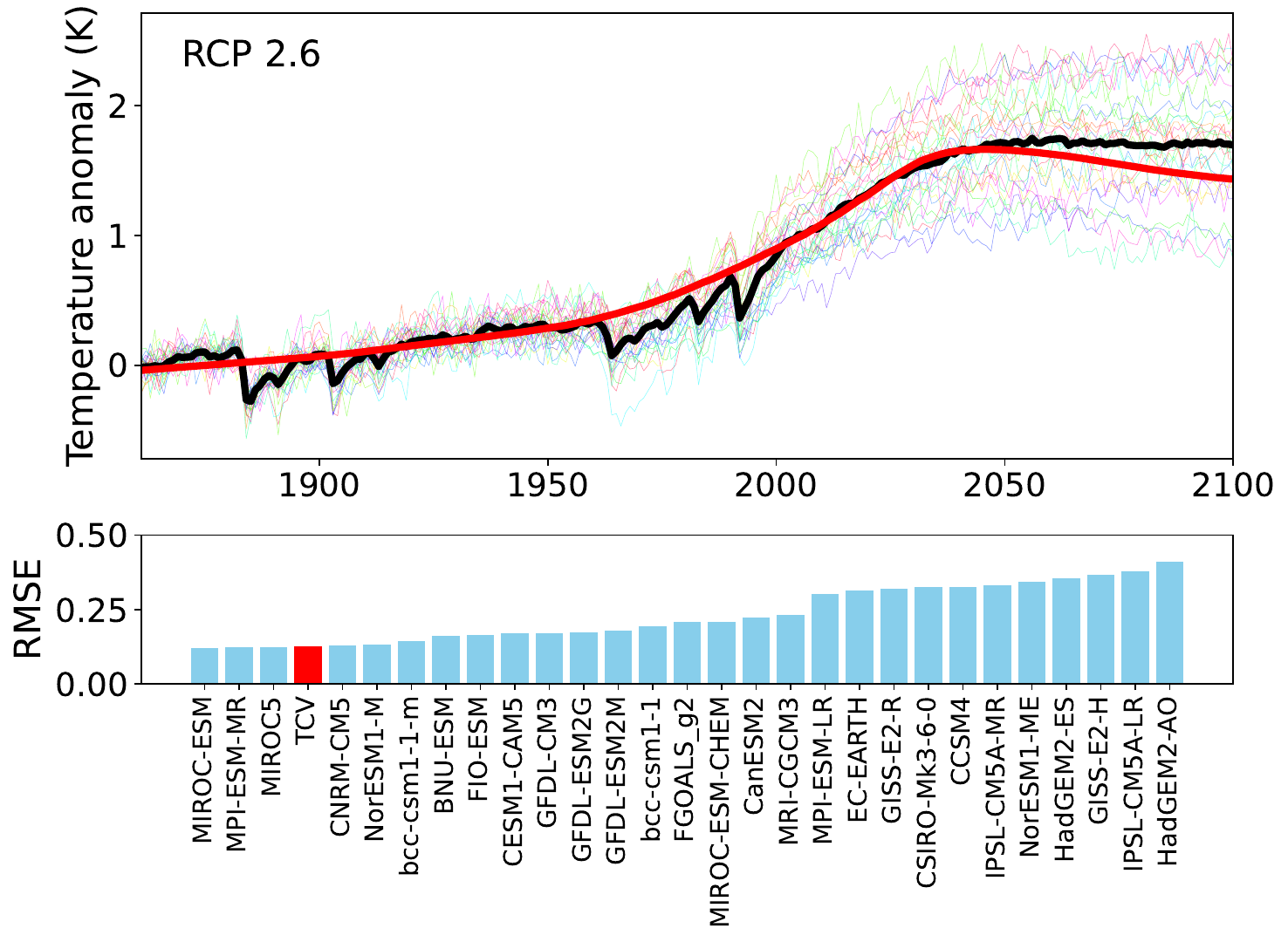} % RCP 26
		\includegraphics[scale=0.3]{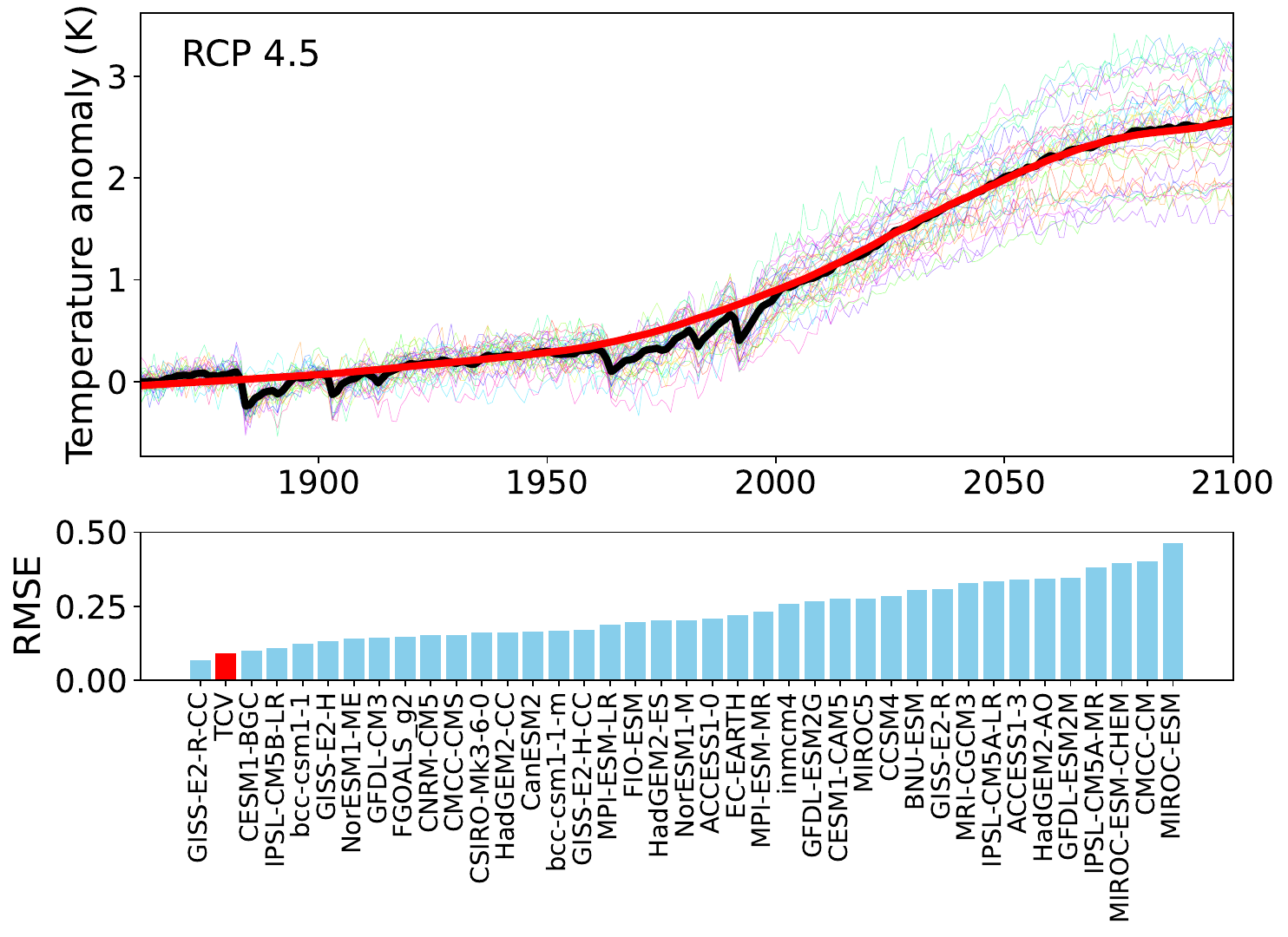} % RCP 45
		
		%%% middle panels %%%
		
		\includegraphics[scale=0.3]{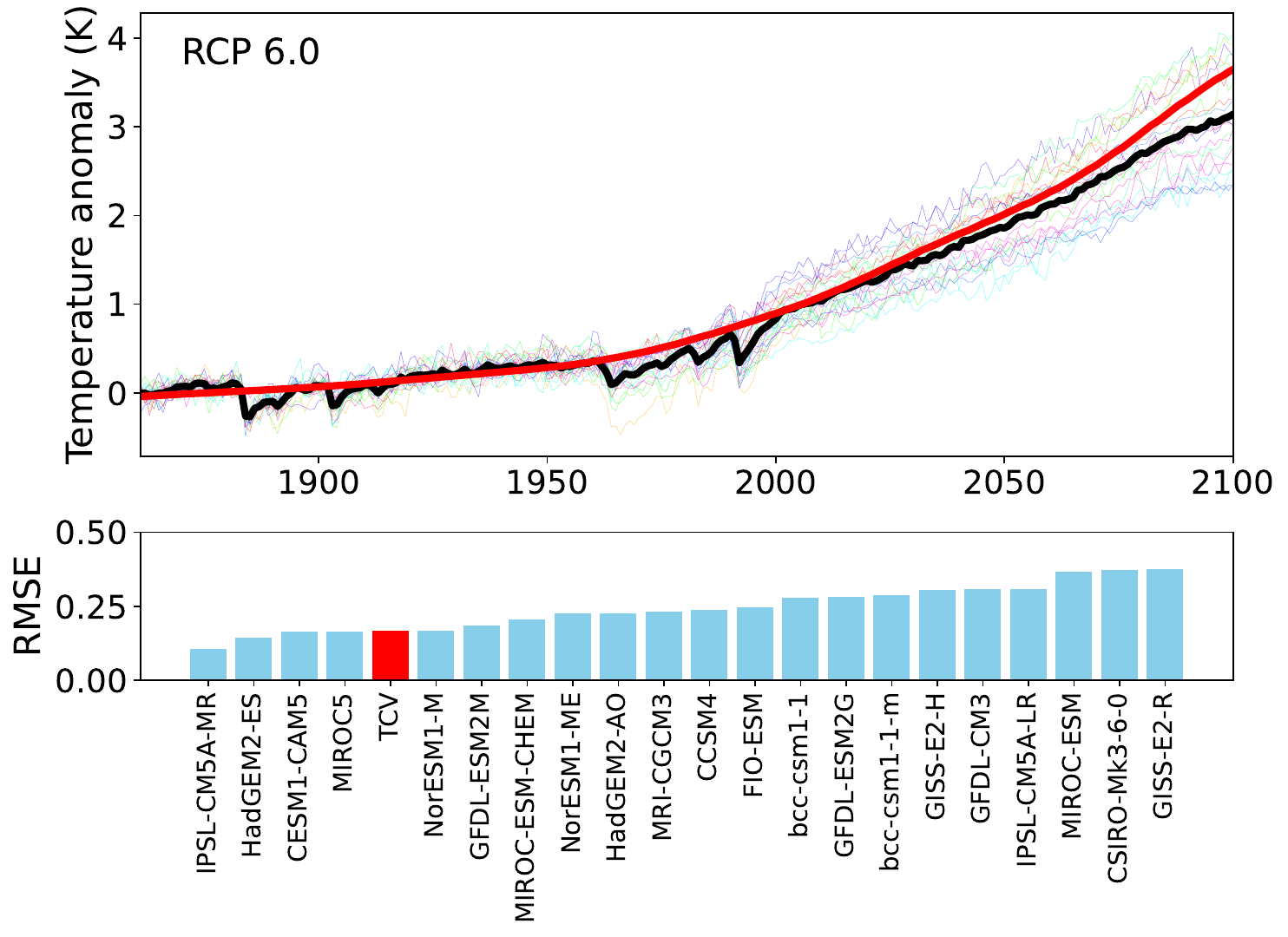} % RCP 60r
		\includegraphics[scale=0.3]{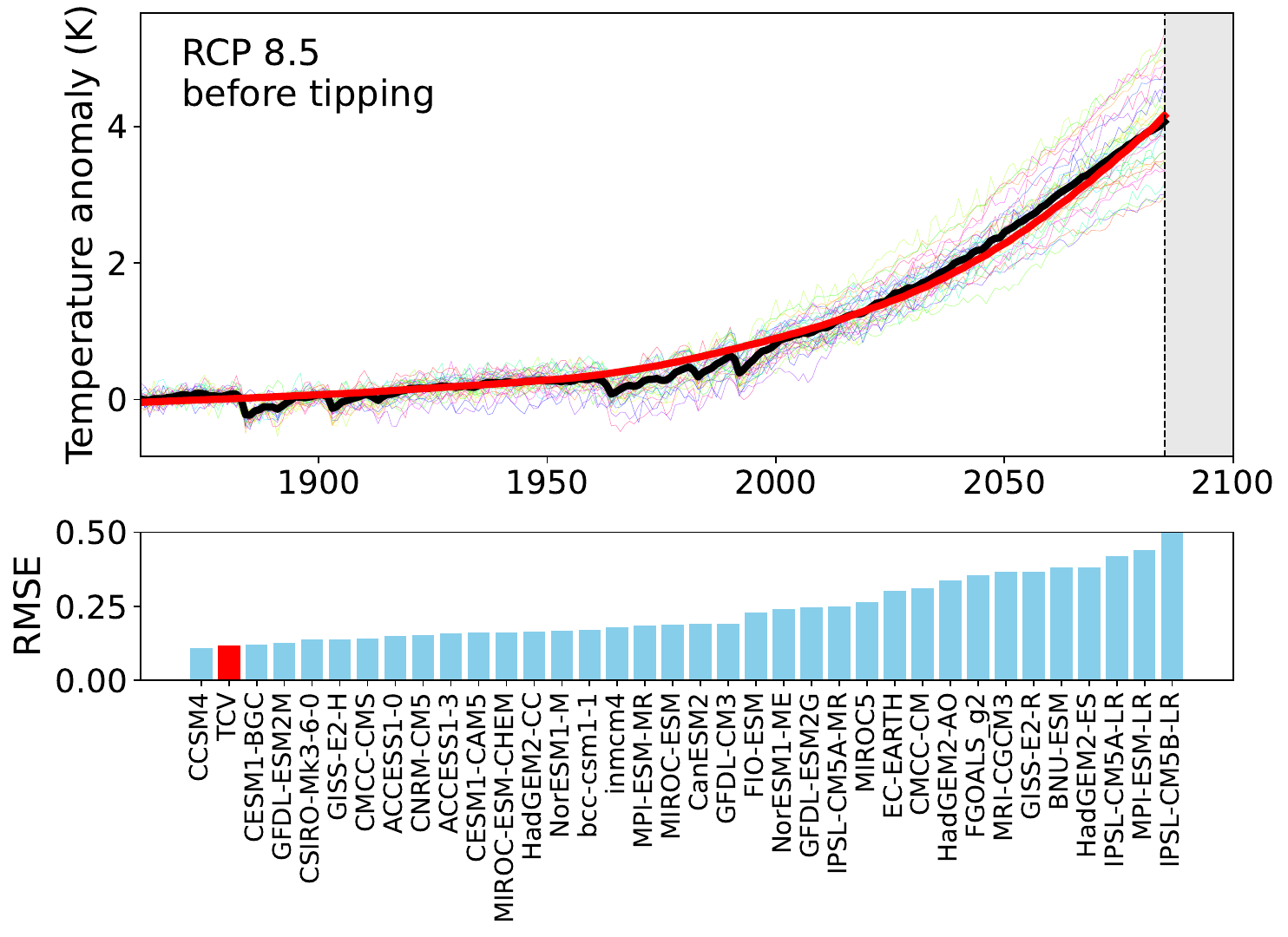} % RCP 85
		
		%%% Adjusting space before the lower panel %%%
		\vspace{0mm} % Adjust the value as needed to reduce space
		
		%%% lower panel %%%
		\centering
		\includegraphics[scale=0.65]{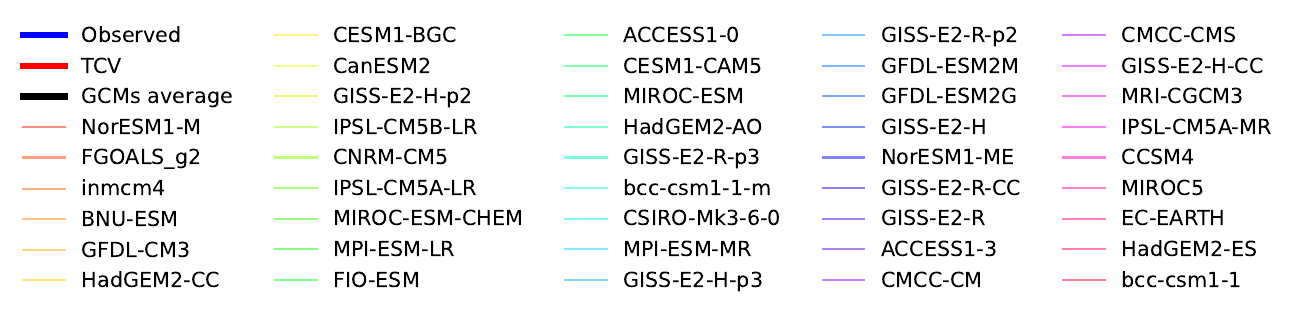} % IPCC Models
		
		\caption{{\footnotesize{\bf TCV model validation with respect to observational data and IPCC-class GCM simulations.} In each panel, the TCV model's temperature simulation (heavy red line) is compared with observational data (heavy blue line), as well as with 42 different IPCC-class GCM models (light colors) subject to four RCP scenarios \parencite{Taylor2012} and with their ensemble mean (heavy black line). The TCV model is benchmarked against individual GCMs using its RMSE  with respect to either (a) observed data or (b--e) the GCM ensemble mean}. }
		\label{Fig:IPCC_model_comparison} % label of the full figure
	\end{figure}

	The accuracy and robustness of the model so obtained is evaluated by comparing its simulations with both observational data and IPCC-class GCM simulations subject to several RCP carbon emission scenarios. Figure~\ref{Fig:IPCC_model_comparison} displays the comparison of the TCV model (in red) with globally and annually averaged observed temperatures (blue) from year 1880 to year 2022 in panel (a); in panels (b)--(d) with the ensemble mean (black) of up to 42 different GCMs subject to RCP 2.6, 4.5, and 6.0, from 1861 to 2100; and in panel (e) for RCP 8.5 up to the point of the global tipping shown in Fig.~\ref{fig:TempEvolRCP85} of the next subsection. 
	
	The difference between the TCV model and  the observed temperatures or the ensemble mean of the 42 GCMs is measured by the root-mean-square error (RMSE). This measure is used to compare the performance of each GCM against observations as well. 
	The TCV model has the 7$^{\rm th}$ lowest RMSE amongst the 42 GCMs when compared with the observed instrumental temperatures. It ranks either as highly as an average GCM or better when compared with the ensemble mean for each of the four RCP scenarios. The Bland-Altman plots \parencite{Bland+A.1999} show that median difference between the TCV model's simulated temperatures and the global mean observational data is lower than 0.1~K; see Fig.~\ref{Fig:Bland-Altman}. Moreover, the Bland-Altman scatter plots in the figure show that the TCV model's global temperature estimates are in agreement with those of the set of the 42 IPCC-class GCMs.

	The carbon fluxes of the TCV model are assessed by comparing them with global estimates from 1959 to 2021 obtained by high-end carbon models included in the Global Carbon Project (GCP) \parencite{Friedlingstein_2022}: 11 models for the terrestrial fluxes and 16 for the ocean fluxes; see Fig.~\ref{Fig:Carbon_robustness_test}. Panels (a) and (b) clearly show the large scatter of GCP model simulations, for both ocean and land fluxes, respectively. In panels (c) and (d), we consider more systematically this scatter and the TCV model performance. The mean of the differences at each point in time between the largest and the smallest fluxes in the GCP models equals 0.85 GtC/yr for the ocean fluxes and 2.83 GtC/yr for the terrestrial fluxes (dashed red line in the panels) vs. the RMSE between the TCV model and GCP models, which is below 0.5 GtC/yr for half the ocean fluxes --- namely for the CNRM, MPI, IPSL, Princeton, and CESM-ETHZ models, see Fig.~\ref{Fig:Carbon_robustness_test}(e) --- and equal or less than 1.6 GtC/yr for 6 of 16 of the land flux models --- namely OCNv2, CLM5.9, JSBACH, ISAM, CABLE-POP, and VISIT models; see Fig.~\ref{Fig:Carbon_robustness_test}(f).
	
	Therefore, the TCV model's simulations and the results of their validation confirm the robustness of its underlying representation of the different biogeochemical processes included in the model. Further statistical validation tests are provided in the Supplementary Material on ``TCV model evaluation: further tests.'' 
	
\begin{figure}[ht!]
		\subfigure[Land albedo for current climate equlibrium]{ \label{fig:TerrAlb_NoMask}
			% label of upper left panel    
			\centering \includegraphics[scale=0.3]{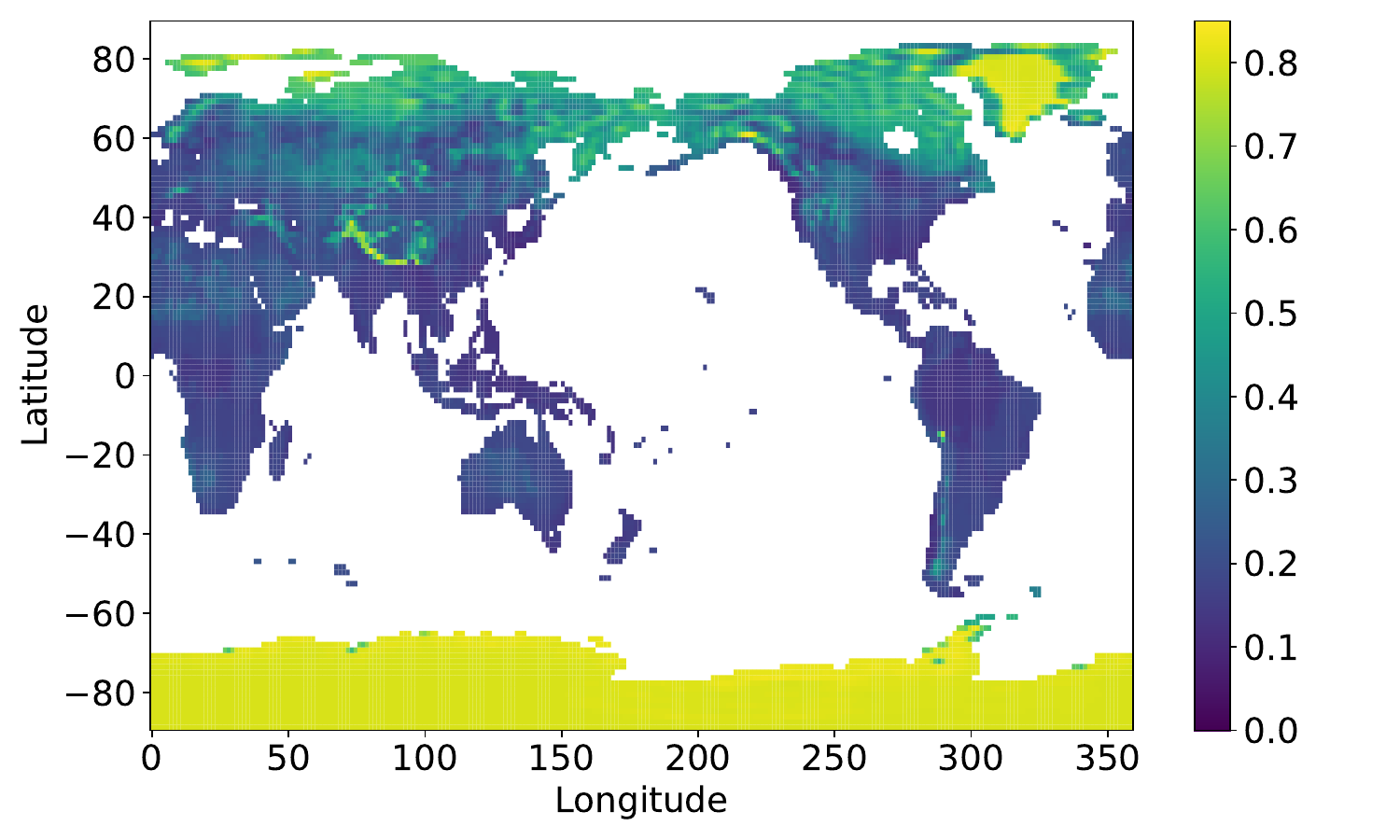}
		} \subfigure[Land albedo after algal darkening]{ \label{fig:TerrAlb_WithMask}
			% label of upper right panel    
			\centering \includegraphics[scale=0.3]{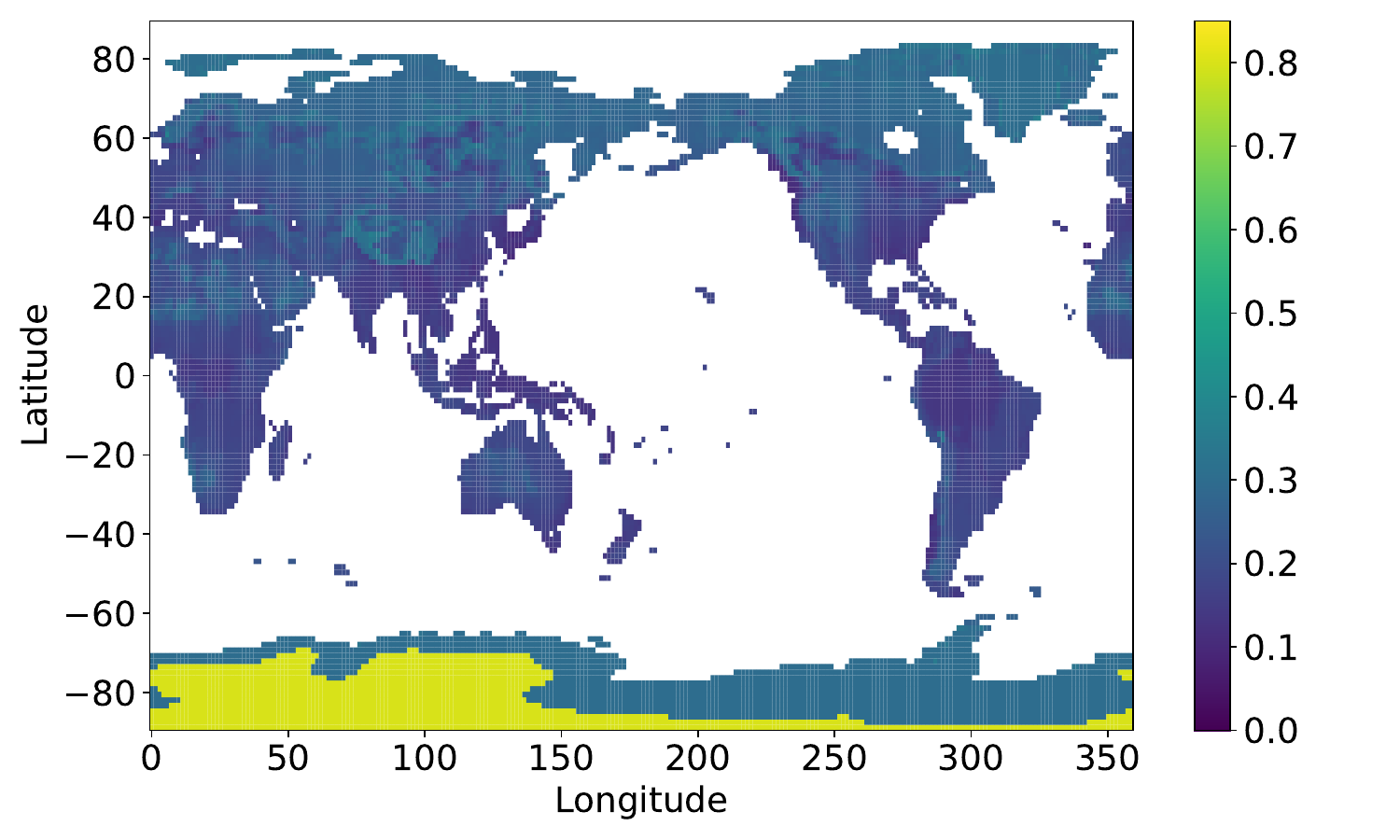}
		}
		
		\caption{{\footnotesize{\bf Global terrestrial surface albedo darkening.}
				Plots of climatological terrestrial albedo with and without algal darkening of ice sheets and midlatitude cryosphere. Grid-level values were computed using raw data from simulations of the CNRM-CM5 climate model coupled with the ISBA-ES snowpack dynamics and hydrology model \parencite{Boone2001,Boone2010} for the 2070--2100 interval, subject to the RCP 8.5 emission scenario. Panel (a) displays grid-level annual climatological terrestrial albedo with no algal darkening; and panel (b) displays grid-level albedo with maximum darkening of land cryosphere due to algal blooms. The minimum and maximum global average terrestrial albedo values used in $\alpha_{{\rm{L}}}(T)$ of Eq.~{\ref{eq:TempEq}} (see Eq.~\ref{eq:alb_terrestrial} in Methods) are obtained after performing a weighted spatial average of gridded data in (a) and (b) accounting for both the latitudinal gradient of annual cumulative radiation and the latitude dependence of gridded areas; see Fig.~ \ref{Fig:Terrestrial-Albedo-Weights} and Methods section.  
		}}
		\label{Fig:Terrestrial-Albedo-Darkening} % label of the full figure
	\end{figure}

	Based on these results, we proceed to assess the effect on global temperature-carbon dynamics of the inclusion of potential terrestrial albedo darkening by glacial algal blooms \parencite{Williamson2019,Williamson2020,Yallop2012} on the land cryosphere, a mechanism that is not currently included in IPCC-class GCMs. To do so, the terrestrial albedo darkening is evaluated based on the use of output from simulations of the CNRM-CM5 climate model coupled with the ISBA-ES snowpack dynamics and hydrology model \parencite{Boone2001,Boone2010}. Figure~\ref{Fig:Terrestrial-Albedo-Darkening} displays the grid-level surface albedo prior to  algal darkening (Fig.~\ref{fig:TerrAlb_NoMask}) and following maximum darkening (Fig.~\ref{fig:TerrAlb_WithMask}). The globally averaged value is computed while accounting for latitudinal gradients (see Fig.~\ref{Fig:Terrestrial-Albedo-Weights}) of annual cumulative radiation and the latitude dependence of gridded areas in order to parametrize the $\alpha_{{\rm max}}^{{\rm L}}$ (Fig.~\ref{fig:TerrAlb_NoMask}) and $\alpha_{{\rm min}}^{{\rm L}}$ (Fig.~\ref{fig:TerrAlb_WithMask}) values of the temperature-dependent terrestrial albedo $\alpha_{{\rm{L}}}(T)$ function (see Eq.~\eqref{eq:alb_terrestrial}) of Eq.~\eqref{eq:TempEq}. Further details are provided in the Methods section.
	
	\subsection*{The TCV-DAE model's two stable steady states} 
	
	\label{ssec:steady}
	
		\begin{figure}[!hb]
			
			\centering \subfigure[Phase portrait for $e=0$~GtC/yr and $V=0$]{ \label{fig:Nullclines-modelx}
			% label of lower left panel
			\centering \includegraphics[scale=0.35]{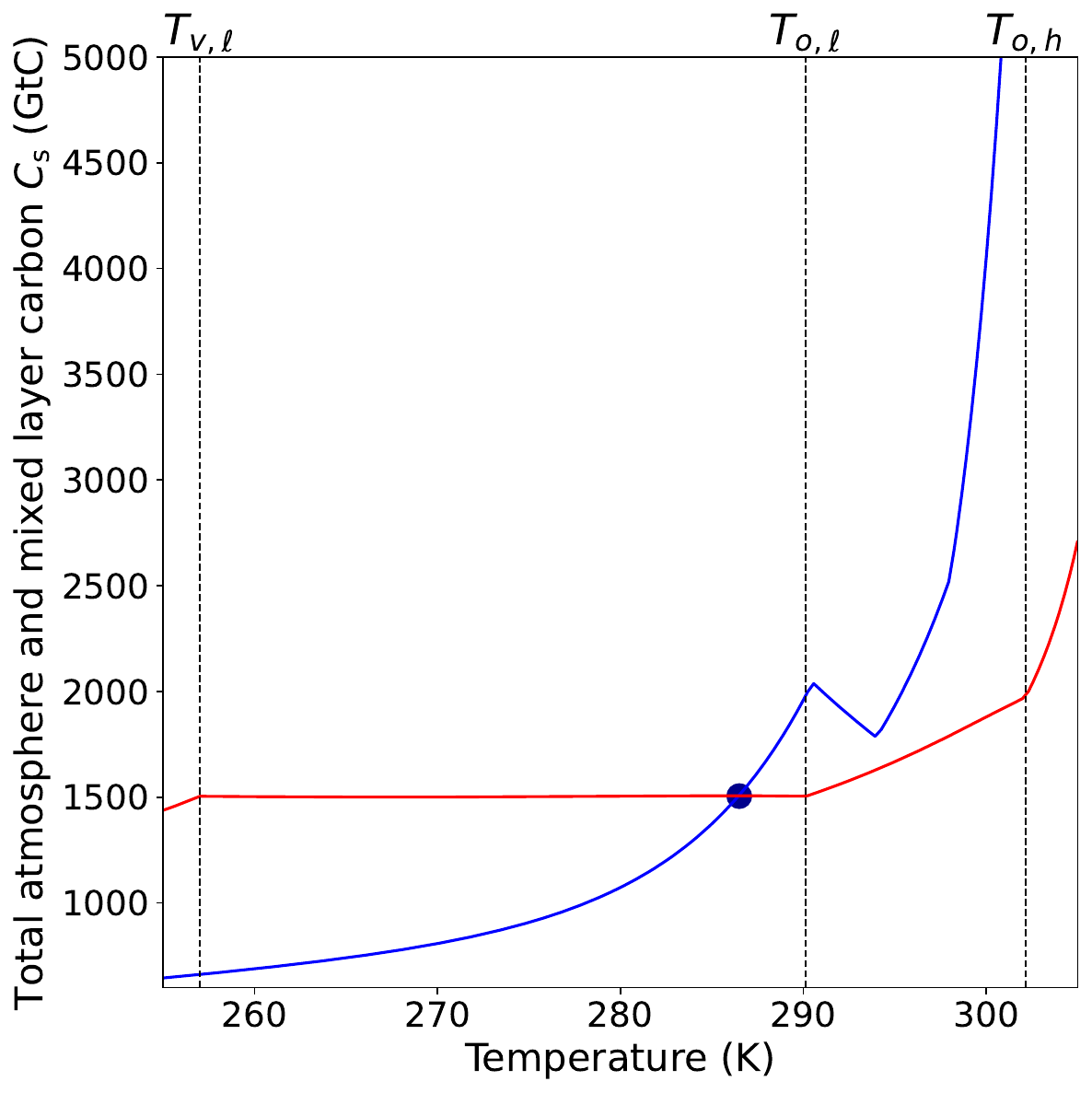}
		}
		%%%%%    lower panels     %%%%%%%%%%%%%%%%
		
		\subfigure[Phase portrait for $e=21$~GtC/yr and $V=0$.]{ \label{fig:Nullclines-modelx-emissions}
			% label of upper left panel    
			\centering \includegraphics[scale=0.35]{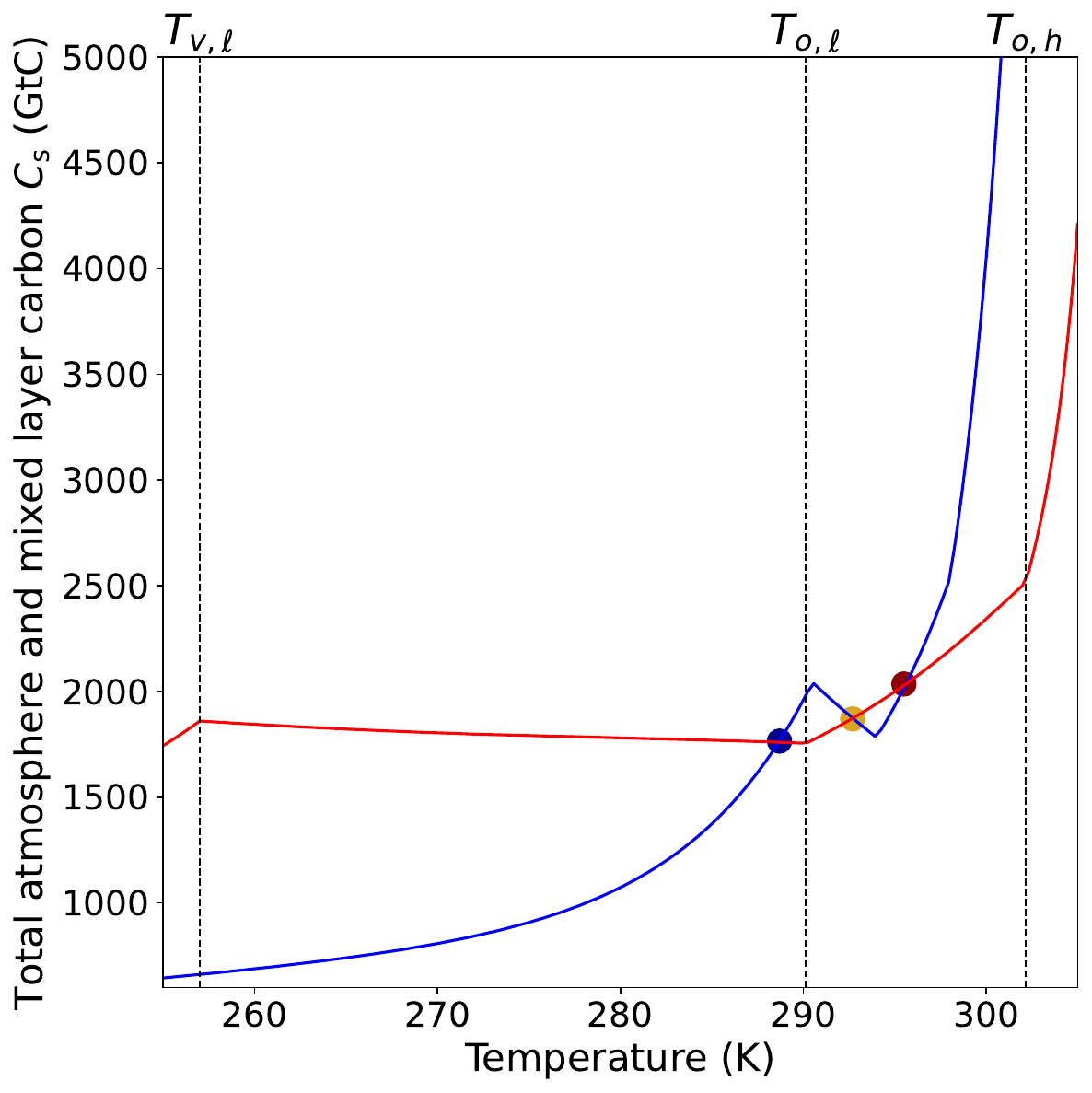}
		} \subfigure[Blow-up of (b) for $e=21$~GtC/yr and $V=0$.]{ \label{fig:Nullcline zoom-emissions-modelx}
			% label of upper right panel    
			\centering \includegraphics[scale=0.35]{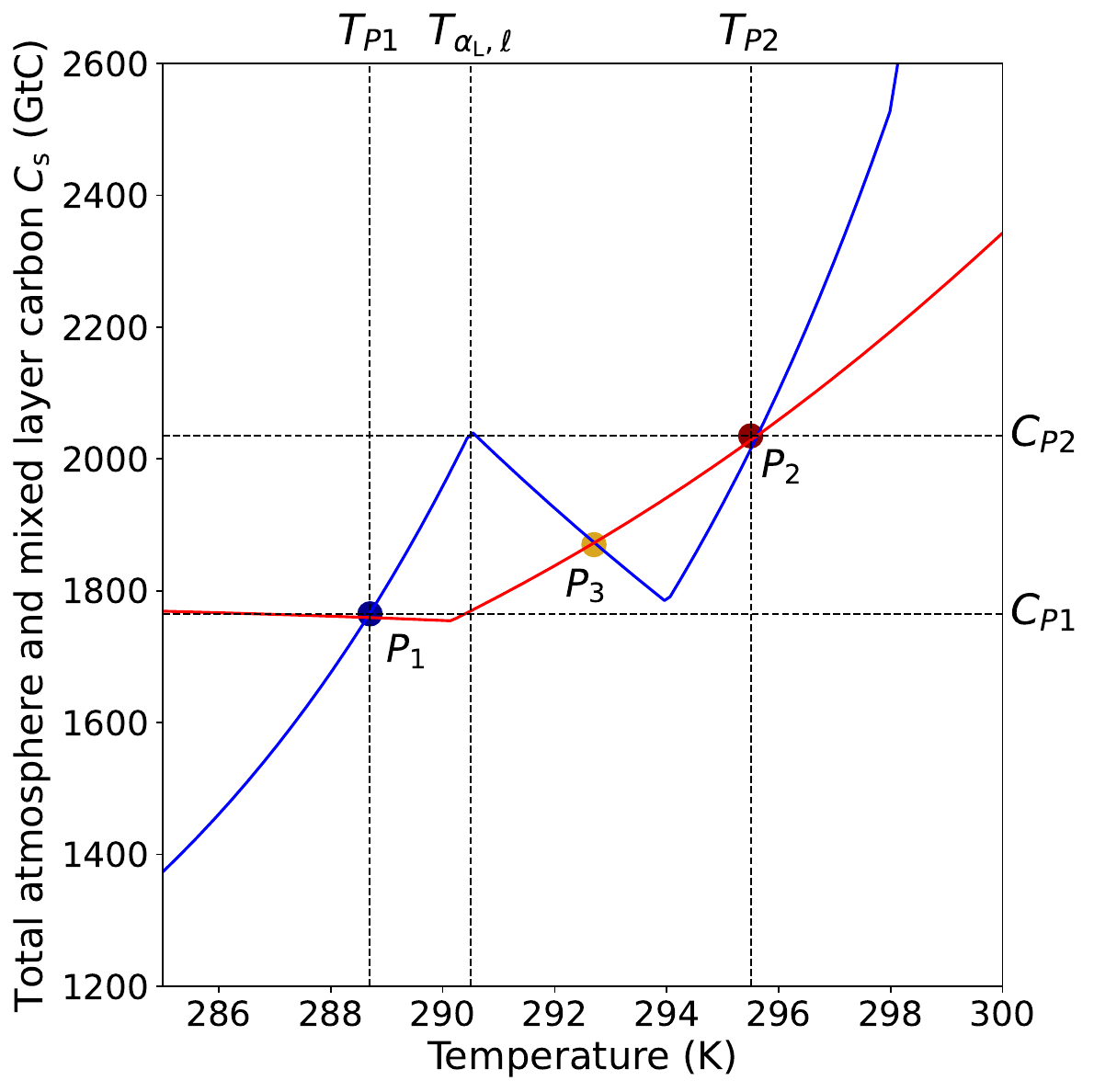}
		}

		\caption{{\footnotesize{}{}{}{\bf Phase portrait of the coupled DAE model \eqref{eq:TC-3box} for two emissions levels, $e = 0$ and $e = 21$, and constant ecosystem degradation level, $V = 0$.} The coordinates are global temperature $T$ in degrees K on the abscissa and total surface carbon $C_{\mathrm{S}}$ in GtC on the ordinate. 
				The temperature % (Eq.~\eqref{eq:TempEq3}) 
				and carbon % (Eq.~\eqref{eq:TotalCarb}) equations 
				nullclines $F_1(T, C_{\mathrm{S}})=0$ and $F_2(T, C_{\mathrm{S}})=0$ that correspond to equations
				\eqref{eq:TempEq3} and  \eqref{eq:TotalCarb}
				are in blue and red, respectively. 
				(a) The phase portrait for preindustrial conditions;  
				(b) the phase portrait for $e = 21$~GtC/yr that illustrates the model's three steady states; and 
				(c) blow-up of area in panel (b) that contains the model's three steady states. 
				The blue-filled circle indicates the stable current climate 
				$P_1$ (left) and the red-filled circle corresponds to the hothouse climate $P_2$ (right), with the unstable fixed point $P_3$ in orange in-between. 
				The light vertical line in all three panels indicates a discontinuity in the derivative of $\alpha_{{\rm {O}}}$ at $T_{\alpha, \ell} = 290$~K that is explained in the Methods section; in panel (c) the temperatures $T$ and carbon stock $C_{\rm S}$ of the two stable steady states are also indicated by light dashed lines. 
%				$e = 21$~GCt/yr that is only reached in RCP8.5 is chosen for panel (b) for illustrative purposes and showcase the coexistence of three steady states. As shown in Fig.~\ref{Fig: Ca-bifurcation} panel (a), we show that for $V=0$ in fact three steady states coexist from $e = 16$~GtC/yr. 
		}}
		\label{Fig: Nullclines-model5} % label of the full figure
	\end{figure}
	
	In the absence of anthropogenic GHG emissions,  i.e., with $e=0$, and for the chosen parameter values in Tables~\ref{tab:param-temp}--\ref{tab:param-ocean-carbon}, the TCV model has only one stable steady state corresponding to our current climate's equilibrium plotted as a blue-filled circle in  Fig. \ref{fig:Nullclines-modelx}. When the level of atmospheric carbon $e$ is increased by anthropogenic annual emissions to a value above 16 GtC/yr, while still using $V = 0$, two stable steady states, separated by an unstable one, co-exist. The two stable steady states are plotted as filled blue and red circles, respectively,
	in Figs.~\ref{fig:Nullclines-modelx-emissions} and \ref{fig:Nullcline zoom-emissions-modelx}. 
%	for $e=21$ and $V=0$.
	
	The present climate is denoted henceforth in the text by $P_{1}=(T_{1},C_{1})=(286.5$~K,
	1,505 GtC) = (13.4 ${^{\circ}}$C, 284~ppm); see Fig. \ref{fig:Nullclines-modelx}. The stable hothouse steady state shown in \ref{fig:Nullcline zoom-emissions-modelx} for $e=21$~GtC/yr and $V=0$ appears for higher levels of carbon emissions and is denoted by $P_{2} = (T_{2},C_{2}) = (295.8$~K, 2,071~GtC) = (22.7 ${^{\circ}}$C, 617~ppm). % and it has real negative eigenvalues 
	 The eigenvalues associated with the linearization of the TCV-DAE model at these two fixed points are all real and negative: for $P_1$ they are given by $(\lambda_{1,P1}=-0.11, \lambda_{2,P1}=-0.07)$, while for $P_2$ they are $(\lambda_{1,P2}=-0.14, \lambda_{2,P2}=-0.05)$. 

	Note that the nullclines in Fig.~\ref{Fig: Nullclines-model5} have discontinuities in the derivatives, because of the piecewise linear character of several terms in the right-hand sides of the ODEs; see Methods section. This property does not invalidate in any way the conclusions we draw from the figure. In particular, all the fixed points lie at the intersections of nullcline segments that are continuously differentiable. Hence the positions of these points in the model's phase-parameter space depend also smoothly on the model's parameters, at all but the bifurcation points themselves, where the one-sided derivatives exist but are unequal between one side and the other.
	
	The unstable fixed point $P_{3}=(T_{3},C_{3})$ that separates  $P_{1}$ from  $P_{2}$ is marked by an orange-filled circle in both panels.
	The corresponding eigenvalues are one positive and one negative: 
	at $e = 21$ GtC/yr they are $(\lambda_{1,P3} = + 0.20, \lambda_{2,P3} = - 0.10)$.

\begin{figure}[!ht]
		\centering \subfigure[Bifurcation diagram with $V=0$]{ \label{fig: BD_K0.0}
			% label of upper left panel    
			\centering \includegraphics[scale=0.35]{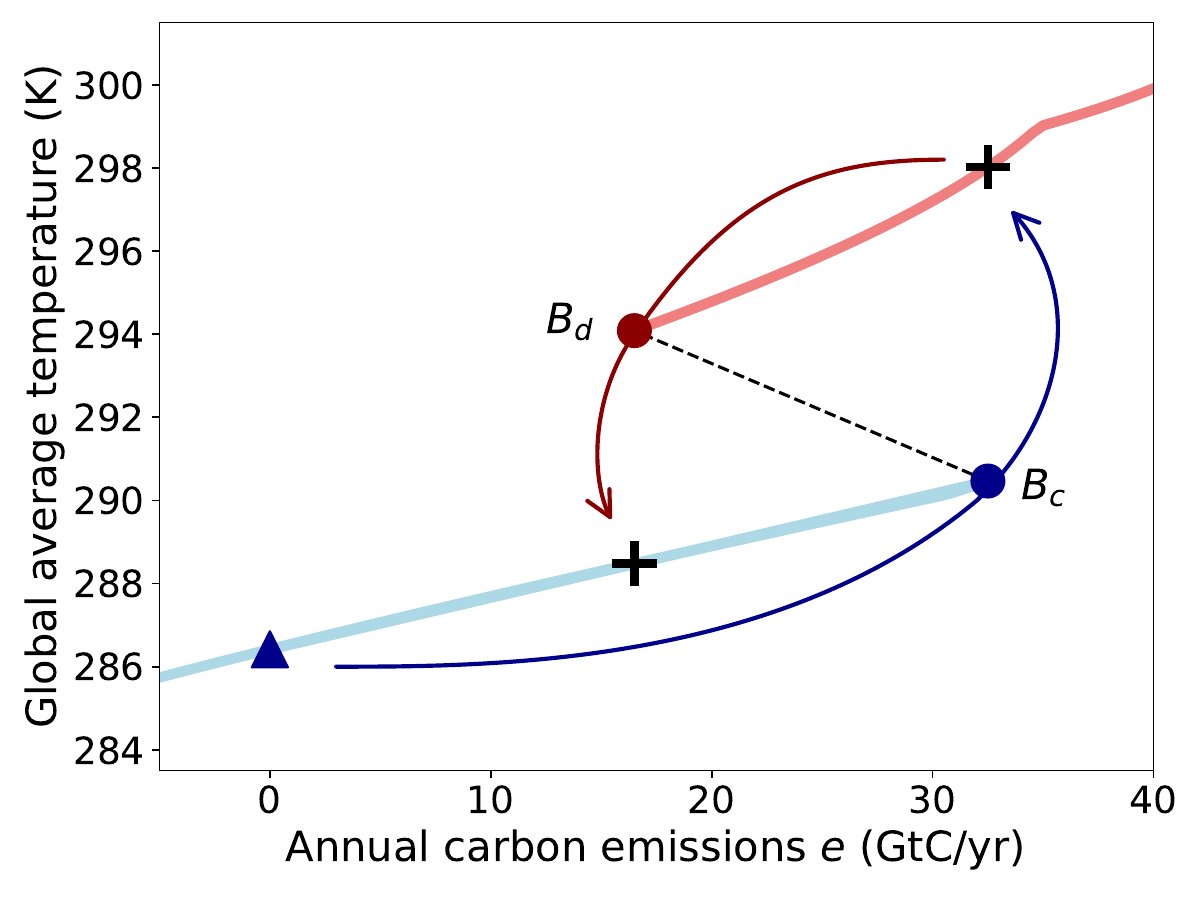}
		} \subfigure[Bifurcation diagram with $V=0.3$]{ \label{fig: BD_K0.5}
			% label of upper right panel    
			\centering \includegraphics[scale=0.35]{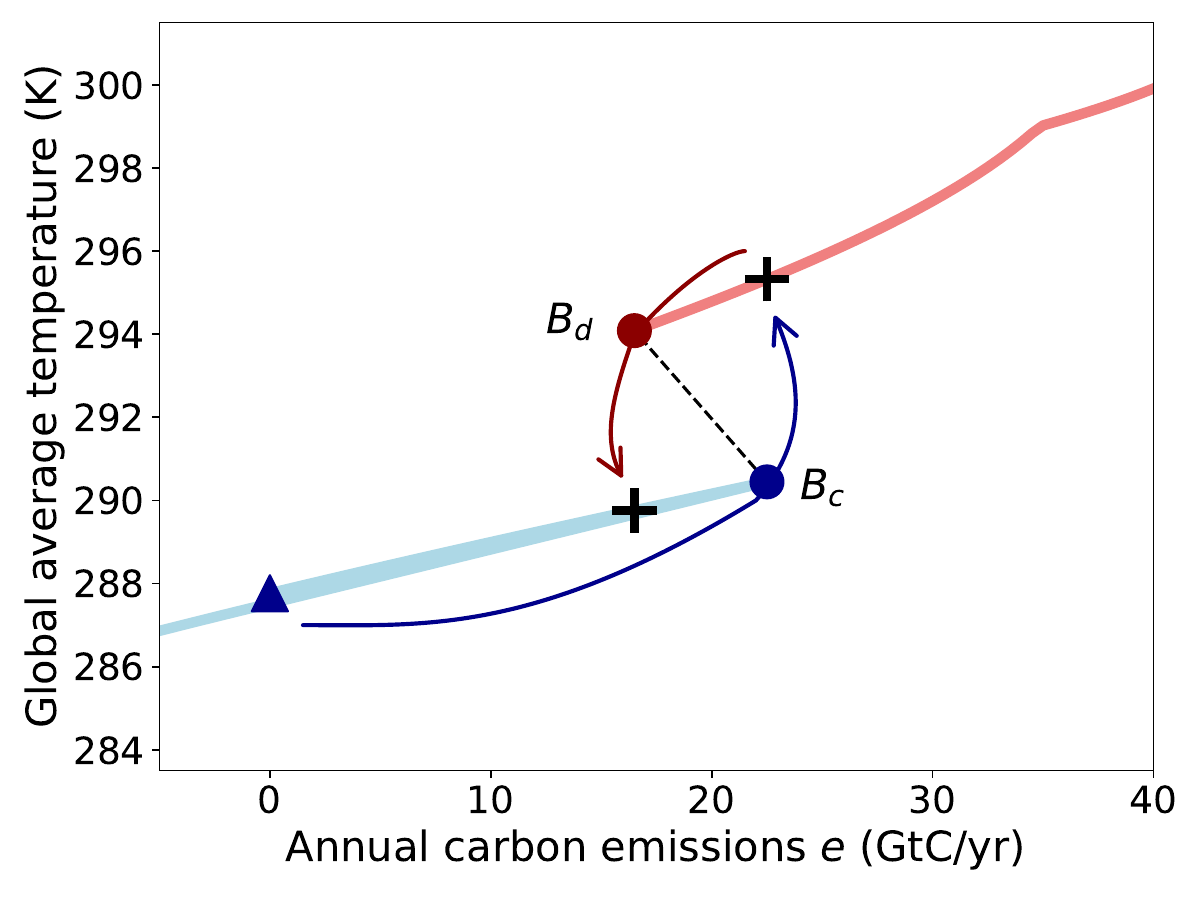}
		}
		
		\caption{{\footnotesize{}{}{}{}{\bf Bifurcation diagrams of the TCV-DAE model with fixed vegetation degradation $V$. }  The two diagrams show the steady states' temperature $T$ as a function of the yearly carbon emissions $e$: (a) for intact ecosystems, $V=0$; and (b) for nonzero ecosystem degradation, $V=0.3$.
			 In both panels, the stable branches are light blue for the current climate and light red for the hothouse; the intermediate, unstable solution branch is dashed black. 
			 The dark blue triangle indicates the preindustrial steady state $P_1$ for the standard 
			 parameter values of Tables~\ref{tab:param-temp}, \ref{tab:param-terrestrial-carbon}, and \ref{tab:param-ocean-carbon}.
			 The bifurcation point of the current climate $P_1$ and intermediate branch $P_3$ is shown as a filled blue circle $B_{\rm c}$ with coordinates $(T_c = 290$~K, $e_c = 31.5$~GtC/yr) for $V = 0$ and $(T_c = 290$~K, $e_c = 21.5)$~GtC/yr for $V = 0.3$, while the other bifurcation point $B_{d}$, shown as a filled red circle, has coordinates $(T_d = 294.2$~K, $e_d = 16.5$~GtC/yr) for both $V=0$ and $V=0.3$. 
			}} 
		\label{Fig: Ca-bifurcation} % label of the full figure
\end{figure}

The yearly carbon equivalent flux of  	
$e = 21$~GCt/yr that is only reached in RCP8.5 was chosen for panels (b) and (c) of Fig.~\ref{Fig: Nullclines-model5} for illustrative purposes, to showcase the coexistence of three steady states. As shown hereafter in Fig.~\ref{Fig: Ca-bifurcation}(a), even for $V=0$ three steady states coexist already from $e = 16$~GtC/yr on. 

	A more complete understanding of the model behavior's parameter dependence is obtained by considering the bifurcation diagrams in Fig.~\ref{Fig: Ca-bifurcation}.
	These show the temperature of the model's steady states and their stability as a function of $e$, considered here as a parameter with prescribed values; $e = 0$ in the diagrams corresponds to the preindustrial climate state, with $T = 286.5$~K \cite{Hawkins2017}.
	
	In Fig.~\ref{Fig: Ca-bifurcation}, we concentrate on the dependence of the current climate $P_1$ and the hothouse $P_2$ on the annual GHG emissions $e$, and plot $T_j = T_j(e)$ for
	${j=1, 2, 3},$ shown as solid light blue and light red lines for the two stable steady states, $j=1,2$, respectively. The unstable intermediate branch that corresponds to $j = 3$ is dashed and black. 
%	We omit the fourth, unstable branch as less relevant and focus on these three branches that correspond 
	
		The bifurcation diagrams herein correspond roughly to the double-fold bifurcation. %displayed in Fig.~\ref{fig:hysteresis}, with the differences explained in the Supplementary Material. 
	%The latter curve is a simplified version of those obtained with full, latitude-dependent EBMs \parencite{Ghil2020, Ghil1976, Heydt2016}.
	The curved arrows --- shown in Fig.~\ref{Fig: Ca-bifurcation} in dark blue for the upward portion of the hysteresis cycle and in dark red for its downward portion --- correspond to the heavy solid segments of the hysteresis cycle.% in Fig.~\ref{fig:hysteresis} that have the same color. 
	The bifurcation points in Fig.~\ref{Fig: Ca-bifurcation} are denoted by $B_c$ and $B_d$, and shown as blue- and red-filled circles, respectively.
	
	As $e$ increases, the steady state corresponding to current climatic conditions disappears at $B_c$, at an emission level $e_c = 31.5$~GtC/yr for $V = 0$, and the system jumps onto the hothouse branch of solutions. Note that, for the more deteriorated land vegetation in  Fig.~\ref{Fig: Ca-bifurcation}(b), this bifurcation is reached at a much lower value of the annual emissions, namely  $e_c = 21.5 < 31.5$~GtC/yr. 
A full return to current conditions once the hothouse state is reached requires a decrease of annual emissions, possibly complemented by new or enhanced carbon sinks, to remove carbon below $e_d = 16.5$~GtC/yr, all the way to the red dots in Figs.~\ref{fig: BD_K0.0} and \ref{fig: BD_K0.5}. This discrepancy between $e$-values for the upward and the downward jump between climate branches is part of the hysteresis cycle associated with the tipping. 

Note that in usual saddle-node bifurcations \parencite{Arnold.ODE.1983,Ghil.Chil.1987,Guckenheimer1983}, where the transition between the stable and the unstable branch is smooth, one of the two eigenvalues of the governing ODE system's linearization goes through zero and changes sign.%; see Fig.~\ref{fig:hysteresis} in the Supplementary Material.
Here, instead, the eigenvalue that distinguishes between the stable nodes all along the branch of the current climate and the saddles all along the unstable branch jumps from a negative to a positive value at the bifurcation point, without crossing zero; see \verb|TCV_video-V=V(t)| video in the Supplementary Material.

The geometry of the nullclines in Figs.~\ref{Fig: Nullclines-model5} and  \verb|TCV_video-V=V(t)| reveals that the merger of the stable and unstable fixed points $P_1$ and $P_3$ occurs when the global temperature $T$ reaches $T_{\alpha_{\rm L}, \ell},$ at which point the  darkening of the cryosphere due to algal blooming starts to increase, causing the percentage of incoming radiation $Q_0$ absorbed by Earth's surface to increase, too. This increase leads to the upward jump of the climate from the current climate branch to the hothouse branch. The leftmost red-filled dot in Fig.~\ref{Fig: Ca-bifurcation}, which occurs at $e_d = 16.5$~GtC/yr for both $V = 0$ and $V = 0.3$, corresponds to the merger of the stable and unstable fixed points $P_2$ and $P_3$ that occurs when the global temperature $T$ reaches $T_{\alpha_{\rm L}, u},$ triggering a jump from the stable hothouse branch of solutions back to the current climate branch.

%\textcolor{blue}{
	%{\mg 
We assessed the robustness of the two–steady‑state regime by using two complementary sensitivity analyses. First, in an one-at-a-time (OAT) sweep, we varied every temperature parameter by $\pm 4$~K and every non‑temperature parameter by $\pm 20\%$ around the values in Tables~\ref{tab:param-temp}, \ref{tab:param-terrestrial-carbon}, and \ref{tab:param-ocean-carbon}; bistability persisted for 37 of 39 parameters, with its sensitivity confined to the terrestrial albedo bounds $\alpha_{\min}^{\mathrm{L}}$ and $\alpha_{\max}^{\mathrm{L}}$. 

Second, we applied a variance‑based Sobol analysis \cite{Sobol2001, Saltelli2007, Jansen1999} to the albedo‑calibration subset given by Eqs.~\eqref{Fig:Terrestrial-Albedo-Darkening} and \eqref{eq:alb_terrestrial}. This analysis addressed the probability of bistability at $V=0$ and showed that $\alpha_{\min}^{\mathrm{L}}$ is the dominant contributor, capturing a ``first-order sensitivity, or main-effect, index" $S_{1} \simeq 0.39$ of the variance on its own, with $\sim 28$–$49$\%, and a ``total-effect index" of $T \simeq 0.79$, including interactions, with $\sim 70$–$89$\%. The second factor, $\alpha_{\max}^{\mathrm{L}}$, exhibits a main effect $S_{1} \simeq 0.16$, with $\sim 9$–$21$\%, and a total effect $T \simeq 0.53$, with $\sim 42$–$63$\%. Fig.~\ref{fig:sobolint} indicates that the interactions-driven sensitivity is driven primarily by pairwise $\alpha_{\min}^{\mathrm L}\!\times\!\alpha_{\max}^{\mathrm L}$ parameters joint variation; the contrast between terrestrial albedo bounds, contributing $\sim 25\%$ of the total $\sim 70$–$89$\%.

The threshold temperatures contribute but little to the sensitivity: for $T_{\alpha_{L},\ell}$, $S_{1} \simeq 0.02$, with $\sim 0$–$5$\%, and $T \simeq 0.20$, with $\sim 15$–$25$\%, while for $T_{\alpha_{L},u}$, $S_{1} \simeq 0$, with $ \simeq 0$–$2$\%, and $T \simeq 0.13$, with $\sim 8$–$18$\%; see Fig.~\ref{fig:sobol}. Thus, the model's albedo bounds, $\alpha_{\min}^{\mathrm{L}}$ and $\alpha_{\max}^{\mathrm{L}}$, themselves exhibit pronounced differences in their main and total effects, $S_{1}$ and $T$, while the temperature thresholds have overall a small impact on the bistability. Together with the OAT results in Fig.~\ref{fig:oat}, this identifies the terrestrial minimum albedo as the primary control on bistability, with a secondary direct role for the maximum albedo.
%}

	\subsection*{Time evolution of the model under anthropogenic forcing}
	\label{ssec:anthro}
	
	So far we considered the behavior of our TCV model for a fixed $e = 0$ and $e = 21$ GtC/yr, as illustrated in Fig.~\ref{Fig: Nullclines-model5}, and for distinct values of $e$, each of which is  fixed in time, as shown in Fig.~\ref{Fig: Ca-bifurcation}. We next subject our coupled TCV-DAE model, using a terrestrial vegetation degradation $V(t)$ that evolves according to equation~\eqref{eq:EnvDegr3}, to anthropogenic forcing that corresponds to a prescribed function $e = e(t)$. We rely here on the RCP scenarios of GHG emissions, measured in carbon-equivalent concentrations, which were used in the most recent IPCC assessment reports \parencite{Meinshausen2011}.
	
\begin{figure}[!ht]
		\subfigure[Global average temperature]{ \label{Fig. temperature dynamics}
			% label of upper left panel    
			\centering \includegraphics[scale=0.315]{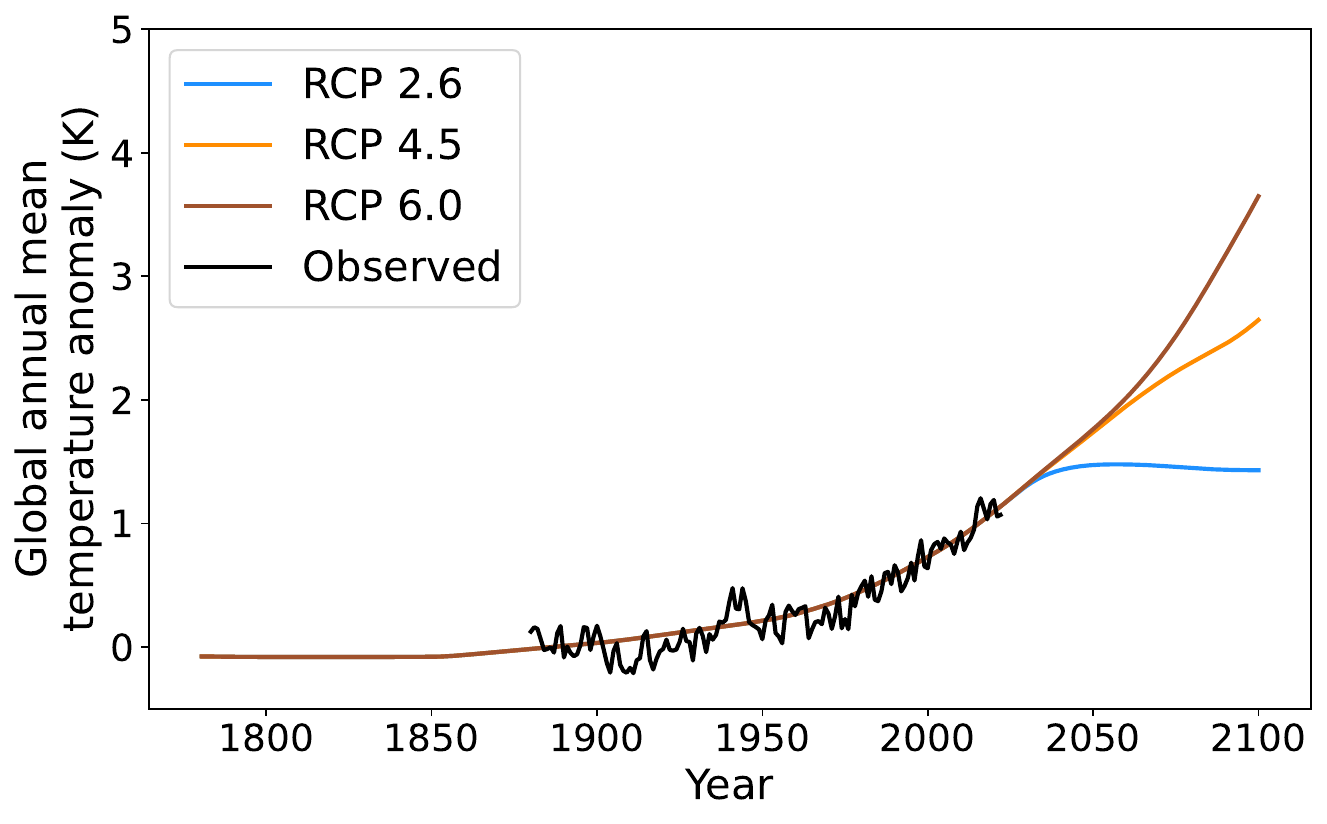}
		} 
		\subfigure[Atmospheric carbon stock]{ \label{fig: Atmospheric_stock_carbon}
			% label of upper right panel    
			\centering \includegraphics[scale=0.29]{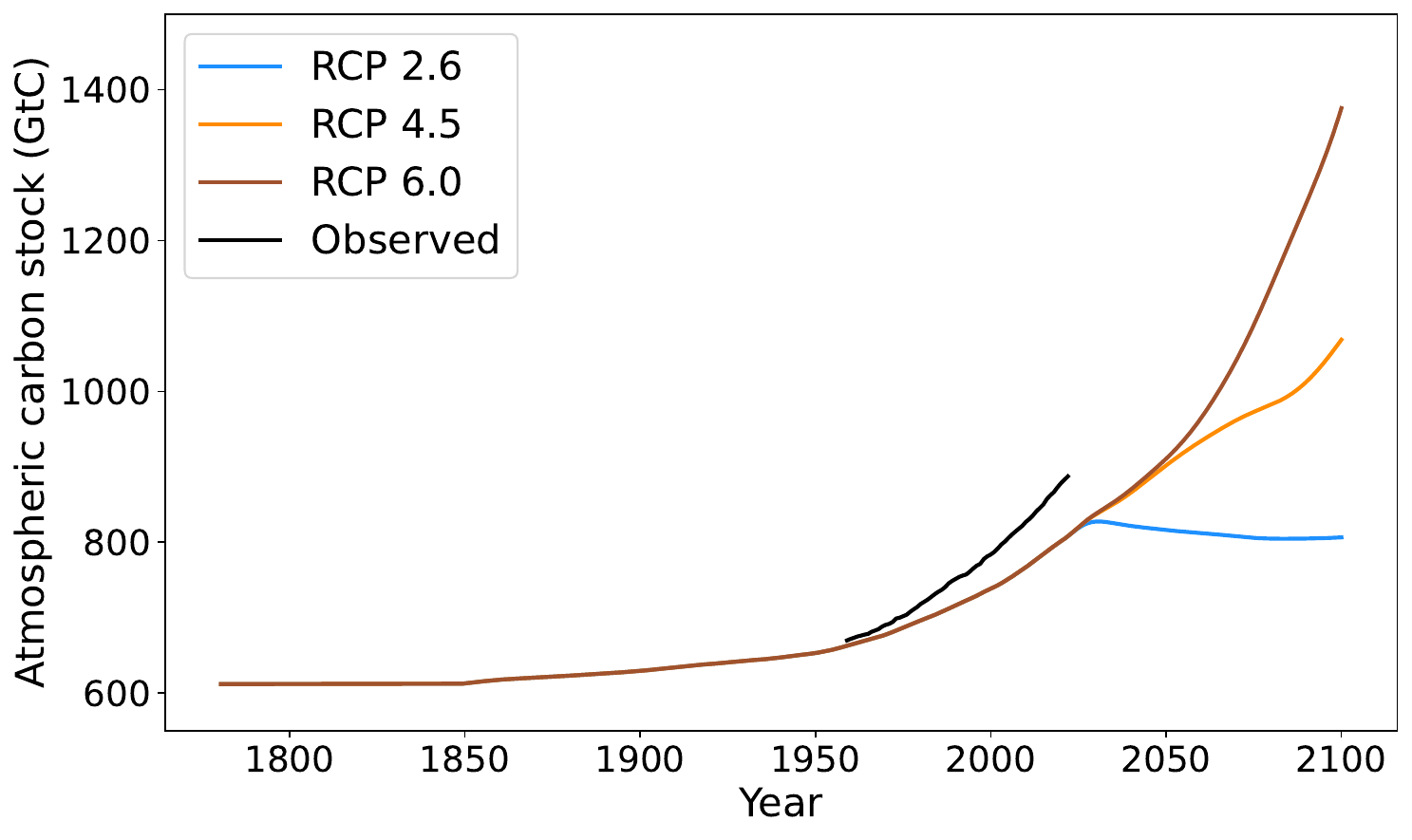}
		}
		
		%%%%%%%%%%%%%%%%%%%%%%%%%%%%%
		%%%%%    lower panels     %%%%%%%%%%%%%%%%
		
		\subfigure[Carbon flux from atmosphere to ocean mixed layer]{ \label{fig:Ocean_sink_carbon}
			% label of upper left panel    
			\centering \includegraphics[scale=0.29]{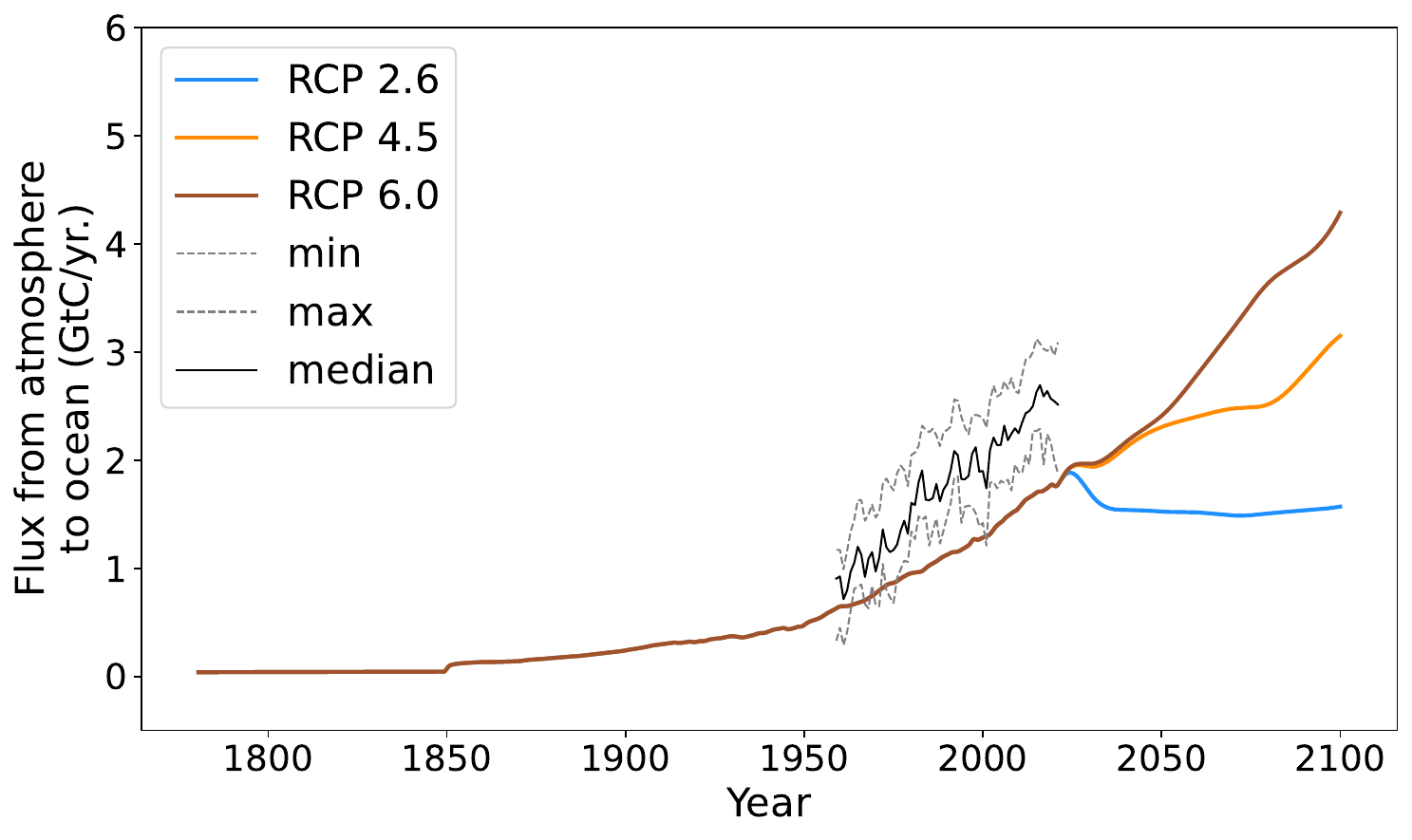}
		} 
		\subfigure[Carbon flux from atmosphere to land vegetation]{ \label{fig: Land_sink_carbon}
			% label of upper right panel    
			\centering \includegraphics[scale=0.29]{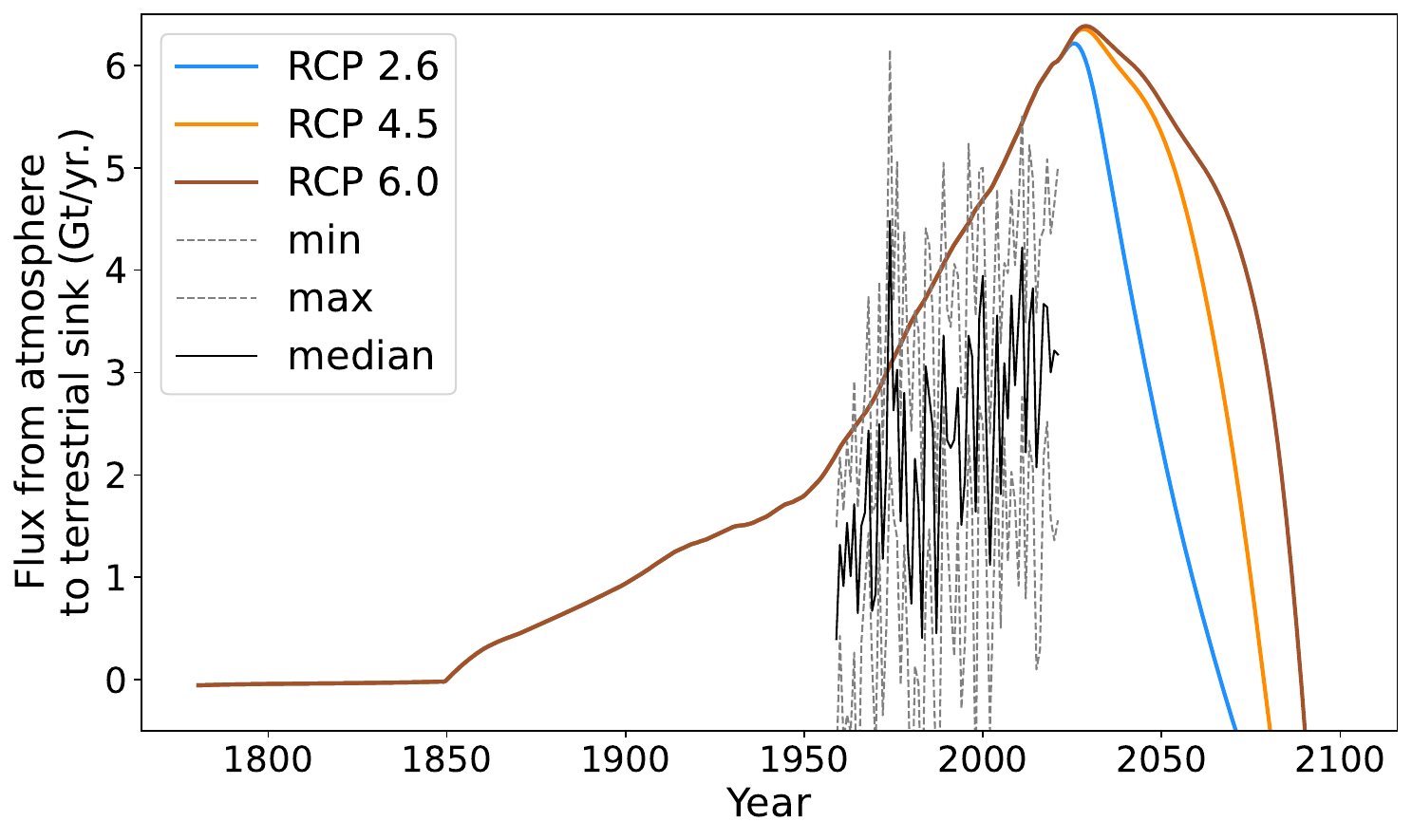}
		}
		
		\caption{{\footnotesize{\bf Evolution of the global temperature anomalies, carbon stocks, and  carbon fluxes in 
					our TCV model with anthropogenic forcing.} The parameter values used are given in Tables~\ref{tab:param-temp}, \ref{tab:param-terrestrial-carbon}, and \ref{tab:param-ocean-carbon}. and the results are shown for Representative Carbon Pathways (RCPs) 
				RCP~2.6 (blue), RCP~4.5 (orange), and RCP~6.0 (brown) \parencite{Meinshausen2011}.
				%Anthropogenic GHG emission scenarios are plotted in Fig.~\ref{Fig. RCP scenarios cumulative}, and the 
				Preindustrial initial state is $(T_0 = 286.5\,{\rm K}, \,C_{A,0} = 589\,{\rm{GtC}})$.
				(a)	Temperature anomaly evolution, with observed anomalies in solid black \parencite{NOAA2023};
				(b) atmospheric carbon stock $C_{\rm A}$, with observed stock in solid black \parencite{Friedlingstein_2022};
				(c) Fluxes $F_{{{\rm A}} \to {{\rm O}} }$ to the ocean mixed layer; and (d) fluxes $F_{{{\rm A}} \to {{\rm L}} }$ to the land vegetation. The minimum (dashed), maximum (dotted) and median (light solid) lines in panels (c) and (d) are based on different observation and model-based estimates of carbon fluxes aggregated in the Global Carbon Budget \parencite{Friedlingstein_2022}.}}
	\label{Fig:TempCarb_Evolution} % label of the full figure
	\end{figure}

	The resulting temperature evolution for each of the RCPs %in Fig.~\ref{Fig. RCP scenarios cumulative}
	are displayed in Fig.~\ref{Fig. temperature dynamics} for the parameter set of Tables~\ref{tab:param-temp} --\ref{tab:param-ocean-carbon}.
	In the more moderate RCPs 2.6, 4.5, and 6.0, the increase in emissions
	leads to increases in the average global temperature as well of atmospheric carbon concentration \parencite{Keeling1976} and carbon fluxes accross the land and ocean sinks that are
	comparable to observational datasets \parencite{Friedlingstein_2022}, as well as to those obtained in high-end model simulations of successive IPCC assessment reports.

	Similarly, the model fluxes of carbon to the separate sinks of the coupled atmosphere-land-ocean system are comparable with observations and with simulations obtained from high-end models, as displayed in aggregated form in the comprehensive Global Carbon Budget \parencite{Friedlingstein_2022}. The fluxes of carbon from the atmosphere to the oceanic mixed layer and to the land vegetation, as well as the evolution of atmospheric carbon stock  are displayed in Figs.~\ref{Fig:TempCarb_Evolution}(b,c,d), respectively.
	
	In Fig.~\ref{Fig. temperature dynamics}, the noticeable plateauing of global temperature at a higher level following anthropogenic GHG forcing is consistent with the recalcitrant warming simulated by more detailed climate models \parencite{Held2010, Pierrehumbert2014}. In the present model, it is the degradation of the land vegetation's carbon sink capacity that underlies this climate commitment; see Methods section.
	
We carried out numerical simulations of model trajectories starting near the present climate $P_1$ 
for anthropogenic emission scenarios given by RCPs 2.6, 4.5,
6.0 and 8.5 \parencite{IIASA2009, Meinshausen2011}, using both the full and the reduced versions of our TCV model.  The model reproduces accurately the trajectory of global average temperatures $T$, as well as global land vegetation and ocean carbon fluxes $F_{{\rm A} \to {\rm L} }$ and $F_{{\rm A} \to {\rm O} }$, based on both observations and high-end, IPCC-class climate and carbon cycle models for the low, moderate, and moderately high RCPs 2.6, 4.5 and 6.0; see Fig.~\ref{Fig:TempCarb_Evolution}. In fact, the differences between the simulations using our model's full and reduced versions are truly minimal.%; see Fig.~\ref{Fig:34EqSystems}.
	
\begin{figure}[ht!]
	
	\centering \subfigure[Regime diagram in the $(T_{\alpha_{\rm L},\ell}, V, e)$ parameter space]
	{ \label{fig:RegimeDiagram}
		% label of lower left panel
		\centering \includegraphics[scale=0.4]{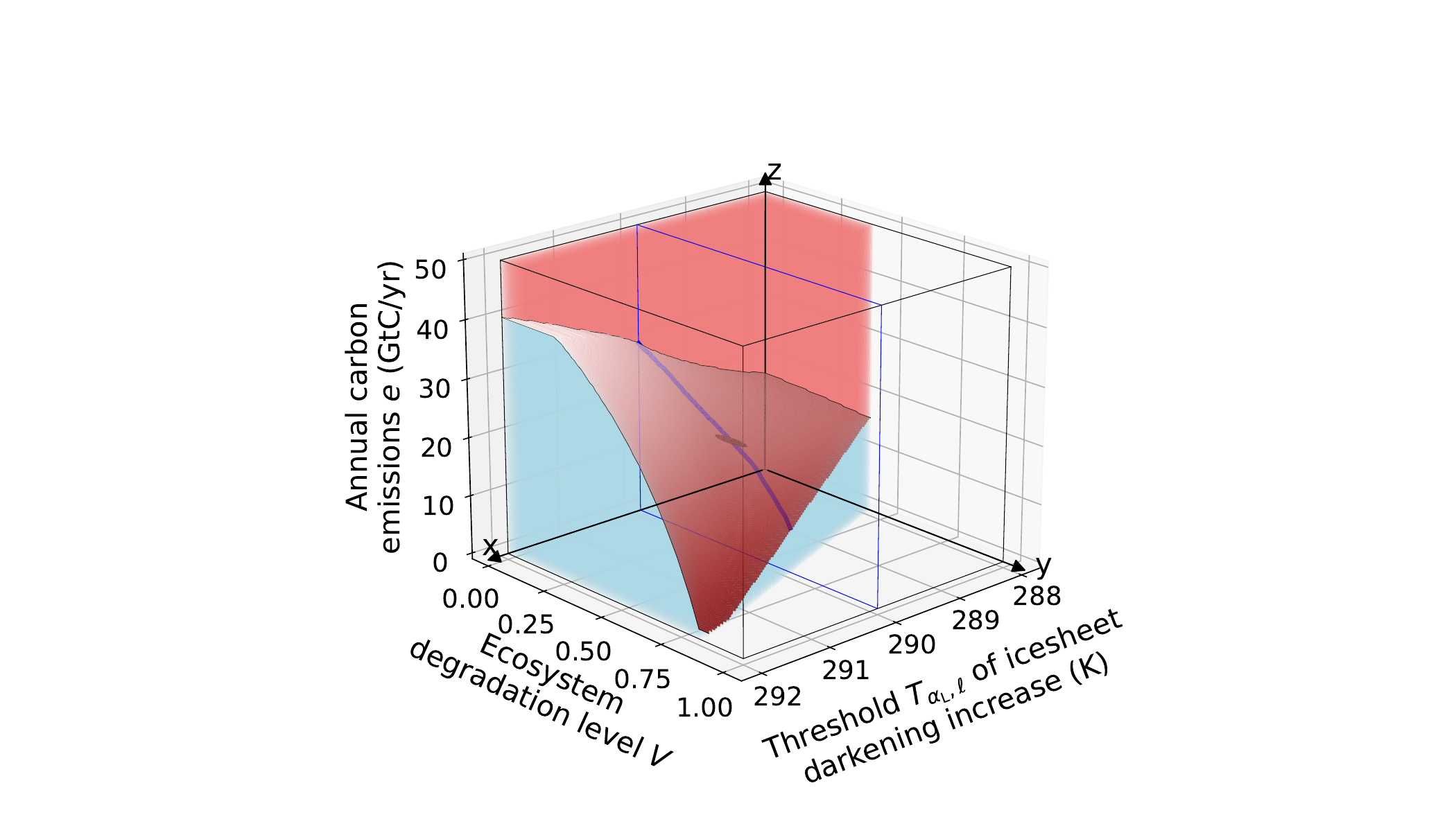}
	}
	%%%%%    lower panels     %%%%%%%%%%%%%%%%
	
	\subfigure[Regime boundary for $T_{\alpha_{\rm L},\ell} = 290$~K]{ \label{fig:K-g_T290}
		% label of upper left panel    
		\centering \includegraphics[scale=0.28]{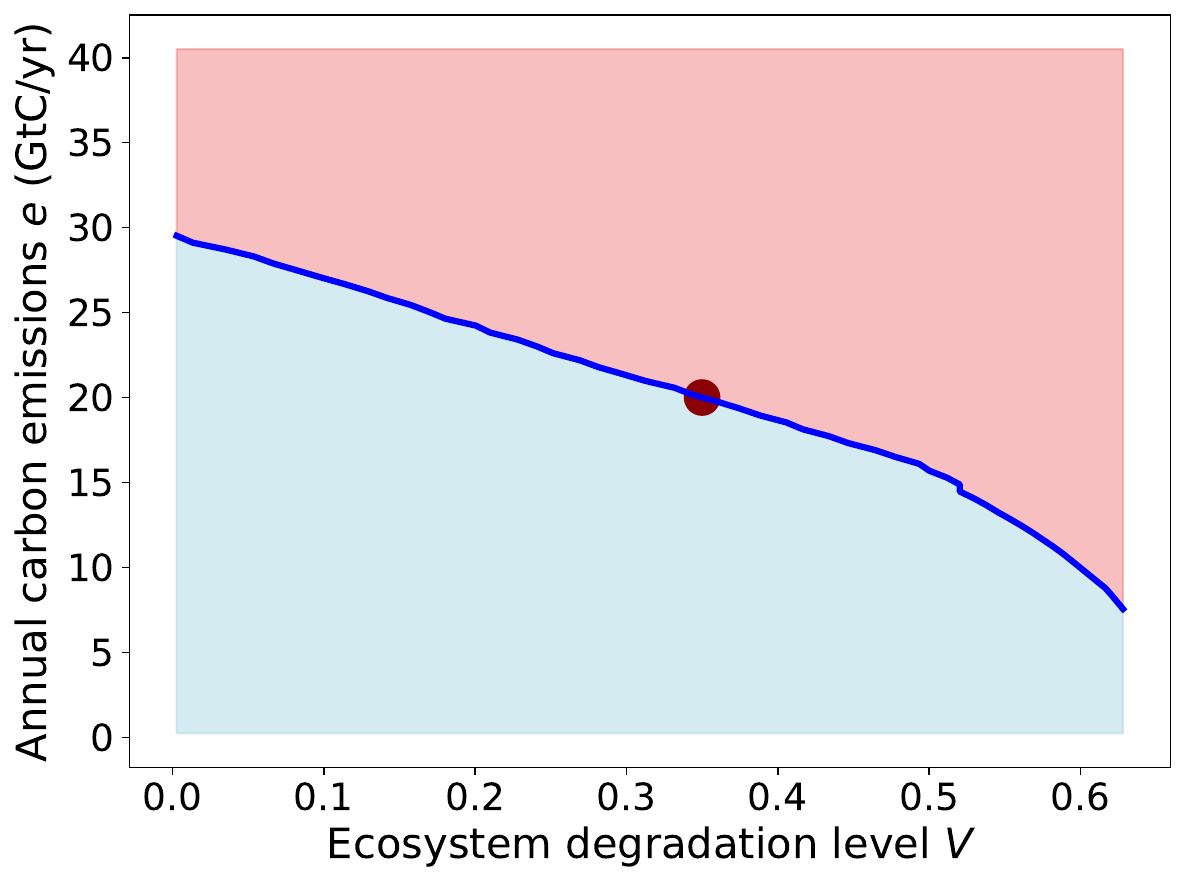}
	} \subfigure[Temperature evolution under RCP 8.5]{ \label{fig:TempEvolRCP85}
		% label of upper right panel    
		\centering \includegraphics[scale=0.28]{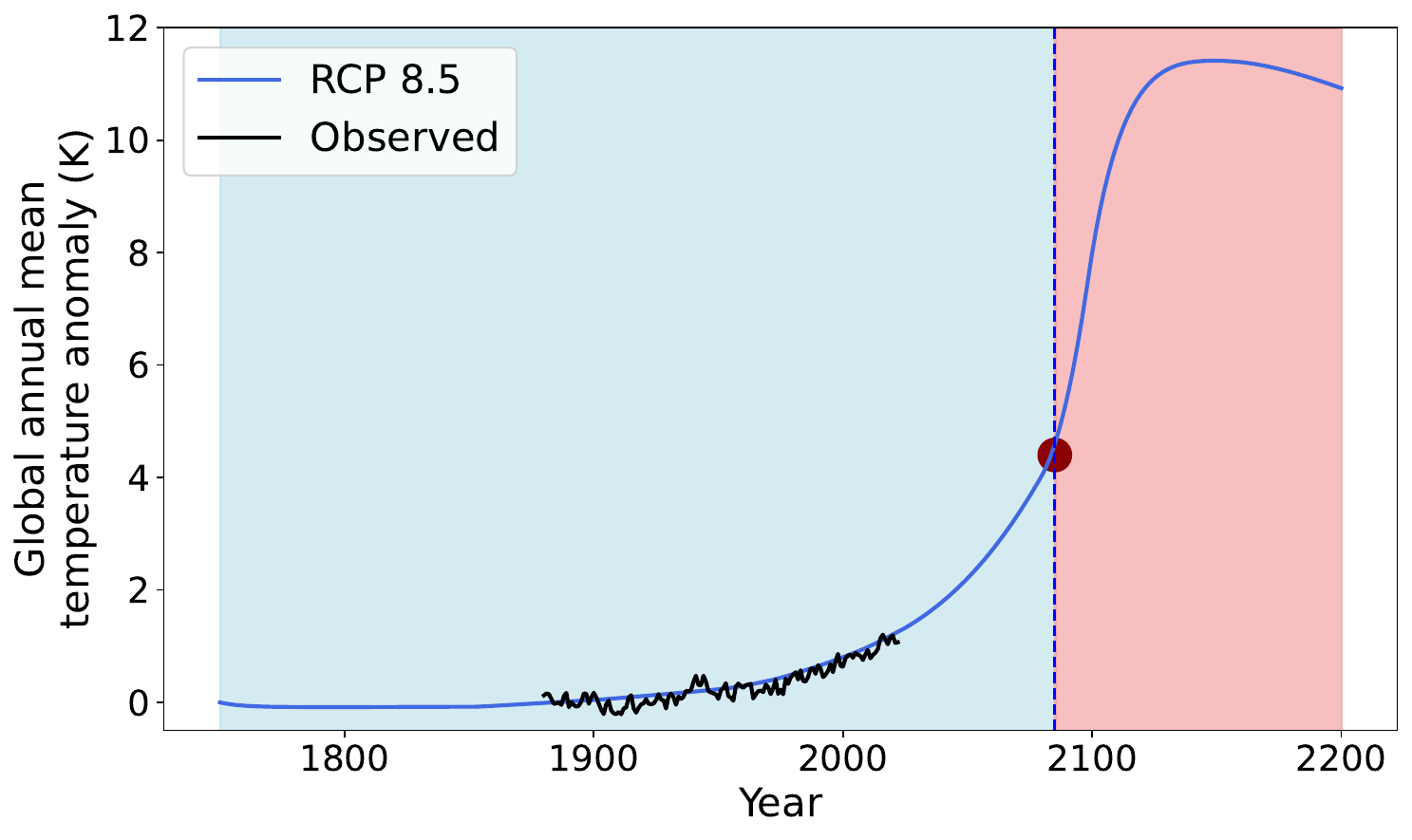}
	}
	
	\caption{{\footnotesize{\bf Regime diagram for the TCV model's shift from
				present climate to hothouse.} (a) Regime diagram in a three-dimensional parameter space. The diagram's axes are ($x$) the threshold temperature $T_{ \alpha_{\rm L},\ell}$ of darkening increase on the global cryosphere; 
				($y$) land vegetation degradation index $V$; and ($z$) annual emissions $e$. 
				The $e$ values for each  ($T_{\alpha_{\rm L},\ell}$,$V$) indicate the threshold of annual emissions beyond which the present climate $P_1$ will shift to the hothouse $P_2$;  see Figs.~\ref{Fig: Nullclines-model5} and Fig.~\ref{Fig: Ca-bifurcation}. 
			    (b) Two-dimensional section of panel (a) at $T_{\alpha_{\rm L},\ell} = 290$~K; the regime boundary  separating in this section the hothouse (red-pink) from the current climate (light blue) steady states is shown in dark blue.
%			\textcolor{blue}{changed the caption of panel (b)}
			(c) Evolution of temperature for the RCP 8.5 emissions scenario, when $V(t)$ evolves according to equation~\eqref{eq:EnvDegr3}. 
			%		$T$--$C$ phase-space trajectory subject to the same emission scenario. 
			In both panels (b) and (c),
			the red-filled circle indicates the occurrence of the TCV model's tipping from the stable $P_1$ present climate via the unstable $P_3$ steady state to the $P_2$ stable hothouse.
	}}
	\label{Fig:bdry} % label of the full figure
\end{figure}

	For the most severe anthropogenic GHG emission scenario, namely RCP 8.5,  %(black line in Fig.~\ref{Fig. RCP scenarios cumulative}) 
	a gradual increase in temperatures from preindustrial
	values leads in a few decades' time from now to an upward jump of roughly 8~K as the Earth transitions to the hothouse steady state, as shown in Fig.~\ref{fig:TempEvolRCP85}. As discussed in connection with Fig.~\ref{Fig: Ca-bifurcation} above, this jump appears to be irreversible unless one considers much augmented carbon sinks, along with carbon capture and storage technologies \parencite{Lackner.ea.2022}. The present TCV model suggests that fairly substantial decreases of $e$ below preindustrial values, denoted here by $e = 0,$ would be required to do so.
	
	The effects of gradual changes in GHG emissions % and the increase of ecosystem degradation $V$ 
	on the increase in temperatures %on the distance separating the current climate steady state from a saddle-node tipping 
	are illustrated in Figs.~\ref{Fig. temperature dynamics} and~\ref{fig:TempEvolRCP85} for the four RCPs considered in recent IPCC assessment reports \parencite{Meinshausen2011}. The first three, more moderate ones, only lead up to 290~K by 2100, in the case of RCP 6.0. The most severe one, RCP 8.5, is shown to be unstoppable on the way to the hothouse, past roughly 291~K.
	
	For a fixed amount of annual carbon emissions $e$, two processes are decisive in the TCV model
	in determining the transition % from the current climate regime 
	to a stable hothouse climate
	First, the %value of the threshold temperature $T_{\alpha, \ell}$ of 
	self-sustained darkening of ice sheets and other terrestrial cryosphere components, due to the glacial micro algae growth on their surface \parencite{Williamson2020,Millar2024,Tedesco2016}. While no ice sheets, mountain glaciers or snow cover terrestrial are explicitly represented in our model, this algal-bloom effect is implicitly captured by including it in the formulation of the land albedo $\alpha_{{\rm{L}}}$ as a function of temperature; see equation~(M.2$^{\prime}$) in the Methods section.
Second, the degradation of the global ecosystem due to the incremental death of the vegetation that is least tolerant to heat stress, and the subsequent decline in global net ecosystem productivity occuring after an initial increase driven by carbon fertlization effect encoded by Eq.~\eqref{Eq.  gc(C)} %are captured in equation~\eqref{eq:EnvDegr3} 
\parencite{Allen2010,Xu_2020,Hammond_2022,Wu_2022};  see "Impact of temperature stress on land vegetation" in the Methods section.
%\cj{This is not completely accurate: the global statue of the ecosystem is given by eq 2b, however the effect on the productivity, through die-off of vulnerable plants, is actually hidden in the flux atmosphere-to-land}

Figure~\ref{fig:RegimeDiagram} is a three-dimensional regime diagram in the $(T_{\alpha_{\rm L}, \ell}, V, e)$ parameter space: it is divided by a two-dimensional surface that separates the current-climate regime below it from the regime above it, in which a hothouse-like state is the only one that is stable.
In the latter regime, global average temperature $T$ is about 10 K higher than the
present-climate regime. Videos  that follow the changes in the model's phase portrait for $V = 0$ and $V = 0.3$ as $e$ increases can be found in the Supplementary Material.

Figure~\ref{fig:K-g_T290} illustrates the boundary (blue solid line) between the current-climate and the hothouse regime for $T_{\alpha_{\rm L}, \ell}=290$~K. Figure~\ref{fig:TempEvolRCP85} shows the evolution of global temperature for the RCP 8.5 emissions scenario. It exhibits a clear increase in slope as the trajectory enters the hothouse world. A video, labeled \verb|TCV_video-V=V(t)|, that illustrates in greater detail the model's behavior in this case is also available in the Supplementary Material. This video is compared there with two videos, labeled \verb|TCV_video-V=0| and \verb|TCV_video-V=0.3|, in which the vegetation degradation index $V$ is kept constant at $V=0$ and $V=0.3$, respectively.

% $(V,e)$-plane; the ($V, g, T_{\alpha_{\rm L},\ell}$) 

The distinction between model behavior in simulations that keep $e$ or $V$ fixed and those that allow the forcing $e(t)$ to evolve as prescribed and the variable $V(t)$ to evolve according to equation~\eqref{eq:EnvDegr3} can only be understood fully in the context of the theory of nonautonomous dynamical systems \parencite{Boers_2022, Ghil2020}. Such an approach, though, requires further numerical studies that go beyond the framework of the current paper.

Figure~\ref{Fig:bdry} allows one to establish safety margins for annual emissions $e(t)$. When a certain threshold temperature  $T_{\alpha_{\rm L}, \ell}$ is attained, algal growth on the surface of the terrestrial cryosphere allowing therewith its darkening becomes self-sustained, leading to the acceleration of the drop of terrestrial albedo (on the $x$-axis of Fig.~\ref{fig:RegimeDiagram}), while a level of ecosystem degradation $V$ has already occurred (on the figure's $y$-axis). For illustration purposes, the 2022 level of scenario-based GHG emissions is $e \simeq 11.6$~GtC/yr, while the estimated $V$ equals 0.09.

Assuming a median $T_{\alpha_{\rm L}, \ell}=290$~K threshold temperature, the buffer distance --- i.e., the maximum amount of annual emissions $e$ that can still be emitted without causing a jump to the hothouse state ---  is of 22.5 GtC/yr.  This buffer distance diminishes until reaching less than 1~GtC/yr in 2085 for RCP~8.5. We note that for higher levels of global vegetation degradation, the TCV model does not exhibit bistability for all $(V, T_{\alpha_{\rm L}, \ell})$-values contained in the orange volume in Fig.~\ref{Fig:bdry}. Instead, warming accelerates and the global temperature still reaches temperatures comparable to hothouse ones. An accelerated warming is initiated by the onset of darkening of the terrestrial cryosphere. 
% {\mg MG $\to$ Erik: Correct?} \textcolor{blue}{yes} \\

	\section{Discussion and conclusions}
	
	\label{sec:con_rem}
	
We formulated a 0-D model for the global temperatures $T$ and either two separate (TCV) or a single (TCV-DAE) carbon reservoir stock, i.e., either atmospheric $C_{\rm A}$ and 
oceanic mixed layer $C_{\rm M}$ or combined atmospheric and mixed-layer $C_{\rm S}$.
A separate equation for the evolution of the ecosystem degradation effect on the system, carried by the variable $V$, complements those for % temperature and carbon.
$T$ and $C$.
The full TCV model appears as the four-ODE system~\eqref{eq:TC-4box}, the TCV-DAE one as system~\eqref{eq:TC-3box}, with its three ODEs and an algebraic equation that states the fast equilibration of $C_{\rm A}$ and $C_{\rm M}$.

This coupled TCV model is highly idealized but it 
does include major physical, chemical and biological processes affecting
the Earth system. It is based on the system's radiation budget 
\parencite{Kallen1979, Rombouts2015, Craf.Kall.1978}, the carbon sink dynamics
of the ocean  \parencite{Lade_2018, Henson2011} and of the land vegetation
\parencite{Potter_1993, Svirezhev1997,Huang2019, Mahecha_2010}, and on
the radiative forcing by anthropogenic GHGs \parencite{Myhre1998}.
The model's parameter values rely on an extensive review of the existing literature; see Tables~\ref{tab:param-temp},  \ref{tab:param-ocean-carbon}, and \ref{tab:param-terrestrial-carbon}.

Both the full and the reduced model include the ice-albedo feedback and the anthropogenic GHG emissions forcing in the energy balance equation~\eqref{eq:TempEq}  or \eqref{eq:TempEq3}, respectively, while the temperature-dependent carbon fluxes between reservoirs are modeled by equations~\eqref{eq:CarbAt} and \eqref{eq:CarbOc} or by \eqref{eq:TotalCarb}, respectively. The two key novelties in our conceptual model are included in the first ODE of both systems~\eqref{eq:TC-4box} and \eqref{eq:TC-3box}, and in the state-of-the-vegetation equations~\eqref{eq:EnvDegr} or \eqref{eq:EnvDegr3}, respectively. They are: (i) the inclusion of the biogeophysical feedback between the algal bloom on ice sheets and the rest of the land cryosphere and the global temperature $T$, via the reduction of the land albedo $\alpha_{{\rm{L}}}$ due to the darkening of the Earth's cryosphere as shown in Fig.~\ref{Fig:Terrestrial-Albedo-Darkening}, especially at high latitudes, that this bloom produces \parencite{Yallop2012,Williamson2020,Millar2024,Tedesco2016}; and (ii) the temperature--vegetation degradation feedback that leads to the land vegetation's die-off as temperatures rise \parencite{Allen2010,Xu_2020,Hammond_2022,Wu_2022}. The temperature-albedo feedback included in the TCV model is of 1.1 W.m$^{-2}$.K$^{-1}$ which is consitent with orders of magnitude of temperature-albedo feedback that could be observed at planetary scale on Earth \parencite{Flanner2011}. These two regional processes and feedback value are described further in the Methods section.
%The model's parameter values rely on an extensive review of the existing literature; see Tables~\ref{tab:param-temp},  \ref{tab:param-ocean-carbon}, and \ref{tab:param-terrestrial-carbon}.

The difference between the full TCV and the reduced TCV-DAE model resides in equation~\eqref{eq:AlEq}, which postulates instantaneous equilibration between the atmospheric and the mixed-layer reservoirs, based on this equilibration being much faster than between the other pairs of reservoirs\parencite{Zeebe2001}.
The DAE model allowed us to carry out the phase-plane analysis illustrated in Fig.~\ref{Fig: Nullclines-model5}, complemented by the bifurcation analysis whose results appear in Fig.~\ref{Fig: Ca-bifurcation}. 

Figure~\ref{Fig: Nullclines-model5} shows clearly that there are three intersections --- $P_1, P_2,$  and $P_3$ --- of the two nullclines given by $F_1(T, C_{\mathrm{S}})=0$ (blue solid line) and $F_2(T, C_{\mathrm{S}}) = 0$ (red solid line), along which the global temperature $T$  and the combined carbon stock  $C_{\mathrm{S}}$ in the atmosphere and the ocean mixed layer, respectively, are constant. These intersections correspond to fixed points or steady states of the DAE system~\eqref{eq:TC-3box}; an additional, unstable fixed point lies above the area of the phase plane in Fig.~\ref{Fig: Nullclines-model5}. The two bifurcation diagrams in Fig.~\ref{Fig: Ca-bifurcation} illustrate the dependence of these three fixed points on the anthropogenic GHG emission rate $e$. The diagram in panel (a) is for $V = 0$, that in (b) is for $V = 0.3$, where $V$ is the degradation level of land vegetation, with $0 \le V \le 1$. 

In these diagrams, the two bifurcation points occur at the lower-right of the figure --- i.e., for positive emission rates $e_c \simeq + 31.5$~GtC/yr when $V = 0$ --- and further to the left, for lower emission rates, at $e_d = 16.5$~GtC/yr for both $V = 0$ and $V = 0.3$. For positive emission rates, though, degradation $V$ of the vegetation affects quite strongly the point at which the climate will jump from a mildly increased temperature to a hothouse one: it is $e_c = 21.5$~GtC/yr for $V = 0.3$ vs. $e_c \simeq 31.5$~GtC/yr for $V = 0$. 
%\textcolor{blue}{here I changed the values to correspond to figs 2 and 3}

Multistability is ubiquitous in the sciences in general \parencite{Parisi.2023} and bistability, in particular, has been established in the climate sciences in the early 1960s \parencite{Ghil2020, Stommel1961, Lorenz1963}; see also  Fig.~\ref{fig:hysteresis} and its discussion in the Supplementary Material. The crucially new result of the present work is that there exists a branch of hothouse solutions, roughly 8~K above the lowermost branch, which corresponds here to the current climate. This stable branch of hothouse steady states does not seem to have been found so far by either conceptual or more detailed models. On the other hand, paleoclimatic evidence of a climate similar to such a hothouse does exist, during the Paleocene-Eocene Thermal Maximum (PETM) \parencite{Zeebe_2013, Kirtland_Turner_2017}. And even the ecosystem deterioration used in our TCV model seems to have manifested itself in the transition to the PETM; see Methods section.

For the highest-emission scenario, RCP 8.5, though, our model exhibits an abrupt tipping to a stable hothouse climate, with the global temperature reaching the value of 299.7 K (26.5~${^{\circ}}$C); see Fig.~\ref{fig:TempEvolRCP85} and the short video \verb|TCV_video-V=V(t)| in the Supplementary Material.  
%\textcolor{blue}{
%{\mg
The saddle-node bifurcation associated with the collision of the current climate stable steady $P_{1}$ with the unstable $P_{3}$ state occurs in 2055 when 1,235 GtC of cumulative carbon emissions have been reached in the RCP 8.5 scenario. In comparison, RCP 6.0 has reached 1,073 GtC cumulative emissions. The $T$-$C$ trajectory of the TCV model diverges from the RCP 8.5 ensemble of IPCC-class ESMs in 2085, when its temperature rises nonlinearly in response to still rising anthropogenic carbon emissions that reach over 28~GtC/yr. %}
%}

 It appears that processes which are regional in nature --- like the effect of glacial algal blooms on ice and snow, which is largely confined to high latitudes in both hemispheres, and the differences in resilience of vegetation to temperature increase, which are particularly relevant to the tropics, where temperatures are and will remain the highest --- can create pathways of global tipping to a stable hothouse. These two biogeophysical mechanisms play a major role in facilitating the transition through saddle-node tipping from the current climate to the hothouse state.

\textcite{Popp.ea.2016} showed a shift to an unrealistic hothouse steady state about 40 K warmer than preindustrial temperatures using an aquaplanet model initialized at roughly 6~K above present global average temperature. \textcite{Eisenman_2024} recently obtained a hothouse Earth in a CESM2 GCM simulation at a CO$_2$ concentration that was 16-fold the present one. However, the change obtained did not occur due to a tipping mechanism but rather as the result of a smooth transition, unlike the tipping that led to a snowball Earth with the same model. It is worth recalling here that 1-D EBMs \parencite{Held.Suarez.1974,Ghil1976,North1981} predicted such a snowball Earth long before geochemists found proofs for its occurrence in Earth's remote past \parencite{Hoffman2002}, while GCMs were only used for the more detailed investigation of such snowballs after their geochemical documentation \parencite{Pierrehumbert.2004}.

We have shown that the TCV model's simulations compare quite favorably with instrumental observations for the period 1880--2022, with global temperature and carbon stock evolutions that are barely distinguishable from those of 42 IPCC-class GCMs that participated in CMIP5; see Fig.~\ref{Fig:IPCC_model_comparison}. Moreover, Bland-Altman tests show that --- for the  1861--2100 comparison interval used in CMIP5 for evaluating the impact of anthropogenic forcing on global temperatures --- the TCV model's estimates are in close agreement with those of the set of the 42 IPCC-class GCMs. The only noticeable difference between the CMIP5 estimates and the TCV ones occurs at the point at which, for RCP 8.5, the latter undergo a jump to the hothouse climate, while the GCMs -- that do not yet include the terrestrial cryosphere darkening mechanism -- do not.

%	\section*{Online content}
	
%	[Not available at the submission stage.]
	
% \noindent {\mg Pls. replace full first names in the Biblio. by initials everywhere.}
	
	\newpage{}
	
	%% \textcolor{black}{\bibliographystyle{../CT-Climate-Paper-April/21jponew}
		%\bibliographystyle{jponew}
		% {\bibliographystyle{naturemag}
			% \bibliography{Clim_GHG}
			
		\appto{\bibsetup}{\DeclareFieldFormat{labelnumberwidth}{#1\adddot}}
			
			\printbibliography[check=onlynew]
			
		\end{refsegment}

		\newpage{}

		\begin{refsegment}

\section*{Methods}
			\label{sec:methods}
			
			\setcounter{equation}{0} 
			\global\long\def\theequation{S\arabic{equation}}%
			\setcounter{table}{0} 
			\global\long\def\thetable{S\arabic{table}}%
			\setcounter{figure}{0} 
			\global\long\def\thefigure{S\arabic{figure}}%
			
\subsection*{Model}
\label{ssec:model}

We describe here the physical, chemical and biological processes that enter the formulation of the full TCV model governed by the system~\eqref{eq:TC-4box}. Its four ODEs govern the evolution in time of global temperatures $T$, two carbon stocks, atmospheric $C_{\rm A}$ %land vegetation $C_{\rm{L}}$, 
and oceanic $C_{\rm M}$,  and a terrestrial vegetation ecosystem degradation variable $V$. The reduced DAE version thereof --- governed by the system~\eqref{eq:TC-3box}, with its three ODEs and an algebraic equation --- is based on the same processes, except for the absence of the carbon flux between the atmosphere and the ocean mixed layer, which is set to zero in equation \eqref{eq:AlEq}, due to the rapid equilibration between these two reservoirs \parencite{Zeebe2001}. Due to this equilibration, $C_{\rm A}$  and $C_{\rm M}$ are replaced by their sum, $C_{\rm S} = C_{\rm A} +  C_{\rm M}$. 

\subsubsection*{Radiative budget and the TCV model's EBM}
Equations~\eqref{eq:TempEq} and \eqref{eq:TempEq3} are identical to each other, and they represent essentially the model's global-temperature evolution subject to natural radiation balance \parencite{Craf.Kall.1978,Ghil.Chil.1987}, which is modified by the anthropogenic GHG forcing $R_{{\rm {a}}}$ as the atmospheric carbon $C_{\rm A}$ increases above the baseline $C_{\rm A,0}$ preindustrial content due to anthropogenic emissions $e(t)$,
\begin{equation}
c\dot{T} = R_{{\rm {i}}} - R_{{\rm {o}}} + R_{{\rm {a}}}.
\end{equation}
Here the incoming radiation $R_{{\rm {i}}}=Q_{0}\left(1-p\alpha_{{\rm {L}}}(T)-(1-p)\alpha_{{\rm {O}}}(T)\right)$ is the product of the solar constant $Q_{0}$ and of the planetary albedo $\alpha_{\rm{pl}}$ \parencite{Budyko.1969,Sellers.1969}, to which the land and the ocean surface contribute in the proportion $p:(1-p)$ \parencite{Kallen1979}, where $p$ is the fraction of land.
The outgoing radiation $R_{{\rm {o}}}$ is given by a Budyko--style linear function of temperature \parencite{Budyko.1969}, $R_{{\rm {o}}}=\kappa\left(T-T_{\kappa}\right)$, with $T_{\kappa}=154$~K, and $c$ is the heat capacity\parencite{Schwartz2007}.
%while the anthropogenic term $R_{{\rm {ant}}}$ 

The ocean albedo  $\alpha_{{\rm {O}}}\left(T\right)$
is given by the piecewise-linear ramp function of Sellers \parencite{Ghil1976, Sellers.1969, Zaliapin2010}.
\begin{equation}
	\alpha_{{\rm O}}\left(T\right)=\begin{cases}
		\begin{array}{ccc}
			\alpha_{{\rm max}}^{{\rm O}}  & {\rm if} & T\leq T_{\alpha_{\rm O},\ell},\\
			\alpha_{{\rm max}}^{{\rm O}} + {\displaystyle {\frac{\alpha_{{\rm min}}^{{\rm O}} - \alpha_{{\rm max}}^{{\rm O}}}{T_{\alpha_{\rm O},u}-T_{\alpha_{\rm O},\ell}}\left(T-T_{\alpha_{\rm O},\ell}\right)}} & {\rm if} & T_{\alpha_{\rm O},\ell}<T\leq T_{\alpha_{\rm O},u},\\
			\alpha_{{\rm min}}^{{\rm O}}  & {\rm if} & T_{\alpha_{\rm O},u}<T.
	\end{array}\end{cases}\label{eq:alb_cases-1}
\end{equation}

Here $\alpha_{{\rm max}}^{{\rm O}} = 0.57$ \parencite{Sellers.1969, Vrese2021} corresponds to a completely sea ice--covered ocean \parencite{NSIDC2023,Pistone2014}
at global temperatures below $T_{\alpha_{\rm O},u}=200$~K, following an authoritative review of Neoproterozoic literature \parencite{Pierrehumbert2011}, 
while an
ice-free ocean has a top-of-atmosphere albedo of ~$\alpha_{{\rm min}}^{{\rm O}}=0.26$ \parencite{Sellers.1969} when the global temperature is higher than $T_{\alpha_{\rm O},\ell}=299$~K.  Between 200~K \parencite{Pierrehumbert2011} and 299~K \parencite{Kallen1979}, the ocean albedo decreases linearly. 
De Vrese et al. \parencite{Vrese2021} studied snowball Earth \parencite{Hoffman1998,Hoffman2002} climate dynamics to determine plausible albedo values using the MPI-ESM climate model and obtained values of planetary albedos of 0.57--0.7. 

Our model's formulation of land albedo $\alpha_{{\rm {L}}}\left(T\right)$ is likewise via a piecewise-linear ramp function, with $\alpha_{{\rm O}}$ replaced by $\alpha_{{\rm {L}}}\left(T\right)$ and distinct tie points at which the derivative of the albedo with respect to temperature is discontinuous. The land albedo decreases from a maximum level $\alpha_{{\rm max}}^{{\rm L}} = 0.4$ for $T_{\alpha_{\rm{L}},\ell} = 290$~K \parencite{Boone2001} to a minimum land albedo of $\alpha_{{\rm min}}^{\rm {L}} = 0.35$ \parencite{Boone2001,Yallop2012} when the global temperature attains $T_{\alpha_{\rm{L}},u}=294$~K and the cryosphere has been completely darkened as explained below. We shall refer to this land albedo model as equation~(M.2$^{\prime}$).

The global temperature  $T_{\alpha_{\rm L},\ell}$ corresponds to the temperature of initiation of self-sustained darkening of the icesheets. It is when this threshold temperature is reached that the algal-blooms--covered surface increases year after year, leading to decreasing albedo until a minimum is reached at $T=T_{\alpha_{\rm L},u}$. It is at this point, in particular, that a maximum extent of surface of the polar ice sheets and the rest of the terrestrial cryosphere (see Fig.~\ref{Fig:Terrestrial-Albedo-Darkening}), is uniformly darkened by algal blooms and provides a suitable environment for these blooms to develop upon \parencite{Williamson2020,Tedesco2016}. The value
of $T_{\alpha_{\rm{L}},\ell} = 290$~K mentioned above is roughly $4$~K higher than the present climate and it represents % an estimate of 
the median of recent estimates for the triggering of self-sustained darkening of major ice sheets \parencite{McKay2022}. 

The break point $T_{\alpha_{\rm L},u} = 294$~K corresponds to a temperature with a maximum surface of terrestrial cryosphere covered in melted ice or snow. At this temperature, we consider the entire midlatitude and mountain glaciers surface as well as the whole Greenland icesheet  \parencite{Box2012} to be suitable for algal growth. While the surface mass balance over this entire 	area --- i.e., the accumulated mass of snow minus the mass lost through melting --- has become negative, in the Antarctica, the ice sheet surface below 2~500~m of altitude\footnote{Elevation data was retrieved from ref. \parencite{Danielson2011}} is in a melt state \parencite{Garbe2020,Garbe2023}, while the area above remains fully frozen and with a high albedo level of 0.8. 

The two ramp functions used herein are clearly a substantial simplification that
is compatible with the overall simplicity of our TCV model.  Such simplifications have been used in several 0-D\parencite{Sellers.1969} or 1-D\parencite{LeTreut1983} EBMs, as well as in fully-fledged 3-D IPCC-class GCMs \parencite{Madsen2022,Winkelmann2011}. %(Eq. \ref{eq:TempEq}). 
%Concerning the ocean albedo, the exact values of $(T_{\alpha_{\rm O},\ell},T_{\alpha_{\rm O},u})$
%and of $(\alpha_{{\rm min}}^{{\rm O}} ,\alpha_{{\rm max}}^{{\rm O}} )$ have been the topic
%of some disagreement in the literature but are not really that relevant
%to the issues at hand \parencite{Ghil.1984}.
			
Concerning the effect of anthropogenic GHG emissions $e(t)$ in Eq.~\ref{eq:CarbAt} on $R_{{\rm {a}}},$ Myhre et al. \parencite{Myhre1998} showed that a logarithmic function captures fairly accurately this effect over the range of CO$_{2}$ concentrations currently experienced. Based on their widely applied work, we take $R_{{\rm {a}}} = a\ln\left(C_{\mathrm{A}}/C_{0}\right)$, with $a$ the GHG radiative forcing constant of CO$_{2}$; see Table~\refeq{tab:param-temp}. 

In order, though, to determine the future evolution of the global temperature $T$ in response to changes in CO$_{2}$ levels, we need to account for several feedback mechanisms that, either directly or indirectly, damp or amplify the radiative forcing of GHG gases outside past instrumental ranges. 

% \noindent {\mg MG got to here on Thurs., July 4th, 2024.} 

\subsection*{Biogeophysical feedbacks that affect planetary albedo and the carbon cycle}

Two mechanisms play a major role in determining the evolution from the current climate $P_1$ 
via the saddle-node bifurcation that causes its tipping to the hothouse state $P_2$. The first one refers to rapid changes of albedo, mostly in high latitudes, and it has to do with the effect of algal blooms on the effective albedo of cryospheric components, i.e., ice sheets, sea ice, mountain glaciers and seasonal snow cover. The latter one arises from the differential degradation of land vegetation as temperature rises.

\subsubsection*{Glacial micro algae darkening of the terrestrial cryosphere}

Cryospheric darkening stems from brown-carbon from more frequent forest fires, black-carbon from anthropogenic incomplete combustion, and—above all—the spread of glacier-adapted algae \parencite{Tedesco2016}. Field data \parencite{Cook2017,Stibal2017} show this algal effect dominates the darkening-driven rise in surface temperature and melt; satellite analyses reveal a strong inverse ice-sheet-albedo–algal-abundance link \parencite{Williamson2020}.

This temperature-amplified algal darkening is poorly represented in state-of-the-art land-surface schemes, yet its impact on the global energy balance must be quantified \parencite{Williamson2019}. We address it with the TCV model, applying algal-bloom darkening only to land. The process starts at $T_{\alpha_{\rm L},\ell}=290$ K, reaches its maximum at $T_{\alpha_{\rm L},u}=294$ K as shown in Eq.~\eqref{eq:alb_terrestrial}.

\begin{equation}
	\alpha_{{\rm L}}\left(T\right)=\begin{cases}
		\begin{array}{ccc}
			\alpha_{{\rm max}}^{{\rm L}}  & {\rm if} & T\leq T_{\alpha_{\rm L},\ell},\\
			\alpha_{{\rm max}}^{{\rm L}} + {\displaystyle {\frac{\alpha_{{\rm min}}^{{\rm L}} - \alpha_{{\rm max}}^{{\rm L}}}{T_{\alpha_{\rm L},u}-T_{\alpha_{\rm L},\ell}}\left(T-T_{\alpha_{\rm L},\ell}\right)}} & {\rm if} & T_{\alpha_{\rm L},\ell}<T\leq T_{\alpha_{\rm L},u},\\
			\alpha_{{\rm min}}^{{\rm L}}  & {\rm if} & T_{\alpha_{\rm L},u}<T.
	\end{array}\end{cases}\label{eq:alb_terrestrial}
\end{equation}

We do not imply bedrock exposure—complete ice-sheet loss requires centuries–millennia \parencite{Robinson_2012,Bochow2023}. Rather, reduced surface-mass balance and elevation enlarge the melt area, allowing algae eventually to colonise the whole ice sheet. The chosen $T_{\alpha_{\rm L},\ell}=290$ K lies in the mid-range for self-sustained melting \parencite{Box2012,Garbe2020,Garbe2023,McKay2022,Boers2021,Robinson_2012,Noel2021}, the ramp that drives tipping at $(T_{c},e_{c})$.

\begin{figure}[ht!]

	\subfigure[Latitudinal weights for area averaging]{ \label{fig:AreaWeights}
		% label of upper left panel    
		\centering \includegraphics[scale=0.3]{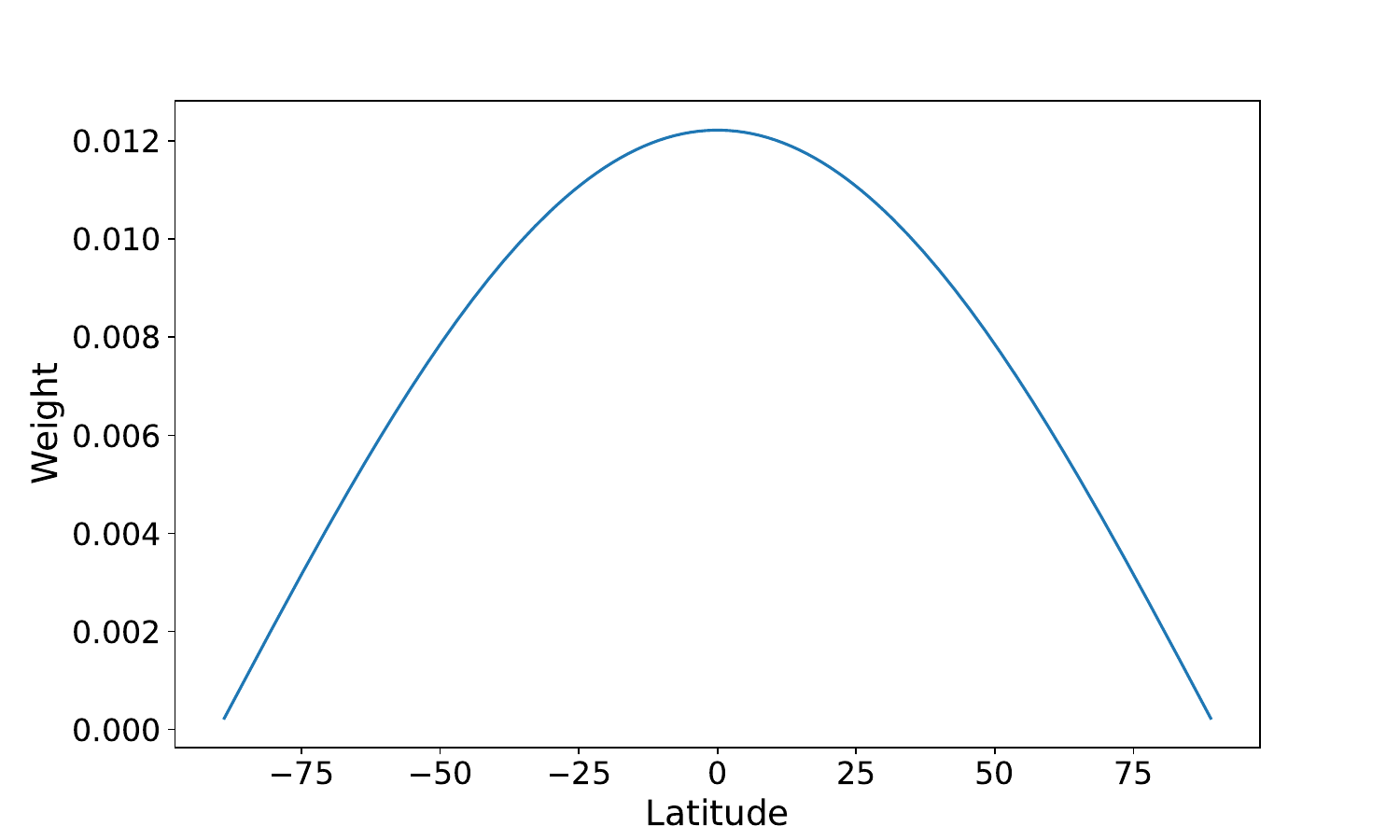}
	} \subfigure[Latitudinal weights accounting to account for annual change of cumulative radiation (Wm$^{-2}$yr$^{-1}$), as a function of latitude]{ \label{fig:RadiationWeights}
		% label of upper right panel    
		\centering \includegraphics[scale=0.3]{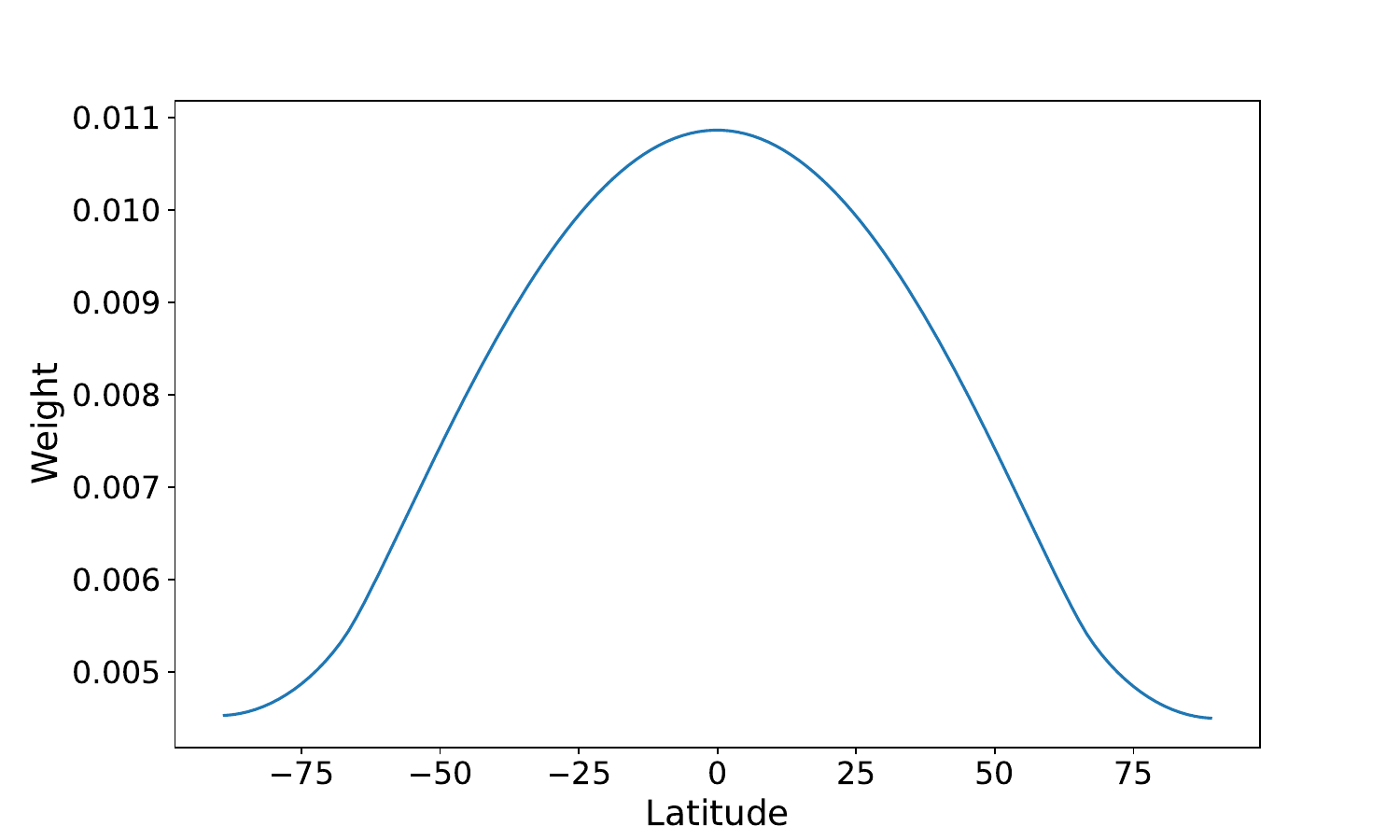}
	}
	
	\caption{{\footnotesize{\bf Global average terrestrial surface albedo computation latitudinal weights.}
			The global averaged albedo is computed by carrying out a weighted sum of grid-level 
			surface albedo, considering both (a) area weighting; and 
			(b) annual cumulative radiation as a function of latitude \parencite{Kopp2023}. %{\mg Erik: Shouldn't this be just Fig.~7, rather than Fig.~S1?!}
	}}
	\label{Fig:Terrestrial-Albedo-Weights} % label of the full figure
\end{figure}

To derive the drop from $\alpha_{\rm max}^{\rm L}$ to $\alpha_{\rm min}^{\rm L}$ used in the TCV model (Fig.\ref{Fig:Terrestrial-Albedo-Darkening}), we analysed daily surface-albedo output from a CNRM-CM5 RCP 8.5 run (0.5° grid) that employs the ISBA-ES snow-and-ice module \parencite{Boone2001,Boone2010}. Focusing on 2070–2100—when $T_{\alpha_{\rm L},\ell}$ is reached in IPCC-class models—we computed two gridded daily climatologies: (i) the raw model albedo and (ii) the same fields after resetting every ice- and snow-covered grid cell to the 0.3 albedo of fully algal-covered surfaces \parencite{Yallop2012}. The resulting global annual-mean terrestrial albedos were then obtained as area- and insolation-weighted averages of the gridded means, accounting for latitude-dependent grid size and incoming radiation (Figs.\ref{Fig:Terrestrial-Albedo-Weights}a,b) \parencite{Kopp2023}. Global terrestrial albedo thus falls by 0.045 (14 \%), giving a temperature-albedo feedback of
$pQ_{0}\Delta\alpha_{\rm L}/(T_{\alpha_{\rm L},u}-T_{\alpha_{\rm L},\ell}) = 0.29\times342.5\times0.045/4 = 1.1$ W m$^{-2}$ K$^{-1}$, consistent with observational estimates \parencite{Flanner2011}.

Because the TCV model closes the top-of-atmosphere (TOA) energy budget, we add cloud contributions. A mean cloud fraction of 0.47 \parencite{Dutta2020} and the cloud-fraction–albedo relation \parencite{Feingold2017} supply $\approx$ 0.17 extra albedo units, giving $\alpha_{\rm max}^{\rm L}=0.40$, $\alpha_{\rm min}^{\rm L}=0.35$, and a clear-sky climatological value of 0.23 \parencite{Jia2022}. Algal darkening of sea-ice—omitted here—would further reduce TOA albedo.

\subsubsection*{Decrease in the land vegetation's gross primary productivity (GPP)}

Excess stress due to rising temperatures affects numerous physiological functions of land vegetation, which can lead to lowering of carbon uptake and storage capacity and, potentially, to increased mortality \parencite{Teskey2014}. In fact, global temperature increase is known to augment the frequency of occurrence and the extent of vegetation die-offs across both the tropics and subtropics \parencite{Allen2010,Williams2012,Mantgem2009,Phillips2009,Zhao2010}. Hammond et al. \parencite{Hammond_2022} document the increase of tree mortality accross each biome globally in response to conditions of hotter droughts.
% In turn, an increasing body of litterature indicates that 
This phenomenon is also linked to the lowering of % the Gross Primary Productivity (GPP) of 
vegetation GPP \parencite{Zhao2010, Wu_2022, Xu_2020, Au_2023}.  The reduction
of the carbon sink capacity of the global land vegetation as temperatures
rise is referred to as the temperature--photosynthesis feedback  \parencite{Ciais_2005}.

\subsection*{Carbon fluxes}	

Here, we describe the terms entering equations~\eqref{eq:CarbAt}, \eqref{eq:CarbOc}, and \eqref{eq:TotalCarb} of the main text.		

\subsubsection*{The land carbon sink}

The carbon  flux between the atmosphere and the land carbon sink is encoded in $F_{\mathrm{A}\rightarrow \mathrm{L}}(T,C_{\mathrm{A}},V)$.  The carbon fluxes are certainly not truly Fickian \parencite{Fick.1855}, i.e., not linearly proportional to the difference in carbon concentrations. By analogy with nonlinear fluxes in other types of transport, for instance of momentum or heat in a fluid  \parencite{CushmanRoisin.2011}, one often assumes such a proportionality and infers from observations or from detailed simulations a coefficient that provides a good approximation to such a proportionality \cite{Potter_1993}.

The flux $F_{\mathrm{A}\rightarrow \mathrm{L}}(T, {\mathcal C_{\rm A}}, V)$ is equal to the difference between the photosynthetic carbon uptake by the land vegetation's $\mathrm{GPP}$ and the release $R_{s}(T)$ of carbon by soils into the atmosphere through respiration, 
%			 In our model, this flux is represented by the following equation:
\begin{equation}
F_{\mathrm{A}\rightarrow \mathrm{L}}(T, {\mathcal C_{\rm A}}, V) = P(T, {\mathcal C_{\rm A}}, V) - R_{s}(T), \label{eq:FluxLand}
\end{equation}			
%			\subsubsection*{Carbon absorption by the land vegetation}
where $P$ is the GPP and the difference on the right-hand side of \eqref{eq:FluxLand} is also called the Net Primary Productivity (NPP). 

We formulated our model for the atmosphere-to-vegetation carbon flux using a carbon simulation model for land vegetation based on the Carnegie-Ames-Stanford approach (CASA)\parencite{Potter_1993} and, in particular, its 0-D version developed by  Svirezhev and von Bloh (1997)\parencite{Svirezhev1997}.  The GPP term in the TCV model's equation~\eqref{eq:CarbAt}
is given by
\begin{equation}
P(T, \mathcal C_{\rm A}, V) = P_{m}g_{T}(T,V)g_{\mathcal C}(\mathcal C_{\rm A}),
\label{Eq.  Chavez-Ghil-Rombouts Cveg}
\end{equation}
where $V$ is the TCV's level of land vegetation degradation and $P_{m}$ is the global maximum potential GPP\parencite{Myneni1995}. %\cj{is vegetation index used anywhere else? Otherwise would change terminology here}

%In Svirezhev and von Bloh \parencite{Svirezhev1997}, the atmospheric carbon concentration $\mathcal C$ evolves according to
%\begin{equation}
%	\dot{\mathcal C}_{\rm A}=-P\left(T, \mathcal C_{\rm A}, N\right)+mN+e(t),
%	\label{Eq.  Sverizhev-Bloh veg productivity-1}
%\end{equation}
%and $N$ is the global vegetation density that we do not account for in our model. In Svirezhev and von Bloh, the land vegetation's  carbon influx into the atmosphere through respiration is $mN$, with $m$ a proportionality constant,  while $e(t)$ represents anthropogenic emissions, as in  equation~\eqref{eq:CarbAt}. Their vegetation productivity $P$ is represented as follows: 			
%\begin{equation}
%P\left(T, \mathcal C_{\rm A}, V\right)=P_{m}g_{T}(T)g_{\mathcal C}(\mathcal C_{\rm A})g_{N}(N). \label{Eq.  Sverizhev-Bloh veg productivity}
%\end{equation}

The multiplicative form of $P$ in \eqref{Eq.  Chavez-Ghil-Rombouts Cveg}%, as well as in \eqref{Eq.  Sverizhev-Bloh veg productivity}, 
follows Liebig's law of the minimum, also called the limiting-factor paradigm \parencite{Sinclair.1994}.  The factor $g_{\mathcal C}( \mathcal C_{\rm A})$ %and $g_{N}(N)$ in ref. \parencite{Svirezhev1997} are 
is monotonically increasing functions of the Michaelis–Menten form that saturate at unity, with $g_{\mathcal C}( \mathcal C_{\rm A})$ capturing the carbon fertilization effect. Given the still poorly understood impact of elevated CO$_2$ concentration on the latter effect, we adopt a simple formulation as follows:
\begin{equation}
g_{\mathcal C}(\mathcal C_{\rm A})=\frac{\mathcal C_{\rm A}}{\mathcal C_{h}+\mathcal C_{\rm A}}. \label{Eq.  gc(C)}
\end{equation}
Here $\mathcal C_{h}$ is half the saturation value of atmospheric carbon concentration, and we use this formula in our own equation~\eqref{Eq.  Chavez-Ghil-Rombouts Cveg}. The value of $\mathcal C_{h}$ is the half saturation atmospheric carbon concentration of vegetation photosynthetic activity \parencite{Haverd2020}. The carbon fertilization effect encoded by $g_{\mathcal C}(\mathcal C_{\rm A})$ is partly responsible for the rise in the flux of carbon from the atmosphere to the land that we observe in our model until ~2050 due to the increase of Gross Primary Productivity it drives (see Figs. \ref{fig: Land_sink_carbon} and \ref{Fig:Carbon_veg_dyn}.

%The factor $g_{T}(T)$ in \eqref{Eq.  Sverizhev-Bloh veg productivity}, though, is a unimodal function with a maximum value of unity for a global optimum argument and two minima of zero at low and high argument values \parencite{Luedeke1995}.
The function $g_{T}(T,V)$%, though, 
in% our own 
equation~\eqref{Eq.  Chavez-Ghil-Rombouts Cveg}, is chosen to reflect a plateau of global optimum temperatures accounting for all biomes rather than a single global
optimum \parencite{Huang2019}, as well as for the impact of the level of degradation of the land vegetation $V$ on $P$.
The precise form of $g_{T}(T,V)$ is given by %\cj{second line of the eq, there is a subscript alpha - this is a typo??}
\begin{equation}
g_{T}\left(T,V\right)=\begin{cases}
	\begin{array}{ccc}
		0 & {\rm if} & T \leq T_{v,\ell}(V),\\
		{\displaystyle \frac{T-T_{v,\ell}(V)}{T_{o,\ell}(V)-T_{v,\ell}(V)}} & {\rm if} & T_{v,\ell}(V)<T\leq T_{o,\ell}(V),\\
		1 & \mathtt{\mathrm{if}} & T_{o,\ell}(V)<T\leq T_{o,u},\\
		{\displaystyle \frac{T-T_{v,u}}{T_{v,u}-T_{o,u}}} & \mathrm{if} & T_{o,u}<T\leq T_{v,u},\\
		0 & \mathrm{if} & T_{v,u}<T.
\end{array}\end{cases}\label{eq:veg-ramp}
\end{equation}

\subsubsection*{Simple model of total ecosystem respiration}

%\noindent \cm{Erik, This subsubsection belongs here, to explain the 2nd term on the RHS of Eq.~\eqref{eq:FluxLand}.}

Soil respiration is the second largest land influx of carbon into the atmosphere
\parencite{Prentice2001}; %\cj{what is the largest then? is it in our model?}
it derives from the respiration of vegetation roots and from the
microbial activity of organic matter decomposition in the soils. Considering only soil respiration, annual flux estimations range from 50--75
\parencite{Crowther_2015} to 90 \parencite{Hashimoto_2015},
or even 108 GtC/year \parencite{Hursh_2016} of carbon per year into the
atmosphere . Several studies \parencite{Davidson_2006, Warner_2019, Huntingford_2017} describe
the different sources of uncertainty concerning the temperature effect on carbon stored in soils and their implications for ESM-based projections.

In our TCV model we use the classic $Q_{10}$ model \parencite{Hoff1898} to represent temperature-dependent soil respiration as follows:
\begin{equation}
R_{s}(T)=R_{S,0}Q_{10,S}{}^{\frac{\Delta T}{10}}; \label{Eq.  Global-Rs-Mahecha}
\end{equation}
here $R_{S,0}$ is the preindustrial respiration value at global temperature $T_0 = 286.5$~K\parencite{Hawkins2017} and $\Delta T=T-T_{0}$. This equation is derived from  the exponential chemical reaction-temperature $R$--$T$ rate \parencite{Hoff1898}, where $R=R_{c}Q_{10}^{\left(T-T_{c}\right)/10}$ with $R$
and $R_{c}$ the reaction rate and baseline rate at $T$ and $T_{c}$,
respectively, and $Q_{10}$ is the relative increase of the ratio $R/R_{c}$
for a 10 K temperature increase. %Bond-Lamberty and Thomson (2010)

%\textcolor{blue}{
%An estimate of $Q_{10}=1.5$ is based on 1,434 observed data points of soil respiration collected from 1989 to 2008 across boreal, temperate, and tropical biomes \parencite{BondLamberty2010}.
%Mahecha et al\parencite{Mahecha_2010} obtain a slightly different value of $Q_{10}=1.4$ \parencite{Mahecha_2010},
%a global-ecosystem estimate of $Q_{10} = 1.4 \pm 0.1$ estimate, 
%based on the analysis of 60 gas flux exchange measurement towers of  the FLUXNET network. The authors concluded that the global-ecosystem estimate does not depend on the biome and it can thus a single value can be used for the purposes of our 0-D TCV model.}

Recent work based on both observational data extracted from the global FLUXNET network and global climate model
simulations forced by emission scenarios of increasing severity \parencite{Niu_2021} indicates that
the cross-biome spread of $Q_{10}$ values should decrease as global
temperature warms converging to a global average of 1.44 in a high
emission scenario. This value is also quite close to the global $Q_{10}$
value of 1.5 that we use.

\subsection*{Carbon absorption by the ocean}

\subsubsection*{Diffusion from the atmosphere to the ocean}

Following Lade et al.'s (2018)\parencite{Lade_2018} two-box ocean carbon model, we represent the change in carbon stock $C_{\rm M}$ in the oceanic mixed layer as:
\begin{equation}
\frac{\mathrm{d}C_{\mathrm{M}}}{\mathrm{d}t}=F_{\mathrm{A}\rightarrow \mathrm{M}}-F_{\mathrm{M}\rightarrow \mathrm{D}}. \label{eq:Ocean-carbon-flux-components-Lade-ea2018}
\end{equation}
Here $F_{\mathrm{A}\rightarrow \mathrm{M}}$ is the flux of carbon from the atmosphere to the mixed layer driven by the gradient of partial pressure $p$ of carbon dioxide, while $F_{\mathrm{M}\rightarrow \mathrm{D}}$ represents the flux of carbon from the upper ocean to the deep ocean through the solubility and biological pumps. 

The equation representing the transport of carbon from the atmosphere to the ocean \parencite{Lade_2018} is:
\begin{equation}
F_{\mathrm{A}\rightarrow \mathrm{M}}=\frac{DC_{\mathrm{M},0}}{rp(C_{\mathrm{M},0}, 0)}\left(C_{\mathrm{A}}-p(C_{\mathrm{M}},\Delta T)\right), \label{eq:Ocean-carbon-Lade-ea2018}
\end{equation}			
with $\Delta T$ being the global temperature anomaly for the atmosphere and near-surface ocean from preindustrial values, and 
\begin{equation}
p(C_{\mathrm{M}},\Delta T)=C_{\mathrm{A},0}\left(\frac{C_{\mathrm{M}}}{C_{\mathrm{M},0}}\right)^{r}\frac{1}{1-D_{T}\Delta T}. \label{eq:Ocean-carbon-partial-pressure-Lade-ea2018}
\end{equation}
Here $C_{\mathrm{M},0}$ is the amount stored in the mixed layer at preindustrial equilibrium\parencite{Sarmiento2002}, $C_{\mathrm{A},0}$
is the carbon stock in the atmosphere, and $r$ is the Revelle buffer factor, equal to the ratio of instantaneous change in CO$_2$ to the change in total dissolved inorganic carbon. Thus, $r$ is a measure of the resistance to atmospheric CO$_2$ being absorbed at the ocean surface \parencite{Sabine_2004,Goodwin2007}. The CO$_2$ solubility here is temperature dependent with a factor $D_{T}=4.23\%$~K$^{-1}$, cf.~ref.
\parencite{Takahashi_1993}.  We note that the Revelle factor, while kept fixed for simplicity, is projected to increase as anthropogenic carbon emissions rise\parencite{Jiang_2019}. The value set by Lade et al that we adopt\parencite{Lade_2018} corresponds to its median value across emission scenarios.

%	\cj{should we point out that if $\Delta T$ is larger than $1/D_T$, this equation breaks down?}

\subsubsection*{Carbon flux due to overturning flow}

Here $F_{\mathrm{M}\rightarrow \mathrm{D}}^P$ represents the flux of carbon from the oceanic mixed layer to the deeper ocean through the solubility pump, with superscript $P$ for the physical, as opposed to $B$ for the biological pump. We follow Lade et al. \parencite{Lade_2018} in writing
\begin{equation}
F_{\mathrm{M}\rightarrow \mathrm{D}}^P=w_{0}(1-w_{T}\Delta T)(C_{\mathrm{M}}-C_{\mathrm{M},0}). \label{eq:Solubility-pump-Lade-ea2018}
\end{equation}
Here $w_{0}$ represents the rate of exchange of water from to mixed layer to the deep ocean, while $w_{T}$ is the weakening of the overturning circulation % that is considered to decrease 
as global temperature $T$ increases.

% \cj{Also here, if $\Delta T>1/w_T$ the flux reverses sign. Should we address this?}

\subsubsection*{Temperature-dependent phytoplankton growth}

$F_{\mathrm{M}\rightarrow \mathrm{D}}^B$ is the second component of $F_{\mathrm{M}\rightarrow \mathrm{D}}$ and represents the flux of carbon from the oceanic mixed layer to the deeper ocean through the the biological pump. This flux is determined both by the amount of phytoplankton growth in the surface ocean and by the fraction of that biomass being exported into the deep ocean. 

A $Q_{10}$ model is used to encode phytoplnakton growth. Eppley\parencite{Eppley1972} used experimental data on a wide range of species of phytoplankton from each of the world's oceans to derive a temperature-dependent maximum growth rate leading to a $Q_{10}$ value of 1.88. However, Eppley-type estimates \parencite{Eppley1972,Bissinger_2008,Regaudie_de_Gioux_2012} relate to maximum growth rates. Instead, field estimates of phytoplankton growth rates allow one to derive estimates accounting for nutrient-limited conditions. Sherman et al (2016)\parencite{Sherman_2016} report a global $Q_{10}$ value of 1.47 $\pm0.08$, based on in situ observations. Based on this, we chose $Q_{10} = 1.5$, as reported in Table~\ref{tab:param-ocean-carbon}.

\subsubsection*{Fraction of phytoplankton transported to the deep ocean}

The carbon flux to the deeper ocean through the sinking of organic carbon is captured through the modelling of primary productivity using a $Q_{10}$-based model \parencite{Eppley1972,Sherman_2016} and the temperature-dependent fraction of biological carbon exported to the deeper ocean by the biopump. We are interested in the deviation of ${\rm{GPP}}_{\mathrm{phyto}}$ from the preindustrial equilibrium ${\rm{GPP}}_{\rm phyto, 0} = p_{0}$, i.e., we consider $\Delta {\rm{GPP}}_{\rm{phyto}}(T)=p_{0} Q_{10}^{\Delta T/10}-p_{0}$. Three alternative models \parencite{Laws2000,Dunne2005,Henson2011} have been proposed to represent the fraction of biological primary production that is preserved from respiration and recycling in the upper ocean and reaches 100 m depth. However, in order to account for a removal of atmospheric carbon from the upper ocean for a duration of more than 100 years, a depth of over 1,000 m has to be reached by the biopump flux.

We relied, therefore, on unpublished data shared by S. Henson and used in Henson et al (2012) \parencite{Henson2012} to derive a fraction of phytoplankton primary productivity exported over a depth of 1,000 m or more. This fraction is referred to as $f_{H}(\Delta T)$ and thus
\begin{equation}
F_{\mathrm{M}\rightarrow \mathrm{D}}^B(T)=f_{H}(\Delta T) \Delta {\rm{GPP}}_{{\rm{phyto}}}(T), \label{eq:Biocarbon-flux}
\end{equation}
with $T$ in Kelvin, and
\begin{equation}
f_{H}(\Delta T) = a_{\rm {exp}}  \exp \left(b_{\rm {exp}}  \left((T + 7.15) - 273.15  \right) \right ). \label{eq:Export_1000m}
\end{equation}
The 7.15 K added to $T$ above represents the sea surface temperature that was used for the measurements yielding these formulas \parencite{Henson2012}.
%\cj{I don't understand this and would be surprised if this 7.15 degrees added is needed. Also kind of by definition, you cannot measure a temperature curve only at one temperature? Erik, it would be good if you could explain it to me.}. 
Here $a_{\rm{exp}} = 0.0196$ and $b_{\rm{exp}}  = 0.0526$. The dataset in question can be found in ref.\parencite{Henson2012}.

\subsection*{Impact of temperature stress on land vegetation }

%\noindent \cm{Erik, This subsubsection belongs here, after the carbon fluxes to the ocean.}

To reflect the growing evidence of the impact of increased climate stress on
land vegetation\parencite{Allen2010,Xu_2020,Hammond_2022},
we introduced equation~\eqref{eq:EnvDegr} that describes the degradation of the terrestrial land vegetation. %encodes a cumulative temperature-dependent stress function into the TCV model. 
The level $V$ of degradation affects the global-level distribution of vegetation through its effect on Eq.~\ref{eq:veg-ramp} representing the temperature response function $g_T(T,V)$ on of the vegetation productivity. 

We repeat below, for the readers' convenience, equation~\eqref{eq:EnvDegr},
\begin{equation}
\dot{V} =  (f(T) - \mu V)(1-V). \label{eq:K-cumulative stress}
\end{equation}
The function $f$ captures the global vegetation mortality rate change in response to an increase in global average temperature $T$ and it is defined by			
\begin{equation}
f(T) = \begin{cases}
	0, &\text{if}\ {\Delta T} \leq 0 ,\\
	a_{\mathrm{d}}\Delta T+b_{\mathrm{d}}{\Delta T}^2, &\text{if}\ {\Delta T} > 0,
\end{cases}
\label{eq:veg_mortality}
\end{equation}

The nonlinear rise in tree‐mortality during hotter droughts is now well established \parencite{Hartmann_2018,Hammond_2022}, as is the parallel decline in biomass production under combined heat and water stress \parencite{RuizBenito2014}. A simple quadratic mortality law, \(f=f(T)\), with linear and nonlinear coefficients \(a_{\mathrm d}\) and \(b_{\mathrm d}\), respectively, reproduces these responses across vegetation types \parencite{Lobell2011,RuizBenito2013}.

With parameters chosen from the species‐specific range in \parencite{RuizBenito2014}, the TCV model yields an average disturbance-induced mortality of \(0.023\) for 2000–2100 under \(\mathrm{RCP}\,8.5\) (Eq.~\eqref{eq:veg_mortality}). This agrees with the LPJmL Dynamic Global Vegetation Model, which returns \(0.02\text{–}0.04\) for the same scenario, depending on forest type \parencite{Pugh2020}.

Cumulative temperature stress, \(V\), evolves toward \(\min\{1,\,f(T)/\mu\}\) and relaxes to zero when \(f(T)=0\), on a timescale \(\mu\) that represents post-disturbance recovery. Satellite data give \(\mu\approx5\text{–}13~\mathrm{yr}\) for most biomes \parencite{Bartels_2016,Goetz_2006,CuevasGonzalez2009}, extending to \(\approx30~\mathrm{yr}\) in boreal forests \parencite{Dobor_2018} and \(>200~\mathrm{yr}\) in tropical reconstructions \parencite{Cole_2014}.

As \(V\) increases, vegetation least tolerant to heat is lost and both the lower bound of optimum temperature, \(T_{o,\ell}(V)\), and the minimum viable temperature, \(T_{v,\ell}(V)\), shift linearly from \(T_{o,\ell}^{-},\,T_{v,\ell}^{-}\) to \(T_{o,\ell}^{+},\,T_{v,\ell}^{+}\) (Eqs.~\eqref{eq:optimum}–\eqref{eq:minimum}). Pre-disturbance limits are from \parencite{Huang2019,Lancaster2020}. Rising \(T\) and \(V\) thus drive thermophilization: cold-adapted C\textsubscript{3} species give way to warm-adapted C\textsubscript{4} (and some warm-adapted C\textsubscript{3}) vegetation \parencite{Nolan2018}. Accordingly, \(T_{v,\ell}^{+}\) and \(T_{o,\ell}^{+}\) are set to the minimum viable and optimum temperatures of C\textsubscript{4} plants \parencite{Sage1999,Still2003}.

\begin{figure}[h!]

\centering

\subfigure[Minimum optimum temperature $T_{o,\ell}$]{ \label{fig:MinOptVeg}
% label of upper left panel    
\centering \includegraphics[scale=0.3]{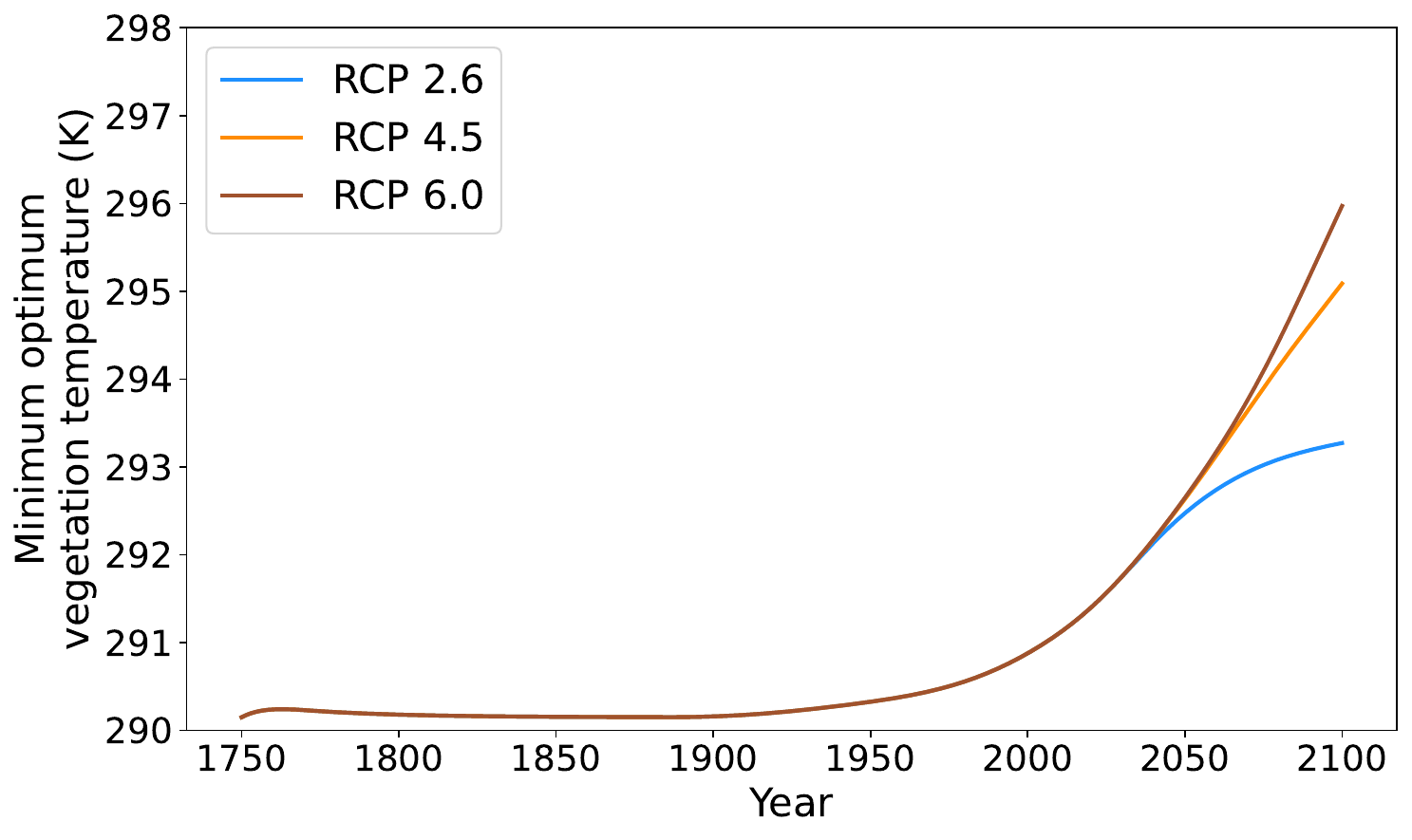}
} \subfigure[Minimum viable temperature $T_{v,\ell}$]{ \label{fig:MinViabVeg}
% label of upper right panel    
\centering \includegraphics[scale=0.3]{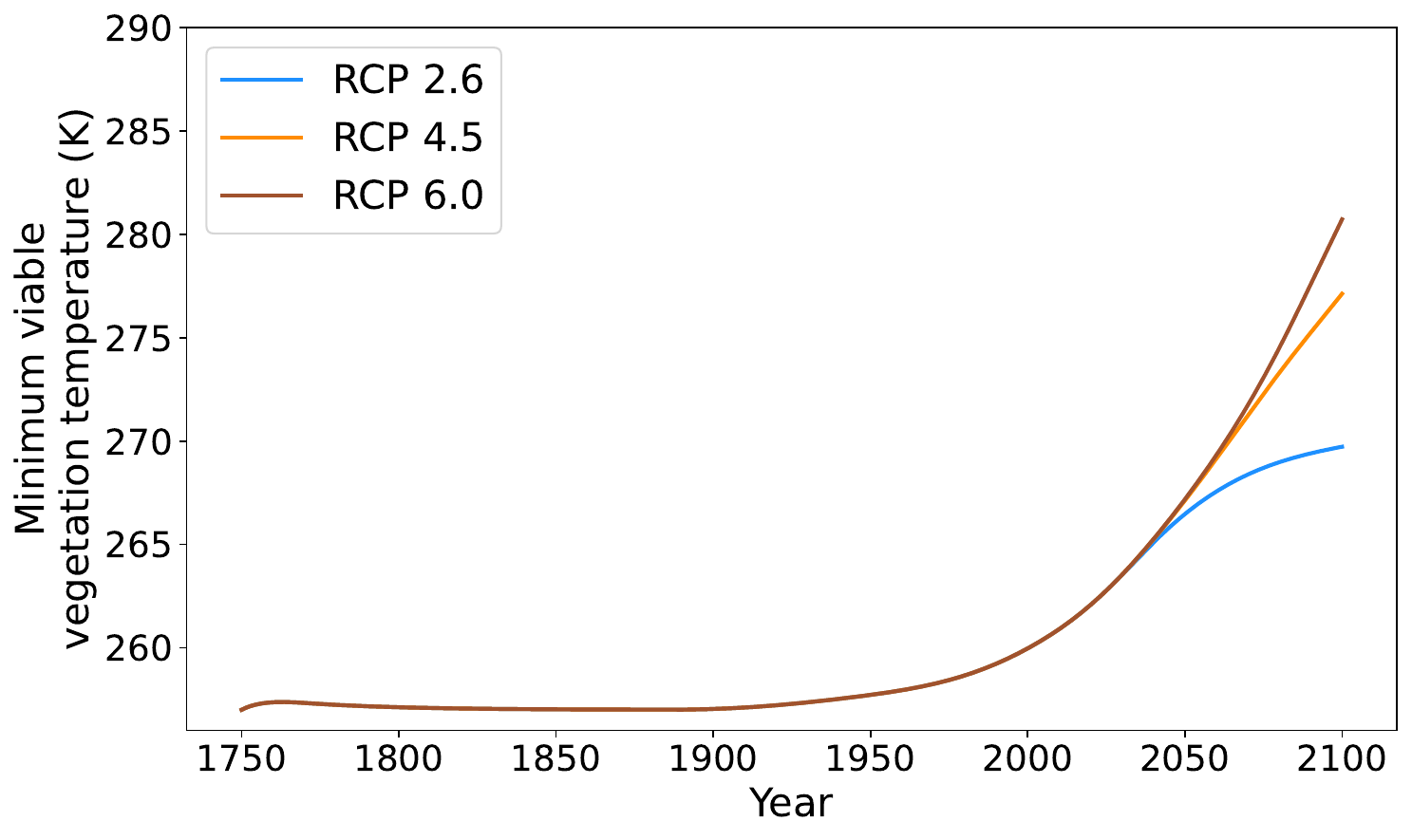}
}

%%%%%%%%%%%%%%%%%%%%%%%%%%%%%
%%%%%   middle panels     %%%%%%%%%%%%%%%%
\subfigure[Pre-disturbance global vegetation productivity dependence on temperature $g_{T}(T,V)$]{ \label{fig: gCa}
% label of upper left panel    
\centering \includegraphics[scale=0.3]{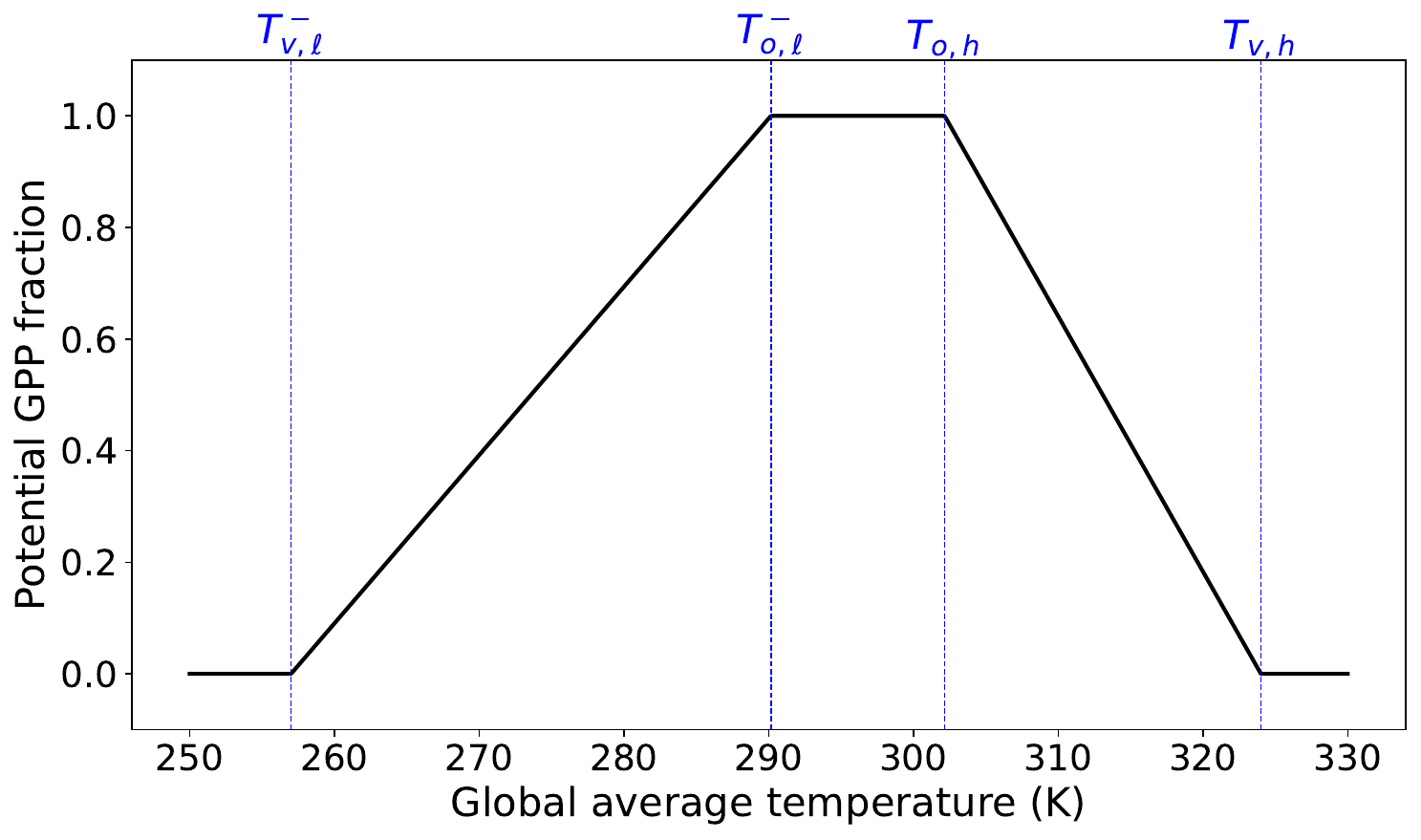}
} \subfigure[Upper limit of global vegetation productivity dependence on temperature $g_{T}(T,V)$]{ \label{fig: gCaSim}
% label of upper right panel    
\centering \includegraphics[scale=0.3]{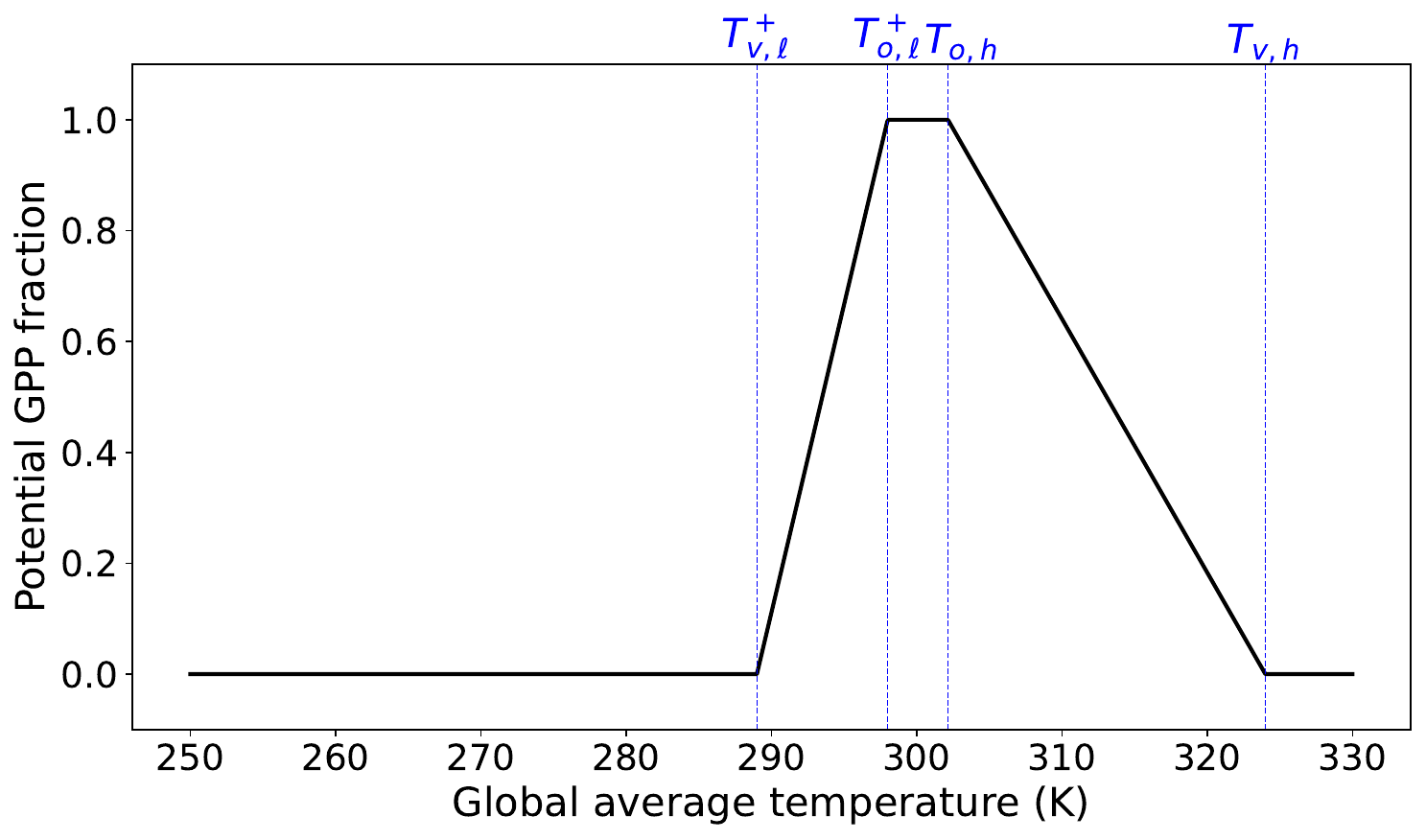}
}

%%%%%%%%%%%%%%%%%%%%%%%%%%%%%
%%%%%    lower panels     %%%%%%%%%%%%%%%%
\subfigure[Vegetation carbon response function $g_{\mathcal C}(\mathcal C)$]{ \label{fig: gT_min2}
	% label of upper left panel    
	\centering \includegraphics[scale=0.3]{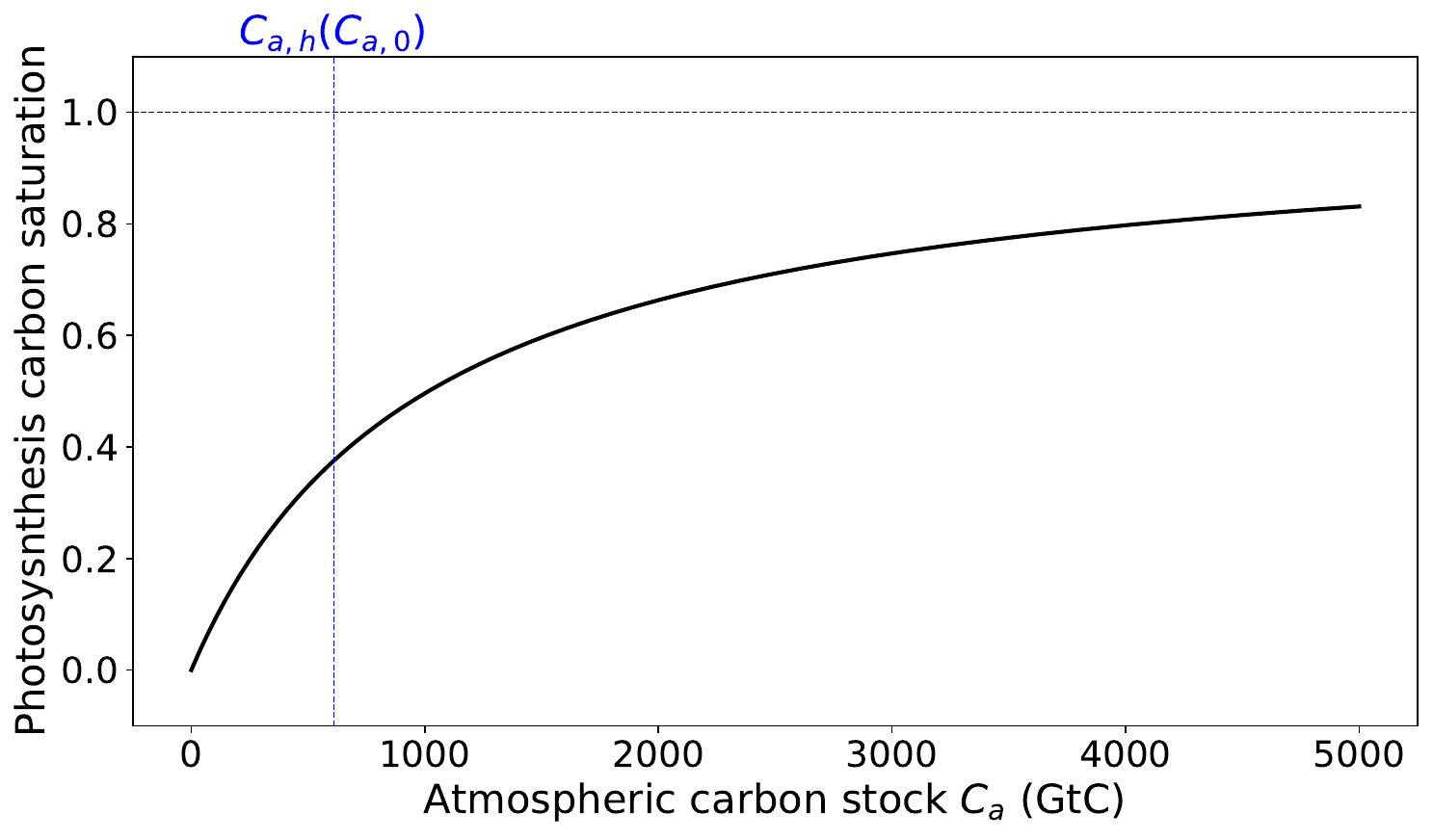}
} \subfigure[Upper limit of global vegetation productivity dependence on temperature $g_{T}(T,V)$]{ \label{fig: gT_max2}
	% label of upper right panel    
	\centering \includegraphics[scale=0.3]{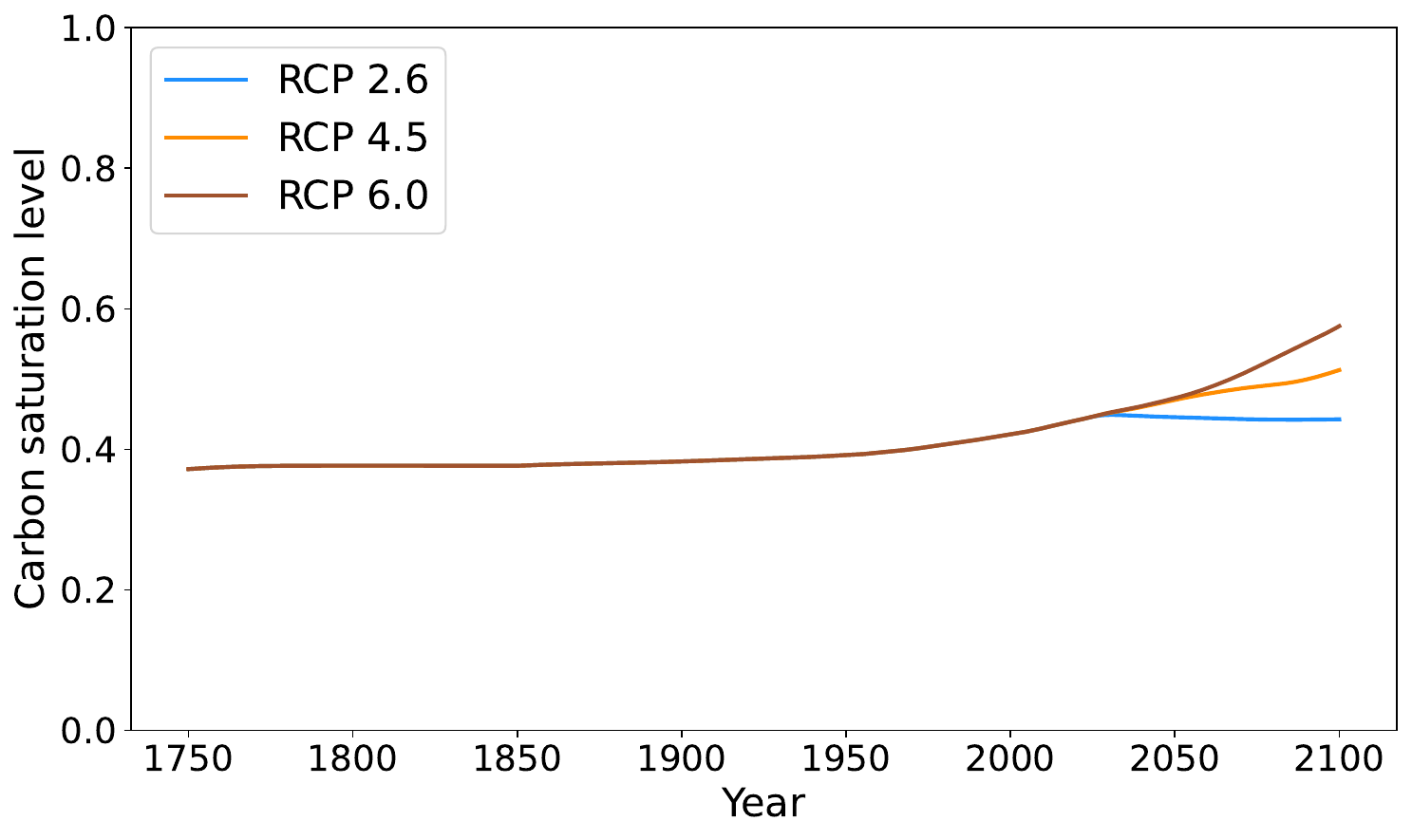}
}

\caption{{\footnotesize{\bf Evolution of the lower optimum and viable 
		temperatures $T_{o,\ell}$ and $T_{v,\ell}$ and of carbon saturation level $g_{\mathcal C}(\mathcal C)$ for vegetation;}
	(a, b) evolution of $T_{o,\ell}$ and $T_{v,\ell}$ simulated with the TCV-DAE model subject to the four
	RCPs 2.6, 4.5, and 6.0; (c) % displays the function 
	global-ecosystem response $g_{T}(T,V)$ of 
	vegetation productivity to global temperature $T$ for global ecosystem degradation $V = 0 $; (d) global-ecosystem response $g_{T}(T,V)$ of 
	vegetation productivity to global temperature $T$ for global ecosystem degradation $V = 1$;  (e) displays the carbon response function of the global vegetation $g_{\mathcal C}(\mathcal C)$ see Eq. \ref{Eq.  gc(C)} and and Table \ref{tab:param-terrestrial-carbon}. $C_{A,h}$ represents the preindustrial carbon atmosphere content; and (e) Evolution of $g_{\mathcal C}(\mathcal C)$ subject to three moderate
	RCPs: RCP 2.6, RCP 4.5, and RCP 6.0. Parameter values that label the vertical dash-dotted lines in panels (c), (d), and (e) are presented in Table~\ref{tab:param-terrestrial-carbon}.} }  %\cj{Panel c would really benefit from a companion: the same plot, but with $V=1$. This would make things much clearer I think}
	%					with initial values from Huang et al. \parencite{Huang2019};
	%					 see equations~\eqref{eq:optimum}, \eqref{eq:minimum},
	%					and Table \ref{tab:param-terrestrial-carbon}. 
	%					\cm{Erik: What does “initial values” mean? The plot is one of $g_T$ as a function of $T$, for $V$ fixed. What is this fixed value of $V$ for which the plot is done? Why include $V$ as an argument in the upper abscissa? Maybe we can discuss this later.}
	\label{Fig:Lower opt veg temp} % label of the full figure
\end{figure}

Figure~\ref{Fig:Lower opt veg temp} illustrates the resulting trajectories of \(T_{v,\ell}(V)\) and \(T_{o,\ell}(V)\) under the three moderate RCP pathways; parameter values are listed in Table~\ref{tab:param-terrestrial-carbon}.

Eqs.~\eqref{eq:optimum} and \eqref{eq:minimum} below:\\
\label{eq:optimum-and-minimum} 
\begin{align}
T_{o,\ell}(V)=\begin{cases}
	\begin{array}{ccc}
		T_{o,\ell}^{-} & {\rm if} & V\leq 0,\\
		{\displaystyle {T_{o,\ell}^{-}+\left(T_{o,\ell}^{+}-T_{o,\ell}^{-}\right)V}} & {\rm if} & 0<V\leq 1,\\
		T_{o,\ell}^{+} & {\rm if} & 1<V;
\end{array}\end{cases}\label{eq:optimum}\\
T_{v,\ell}(V)=\begin{cases}
	\begin{array}{ccc}
		T_{v,\ell}^{-} & {\rm if} & V\leq 0,\\
		{\displaystyle {T_{v,\ell}^{-}+\left(T_{v,\ell}^{+}-T_{v,\ell}^{-}\right)V}} & {\rm if} & 0<V\leq 1,\\
		T_{v,\ell}^{+} & {\rm if} & 1<V.
\end{array}\end{cases}\label{eq:minimum}
\end{align}

Figure~\ref{Fig:Lower opt veg temp} displays the evolutions of $T_{v,\ell}(V)$ and $T_{o,\ell}(V)$
subject to the three more moderate RCP emission scenarios, as well as their dependence on global temperatures $T$.
Parameter values are given in Table~\ref{tab:param-terrestrial-carbon}.

\subsection*{ Reduction of the TCV model to the TCV-DAE model}
%			\cj{The reduction is four to three, with then taking $V$ constant leading to the 2 ODE + 1 AE system, maybe we should change the title of this section \cm{Erik, Pls. see suggestion at {\bf Diffusion from the atmosphere to the ocean}, Eq.~(M.13), about notation.}

To reduce our TCV model governed by the four-ODE system~\eqref{eq:TC-4box} to the TCV-DAE model governed by the mixed, differential-and-algebraic equation system~\eqref{eq:TC-3box}, we first introduce a variable $C_{\rm S}$ 
that accounts for the total amount of carbon in the atmosphere as well as for the dissolved inorganic carbon in the upper ocean, 
% \cj{this is the single occurence of 'DIC' in the paper, we should either define it or otherwise just call it 'carbon in the mixed layer'}, 
with $C_{\rm S} = C_{\mathrm{A}}+C_{\mathrm{M}}$. The evolution of $C_\mathrm{S}$ satisfies 
\begin{equation}
	\dot{C_{\rm S}} =  \dot{C}_{\mathrm{A}}+\dot{C}_{\mathrm{M}} = - F_{\mathrm{A} \to {\rm L}}(T,C_{\mathrm{A}},V)-F_{\mathrm{M}\rightarrow \mathrm{D}}^P(T) - F_{\mathrm{M}\rightarrow \mathrm{D}}^B(T)\label{eq:total-carbon-flux} + e(t).
\end{equation}		
% \cm{Erik: Pls. change signs.}

Defining this new variable leads to a system of ODEs for $T$, $C_\mathrm{S}$ and either $C_\mathrm{A}$ or $C_\mathrm{M}$ that is equivalent to the original one. In order to reduce this system, we make the assumption that there is fast equilibration between the atmospheric carbon and the mixed-layer carbon \parencite{Zeebe2001},
% \cj{why this ref? does it also do that or in any way justify this?}, 
i.e. $F_{{\rm A} \to {\rm M}} = 0$ in equation~\eqref{eq:Ocean-carbon-Lade-ea2018}. Using this assumption,  equation~\eqref{eq:Ocean-carbon-partial-pressure-Lade-ea2018} yields $C_\mathrm{A} = C_{{\rm A},0}\left(C_{\mathrm{M}}/C_{\mathrm{M,0}}\right)^{r}1/(1-D_{T}\Delta T)$ which leads to $C_\mathrm{S} = C_\mathrm{A}+C_\mathrm{M} = C_{{\rm A},0}\left(C_{\mathrm{M}}/C_{\mathrm{{\rm M},0}}\right)^{r}1/(1-D_{T}\Delta T) + C_\mathrm{M}$.			
%\begin{equation}
%	C_\mathrm{A} = C_{{\rm A},0}\left(\frac{C_{\mathrm{M}}}{C_{\mathrm{M,0}}}\right)^{r}\frac{1}{1-D_{T}\Delta T},
%\end{equation}				
% \noindent \cm{Erik, Please check. There was a $\beta_1$ in the text before, which Jan thinks is incorrect.}\\
%which leads in turn to 
%\begin{equation}
%	C_\mathrm{S} = C_\mathrm{A}+C_\mathrm{M} = C_{{\rm A},0}\left(\frac{C_{\mathrm{M}}}{C_{\mathrm{{\rm M},0}}}\right)^{r}\frac{1}{1-D_{T}\Delta T} + C_\mathrm{M}.
%\end{equation}

The latter equation provides an implicit definition of $C_\mathrm{M}$, and hence of $C_\mathrm{A}$, as a function of $C_\mathrm{S}$. Over the range of temperatures and carbon stocks we are looking at, the relationship between $C_\mathrm{M}$ and $C_\mathrm{S}$ is a smooth one-to-one function, say $C_\mathrm{M} = {\mathcal F}(C_{\mathrm S})$. In principle, one could invert the relation and write $C_\mathrm{M} = {\mathcal F}^{-1}(C_\mathrm{S}) = {\mathcal G}(C_\mathrm{S})$, plug ${\mathcal G}(C_\mathrm{S})$ into sytem~\eqref{eq:TC-4box} and reduce it therewith  to three standard ODEs for $T, C_{\rm S}$ and $V$. This means that the stability of the steady states can be computed numerically just as one would for an ODE system, keeping in mind the definition of $C_\mathrm{M}$. 

In practice, however, this inversion can only be carried our analytically for $r = 1$ or $r = 2$, i.e., when  ${\mathcal F}(C_{\mathrm S})$ is a linear or a quadratic function. Even for $r = 3$ and $r = 4$, it is well known that analytic solutions exist but are too complicated to be useful in actual calculations. Table~\ref{tab:param-ocean-carbon} shows that $r$ is both larger than 4 and fractional, so that 
%However, for values of the Revelle factor $r$ that are not small enough, \cm{Erik, Jan, This seems to contradict the statement above that we can do it for the range of variables we're interested in. Pls. explain.} 
we cannot, in general, write $C_\mathrm{M}$ explicitly as a function of $C_\mathrm{S}$. 
%I just meant that if $r = 1$ or $r = 2$, you can actually write $C_M$ explicitly as a function of $C_S$, i.e. using the solution of a quadratic. But if r is higher or not an integer, it becomes quite impossible (technically it could work for r=3 or r=4 but it's very unwieldy). If this is too confusing we could just write “Generally, we cannot write $C_M$ explicitly as a function of $C_S$”.
Therefore, we consider the system formally as a DAE in Eq.~\eqref{eq:TC-3box}.
%\begin{subequations}				 
%	\begin{align}
%		& c\dot{T} = Q_{0}\left(1-p\alpha_{{\rm L}}(T)-(1-p)\alpha_{{\rm  O}}(T)\right)-\kappa\left(T-T_{\kappa}\right) + a\ln\left(\frac{C_{\mathrm{S}}-C_\mathrm{M}}{C_{0}}\right),\\
%		& \dot{C_{\rm S}} = -F_{\mathrm{A}\rightarrow \mathrm{L}}(T,C_{\mathrm{S}}-C_{\mathrm{M}},V)-F_{\mathrm{M}\rightarrow \mathrm{D}}^P(T)-F_{\mathrm{M}\rightarrow \mathrm{D}}^B(T) + e(t),\\
%		&\dot{V} = -\mu V(1-V)+f(T)(1-V), \\
%		& 0 = C_{\mathrm{A},0}\left(\frac{C_{\mathrm{M}}}{C_{\mathrm{M,0}}}\right)^{r}\frac{1}{1-D_{T}\Delta T} + C_\mathrm{M} - C_\mathrm{S}.
%	\end{align}
%\end{subequations}	
For constant $V$, this becomes a system of two ODEs and one algebraic equation, which is used for the phase-plane studies in the main text.

			\subsection*{Code availability}

\label{ssec:code} %\cm{(Part of) this subsection could also be turned into an Appendix.}

The code and data used to produce the numerical results and figures is available online on a Zenodo repository.

\printbibliography[check=onlynew]

@Article{Barnett2021,
  author   = {Barnett, M. and Brock, W. A. and Hansen, L. P.},
  journal  = {The Review of Financial Studies},
  title    = {Pricing uncertainty induced by climate change},
  year     = {2021},
  number   = {3},
  pages    = {1024-1066},
  volume   = {33},
  url      = {http://dx.doi.org/10.2139/ssrn.3008833},
}

@Article{Ghil1976,
  author  = {Ghil, M.},
  title   = {Climate stability for a {Sellers-type model}},
  journal = {Journal of the Atmospheric Sciences},
  year    = {1976},
  volume  = {33},
  pages   = {3-20},
}

@Book{Ghil.Chil.1987,
  author        = {Ghil, M. and Childress, S.},
  publisher     = {Springer Science+Business Media},
  title         = {{Topics in Geophysical Fluid Dynamics: Atmospheric Dynamics, Dynamo Theory, and Climate Dynamics; reprinted as an eBook in 2012}},
  year          = {1987},
  address       = {Berlin/Heidelberg},
  shorttitle    = {Topics in Geophysical Fluid Dynamics},
}

@Article{Keeling1976,
  author  = {Keeling, C.D. and Bacastow, R.B. and Bainbridge, A.E. and Ekdhal, C.A. and Guenther, P.R. and Waterman, L.S.},
  title   = {Atmospheric carbon dioxide variations at Mauna Loa Observatory, Hawai},
  journal = {Tellus},
  year    = {1976},
  volume  = {28},
  pages   = {538-551},
}

@Article{Matthews2009,
  author  = {Matthews, H. D. and Gillett, N. P. and Stott, P. A. and Zickfeld, K.},
  title   = {The proportionality of global warming to cumulative carbon emissions},
  journal = {Nature},
  year    = {2009},
  volume  = {459},
  pages   = {829-832},
}

@Article{Pierrehumbert2014,
  author  = {Pierrehumbert, R. T.},
  title   = {Short-lived climate pollution},
  journal = {Annual Review of Earth and Planetary Science},
  year    = {2014},
  volume  = {42},
  pages   = {341-379},
}

@Article{Friedlingstein2015,
  author  = {Friedlingstein, P.},
  journal = {Philosophical Transactions of the Royal Society A},
  title   = {Carbon cycle feedbacks and future climate change},
  year    = {2015},
  pages   = {1--14},
  volume  = {373},
}

@Article{Lenton.tip.2008,
  author        = {Lenton, T.~M. and Held, H. and Kriegler, E. and Hall, J.~W. and Lucht, W. and Rahmstorf, S. and Schellnhuber, H.~J.},
  title         = {Tipping elements in the {Earth's climate system}},
  journal       = {Proceedings of the National Academy of Sciences USA},
  year          = {2008},
  volume        = {105},
  pages         = {1786--1793},
}

@Article{Lorenz1963,
  author        = {Lorenz, E. N.},
  title         = {Deterministic nonperiodic flow},
  journal       = {Journal of the Atmospheric Sciences},
  year          = {1963},
  volume        = {20},
  pages         = {130--141},
}

@Article{Friedling2015,
  author  = {Friedlingstein, P.},
  journal = {Philosophical Transactions of the Royal Society A: Mathematical, Physical and Engineering Sciences},
  title   = {Carbon cycle feedbacks and future climate change},
  year    = {2015},
  number  = {2054},
  pages   = {20140421},
  volume  = {373},
}

@Article{Schwartz2007,
  author  = {Schwartz, S. E.},
  journal = {Journal of Geophysical Research},
  title   = {{Heat capacity, time constant, and sensitivity of Earth's climate system}},
  year    = {2007},
  pages   = {1-12},
  volume  = {112},
}

@Article{Zaliapin2010,
  author  = {Zaliapin, I. and Ghil, M.},
  title   = {Another look at climate sensitivity},
  journal = {Nonlinear Processes in Geophysics},
  year    = {2010},
  volume  = {17},
  pages   = {113-122},
}

@Article{Myhre1998,
  author  = {Myhre, G. and Highwood, E. J. and Shine, K. P. and Stordal, F.},
  title   = {New estimates of radiative forcing due to well mixed greehouse gases},
  journal = {Geophysical Research Letters},
  year    = {1998},
  volume  = {25},
  number  = {14},
  pages   = {21715-2718},
}

@Article{Rombouts2015,
  author  = {Rombouts, J. and Ghil, M},
  title   = {Oscillations in a simple climate-vegetation model},
  journal = {Nonlinear Processes in Geophysics},
  year    = {2015},
  volume  = {22},
  pages   = {275-288},
}

@Article{Kallen1979,
  author  = {K\"all\'en, E. and Crafoord, C. and Ghil, M.},
  title   = {Free oscillations in a climate model with ice-sheet dynamics},
  journal = {Journal of the Atmospheric Sciences},
  year    = {1979},
  volume  = {36},
  number  = {12},
  pages   = {2292-2303},
}

@Article{Sellers.1969,
  author        = {Sellers, W. D.},
  title         = {A global climatic model based on the energy balance of the {Earth atmosphere}},
  journal       = {Journal of Applied Meteorology},
  year          = {1969},
  volume        = {8},
  pages         = {392--400},
}

@Article{Budyko.1969,
  author  = {Budyko, M. I.},
  journal = {{Tellus}},
  title   = {The effect of solar radiation variations on the climate of the {Earth}},
  year    = {1969},
  pages   = {611--619},
  volume  = {21},
}

@Book{Broecker.Peng.1982,
  author    = {Broecker, W. S. and Peng, T.-H.},
  publisher = {Lamont-Doherty Geological Observatory, Columbia University, Palisades, New York},
  title     = {{Tracers in the Sea}},
  year      = {1982},
}

@Article{Meinshausen2011,
  author  = {Meinshausen, M. and Smith, S. J. and Calvin, K. and Daniel, J. S. and Kainuma, M. L. T. and Larmarque, J. J. and Matsumoto, K. and Montzka, S. A. and Raper, S. C. B. and Riahi, K. and Thomson, A. and Velders, G, J. M. and van Vuuren, D. P. P.},
  journal = {Climatic Change},
  title   = {{The RCP greenhouse gas concentrations and their extentions from 1765 to 2300}},
  year    = {2011},
  pages   = {213-241},
  volume  = {109},
}

@Article{Ghil2019,
  author  = {Ghil, M.},
  title   = {A century of nonlinearity in the geosciences.},
  journal = {Earth and Space Science},
  year    = {2019},
  volume  = {6},
  pages   = {1007-1042},
}

@Article{Ghil2020,
  author  = {Ghil, M. and Lucarini, V.},
  title   = {The physics of climate variability and climate change},
  journal = {Reviews of Modern Physics},
  year    = {2020},
  volume  = {92},
  number  = {3},
  pages   = {1-77},
}

@Article{Lenton2019,
  author  = {Lenton, T. M. and Rockström, J. and Gaffney, O. and Rahmstorf, S. and Richardson, K. and Steffen, W. and Schellnhuber, H. J.},
  title   = {Climate tipping points - too risky to bet against},
  journal = {Nature},
  year    = {2019},
  volume  = {575},
  pages   = {592-595},
}

@Article{Slingo2011,
  author  = {Slingo, J. and Palmer, T. M.},
  journal = {Philosophical Transactions of the Royal Society A},
  title   = {Uncertainty in weather and climate prediction.},
  year    = {2011},
  pages   = {4751-4767},
  volume  = {369},
}

@Book{Guckenheimer1983,
  author    = {Guckenheimer, J. and Holmes, P.},
  publisher = {Springer-Verlag},
  title     = {{Nonlinear Oscillations, Dynamical Systems and Bifurcations of Vector Fields}},
  year      = {1983},
  address   = {Berlin},
}

@Article{Stommel1961,
  author  = {Stommel, H.},
  title   = {Thermohaline convection with two stable regimes of flow},
  journal = {Tellus},
  year    = {1961},
  volume  = {13},
  pages   = {224-230},
}

@Article{LeTreut1983,
  author  = {Le Treut, H. and Ghil, M.},
  title   = {Orbital forcing, climatic iinteraction, and glaciation cycles.},
  journal = {Journal of Geophysical Research Oceans},
  year    = {1983},
  volume  = {88},
  pages   = {5167-5190},
}

@Article{Ghil.2001,
  author        = {Ghil, M.},
  title         = {Hilbert problems for the geosciences in the 21st century},
  journal       = {Nonlinear Processes in Geophysics},
  year          = {2001},
  volume        = {8},
  number        = {4/5},
  pages         = {211--211},
  doi           = {10.5194/npg-8-211-2001},
}

@Article{Held.gap.2005,
  author        = {Held, I. M.},
  title         = {The gap between simulation and understanding in climate modeling},
  journal       = {Bulletin of the American Meteorological Society},
  year          = {2005},
  volume        = {{86}},
  pages         = {1609--1614},
  doi           = {10.1175/bams-86-11-1609},
}

@Article{North1981,
  author  = {North, G. R. and Cahalan, R. F. and Coakley, J. A.},
  title   = {Energy balance climate models},
  journal = {Reviews of Geophysics and Space Physics},
  year    = {1981},
  volume  = {19},
  pages   = {19-121},
}

@Article{Craf.Kall.1978,
  author  = {Crafoord, C. and K{\"a}ll{\'e}n, E.},
  journal = {Journal of the Atmospheric Sciences},
  title   = {A note on the condition for existence of more than one steady-state solution in {Budyko-Sellers type models}},
  year    = {1978},
  number  = {6},
  pages   = {1123--1125},
  volume  = {35},
}

@Book{Arnold.ODE.1983,
  author        = {Arnold, V.},
  publisher     = {Springer},
  title         = {{Geometric Methods in the Theory of Ordinary Differential Equations}},
  year          = {1983},
  address       = {New York/Berlin},
  pages         = {334 pp.},
}

@Article{Held.Suarez.1974,
  author  = {Held, I. M. and Suarez, M. J.},
  journal = {Tellus},
  title   = {Simple albedo feedback models of the ice caps},
  year    = {1974},
  pages   = {613--629},
  volume  = {26},
}

@Article{Pierrehumbert.2004,
  author    = {Pierrehumbert, R. T.},
  title     = {High levels of atmospheric carbon dioxide necessary for the termination of global glaciation},
  journal   = {Nature},
  year      = {2004},
  volume    = {429},
  number    = {6992},
  pages     = {646},
  publisher = {Nature Publishing Group},
}

@Article{Palmer.Stevens.2019,
  author    = {Palmer, T. and Stevens, B.},
  journal   = {Proceedings of the National Academy of Sciences},
  title     = {The scientific challenge of understanding and estimating climate change},
  year      = {2019},
  number    = {49},
  pages     = {24390--24395},
  volume    = {116},
  publisher = {National Acad. Sciences},
}

@Article{Schneider.Dick.1974,
  author  = {Schneider, S. H. and Dickinson, R. E.},
  title   = {{Climate modelling}},
  journal = {Reviews of Geophysics and Space Physics},
  year    = {1974},
  volume  = {25},
  pages   = {447--493},
}

@Article{Held2010,
  author  = {Held, I. M. and Winton, M. and Takahashi, K. and Delworth, T. and Zeng, F. and Valis, G. K.},
  journal = {Journal of Climate},
  title   = {Probing the fast and slow components of global warming by returning abruptly to preindustrial forcing},
  year    = {2010},
  pages   = {2418-2427},
  volume  = {23},
}

@Article{Goldblatt2012,
  author  = {Goldblatt, C. and Watson, A. J.},
  title   = {The runaway greenhouse: implications for future climate change, geoengineering and planetary atmonsphere},
  journal = {Philosophical Transactions of the Royal Society A},
  year    = {2012},
  volume  = {370},
  pages   = {4197-4216},
}

@Article{Huang2019,
  author  = {Huang, M. and Piao, S. and Ciais, P. and Penuelas, J. and Wang, X. and Keenan, T. F. and Peng, S. and Berry, J. A. and Wang, K. and Mao, J. and Alkama, R. and Cescatti, A. and Cuntz, M. and de Deurwaerder, H. and Gao, M. and He, Y. and Liu, Y. and Luo, Y. and Myneni, R. B. and Niu, S. and Shi, X. and Yuan, W. and Verbeeck, H. and Wang, T. and Wu, J. and Janssens, I. A.},
  journal = {Nature Ecology and Evolution},
  title   = {Air temperature optima of vegetation productivity across global biomes},
  year    = {2019},
  pages   = {772-779},
  volume  = {3},
}

@Article{Phillips2009,
  author  = {Phillips, O. L. and Aragao, L. E. C. and Lewis, S. L. and Fisher, J. B. and Lloyds, J. and Lopez-Gonzalez, G. and Malhi, Y. and Monteagudo, A. and Peacock, J. and Quesada, C. A. and van der Hejden, G. and Almeida, S. and Amaral, I. and Arroyo, L. and Aymard, G. and Baker, T. R. and Banki, O. and Blanc, L. and Bonal, D. and Brando, P. and Chave, J. and Alves de Oliveira, A. and Cardoso, N. D. and Czimczik, C. I. and Feldpausch, T. R. and Freitas, M. A. and Gloor, E. and Higuchi, N. and Jimenez, E. and Lloyd, G. and Meir, P. and Mendoza, C. and Morel, A. and Neill, D. A. and Nepstad, D. and Patino, S. and Penuela, M. C. and Prieto, A. and Ramirez. F. and Schwarz, M. and Silva, J. and Silveira, M. and Thomas, A. S. and ter Steefe, H. and Stropp, J. and Vasquez, R. and Zelazowski, P. and Alvarez Davila, E. and Andelman, S. and Andrade, A. and Chao, K-J. and Erwin, T. and Di Fiore, A. and Honorio C., E. and Keeling, H. and Killeen, T. J. and Laurance, W. F. and Pena Cruz, A. and Pitman, N. C. A. and Nunes Vargas, P. and Ramirez-Angulo, H. and Rudas, A. and Salamao, R. and Silva, N. and Terborgh, J. and Torres-Lezama, A.},
  journal = {Science},
  title   = {Drought sensitivity of the Amazon rainforest},
  year    = {2009},
  pages   = {1344-1347},
  volume  = {323},
}

@Article{Mantgem2009,
  author  = {van Mantgem, P. J. and Stephenson, N. L. and Byrne, J. C. and Daniels, L. D. and Franklin, J. F. and Fule, P. Z. and Harmon, M. E. and Larson, A. J. and Smith, J. M. and Taylor, A. H. and Veblen, T. T.},
  journal = {Science},
  title   = {Widespread increase of tree mortality rates in the Western United States},
  year    = {2009},
  pages   = {521-524},
  volume  = {323},
}

@Article{Williams2012,
  author  = {Williams, A. P. and Allen, C. D. and Macalady, A. K. and Griffin, D. and Woodhouse, C. A. and Meko, D. M. and Swetnam, T. W. and Rauscher, S. A. and Seager, R. and Grissino-Mayer, H. D. and Dean, J. S. and Cook, E. R. and Gandodagamage, C. and Cai, M. and McDowell, N. G.},
  journal = {Nature Climate Change},
  title   = {Temperature as a potential driver of regional forest drought stress and tree mortality},
  year    = {2012},
  pages   = {292-297},
  volume  = {3},
}

@Article{Allen2010,
  author  = {Allen, D. A. and Macalady, A. K. and Chenchoui, H. and Bachelet, D. and MacDowell, N. and Vennetier, M. and Kitzberger, T. and Rigling, A. and Breshears, D. D. and Hogg, E. H. T. and Gonzalez, P. and Fensham, R. and Zhang, Z. and Castro, J. and Demidova, N. and Lim, J-H. and Allard, G. and Running, S. W. and Semerci, A. and Cobb. N.},
  journal = {Forest Ecology and Management},
  title   = {A global overview of drought and heat-induced tree mortality reveals emerging climate change risks for forests},
  year    = {2010},
  pages   = {660-684},
  volume  = {259},
}

@Article{Zhao2010,
  author  = {Zhao, M. and Running, S. W.},
  journal = {Science},
  title   = {Drought-induced reduction in global terrestrial net primary production from 2000 through 2009},
  year    = {2010},
  pages   = {940-943},
  volume  = {329},
}

@Book{Lackner.ea.2022,
  editor    = {Lackner, M. and Sajjadi, B. and Chen, W.-Y.},
  publisher = {Springer Nature},
  title     = {{Handbook of Climate Change Mitigation and Adaptation}},
  year      = {2022},
  edition   = {{3rd}},
  doi       = {10.1007/978-3-030-72579-2},
}

@Article{Popp.ea.2016,
  author    = {Popp, M. and Schmidt, H. and Marotzke, J.},
  journal   = {Nature Communications},
  title     = {Transition to a moist greenhouse with {CO}2 and solar forcing},
  year      = {2016},
  number    = {1},
  volume    = {7},
  doi       = {10.1038/ncomms10627},
  publisher = {Springer Science and Business Media {LLC}},
}

@Article{Steffen.ea.2018,
  author    = {W. Steffen and J. Rockström and K. Richardson and T. M. Lenton and C. Folke and D. Liverman and C. P. Summerhayes and A. D. Barnosky and S. E. Cornell and M. Crucifix and J. F. Donges and I. Fetzer and S. J. Lade and M. Scheffer and R. Winkelmann and H. J. Schellnhuber},
  journal   = {Proceedings of the National Academy of Sciences},
  title     = {{Trajectories of the Earth system in the Anthropocene}},
  year      = {2018},
  number    = {33},
  pages     = {8252--8259},
  volume    = {115},
  doi       = {10.1073/pnas.1810141115},
  publisher = {Proceedings of the National Academy of Sciences},
}

@Electronic{IIASA2009,
  author       = {IIASA},
  howpublished = {Online},
  title        = {RCP Database},
  url          = {http://www.iiasa.ac.at/web-apps/tnt/RcpDb},
  year         = {2009},
}

@Article{McKay2022,
  author    = {D. I. {Armstrong McKay} and A. Staal and J. F. Abrams and R. Winkelmann and B. Sakschewski and S. Loriani and I. Fetzer and S. E. Cornell and J. Rockström and T. M. Lenton},
  journal   = {Science},
  title     = {{Exceeding 1.5~deg C global warming could trigger multiple climate tipping points}},
  year      = {2022},
  number    = {6611},
  volume    = {377},
  doi       = {10.1126/science.abn7950},
  publisher = {American Association for the Advancement of Science ({AAAS})},
}

@Article{Friedlingstein_2022,
  author  = {P. Friedlingstein and M. O{\textquotesingle}Sullivan and M. W. Jones and others},
  journal = {Earth System Science Data},
  title   = {{Global Carbon Budget 2022}},
  year    = {2022},
  number  = {11},
  pages   = {4811--4900},
  volume  = {14},
  doi     = {10.5194/essd-14-4811-2022},
}

@Article{Robinson_2012,
  author    = {A. Robinson and R. Calov and A. Ganopolski},
  journal   = {Nature Climate Change},
  title     = {Multistability and critical thresholds of the Greenland ice sheet},
  year      = {2012},
  number    = {6},
  pages     = {429--432},
  volume    = {2},
  doi       = {10.1038/nclimate1449},
  publisher = {Springer Science and Business Media {LLC}},
}

@Article{Svirezhev1997,
  author  = {Svirezhev, Y. M. and von Bloh, W.},
  journal = {Ecological Modelling},
  title   = {Climate, vegetation, and global carbon cycle: the simplest zero-dimensional model},
  year    = {1997},
  pages   = {79-95},
  volume  = {101},
}

@Article{BondLamberty2010,
  author  = {Bond-Lamberty, B. and Thomson, Al.},
  journal = {Nature},
  title   = {Temperature-associated increases in the global soil respiration record},
  year    = {2010},
  pages   = {579-585},
  volume  = {464},
}

@Article{Hashimoto_2015,
  author    = {S. Hashimoto and N. Carvalhais and A. Ito and M. Migliavacca and K. Nishina and M. Reichstein},
  journal   = {Biogeosciences},
  title     = {Global spatiotemporal distribution of soil respiration modeled using a global database},
  year      = {2015},
  number    = {13},
  pages     = {4121--4132},
  volume    = {12},
  doi       = {10.5194/bg-12-4121-2015},
  publisher = {Copernicus {GmbH}},
}

@Article{Davidson_2006,
  author    = {E. A. Davidson and I. A. Janssens},
  journal   = {Nature},
  title     = {Temperature sensitivity of soil carbon decomposition and feedbacks to climate change},
  year      = {2006},
  number    = {7081},
  pages     = {165--173},
  volume    = {440},
  doi       = {10.1038/nature04514},
  publisher = {Springer Science and Business Media {LLC}},
}

@Article{Crowther_2015,
  author    = {T. W. Crowther and S. M. Thomas and D. S. Maynard and P. Baldrian and K. Covey and S. D. Frey and L. T. A. van Diepen and M. A. Bradford},
  journal   = {Proceedings of the National Academy of Sciences},
  title     = {Biotic interactions mediate soil microbial feedbacks to climate change},
  year      = {2015},
  number    = {22},
  pages     = {7033--7038},
  volume    = {112},
  doi       = {10.1073/pnas.1502956112},
  publisher = {Proceedings of the National Academy of Sciences},
}

@Article{Warner_2019,
  author    = {D. L. Warner and B. Bond-Lamberty and J. Jian and E. Stell and R. Vargas},
  journal   = {Global Biogeochemical Cycles},
  title     = {Spatial Predictions and Associated Uncertainty of Annual Soil Respiration at the Global Scale},
  year      = {2019},
  number    = {12},
  pages     = {1733--1745},
  volume    = {33},
  doi       = {10.1029/2019gb006264},
  publisher = {American Geophysical Union ({AGU})},
}

@Article{Hursh_2016,
  author    = {A. Hursh and A. Ballantyne and L. Cooper and M. Maneta and J. Kimball and J. Watts},
  journal   = {Global Change Biology},
  title     = {The sensitivity of soil respiration to soil temperature, moisture, and carbon supply at the global scale},
  year      = {2016},
  number    = {5},
  pages     = {2090--2103},
  volume    = {23},
  doi       = {10.1111/gcb.13489},
  publisher = {Wiley},
}

@Book{Prentice2001,
  author    = {Prentice, I. C. et al.},
  publisher = {Cambridge University Press},
  title     = {Climate Change 2001: The Scientific Basis},
  year      = {2001},
}

@Book{Hoff1898,
  author    = {van't Hoff, J. H.},
  editor    = {translated by Lehfeldt, R. A.},
  publisher = {Edward Arnold},
  title     = {Lectures on Theoretical and Physical Chemistry. Part 1: Chemical Dynamics},
  year      = {1898},
}

@Article{Huntingford_2017,
  author    = {C. Huntingford and O. K. Atkin and A. Martinez-de la Torre and L. M. Mercado and M. A. Heskel and A. B. Harper and K. J. Bloomfield and O. S. O'Sullivan and P. B. Reich and K. R. Wythers and E. E. Butler and M. Chen and K. L. Griffin and P. Meir and M. G. Tjoelker and M. H. Turnbull and S. Sitch and A. Wiltshire and Y. Malhi},
  journal   = {Nature Communications},
  title     = {Implications of improved representations of plant respiration in a changing climate},
  year      = {2017},
  number    = {1},
  volume    = {8},
  doi       = {10.1038/s41467-017-01774-z},
  publisher = {Springer Science and Business Media {LLC}},
}

@Article{Niu_2021,
  author    = {B. Niu and X. Zhang and S. Piao and I. A. Janssens and G. Fu and Y. He and Y. Zhang and P. Shi and E. Dai and C. Yu and J. Zhang and G. Yu and M. Xu and J. Wu and L. Zhu and A. R. Desai and J. Chen and G. Bohrer and C. M. Gough and I. Mammarella and A. Varlagin and S. Fares and X. Zhao and Y. Li and H. Wang and Z. Ouyang},
  journal   = {Science Advances},
  title     = {Warming homogenizes apparent temperature sensitivity of ecosystem respiration},
  year      = {2021},
  number    = {15},
  volume    = {7},
  doi       = {10.1126/sciadv.abc7358},
  publisher = {American Association for the Advancement of Science ({AAAS})},
}

@Article{Ciais_2005,
  author    = {P. Ciais and M. Reichstein and N. Viovy and A. Granier and J. Og{\'{e}}e and V. Allard and M. Aubinet and N. Buchmann and C. Bernhofer and A. Carrara and F. Chevallier and N. De Noblet and A. D. Friend and P. Friedlingstein and T. Grünwald and B. Heinesch and P. Keronen and A. Knohl and G. Krinner and D. Loustau and G. Manca and G. Matteucci and F. Miglietta and J. M. Ourcival and D. Papale and K. Pilegaard and S. Rambal and G. Seufert and J. F. Soussana and M. J. Sanz and E. D. Schulze and T. Vesala and R. Valentini},
  journal   = {Nature},
  title     = {Europe-wide reduction in primary productivity caused by the heat and drought in 2003},
  year      = {2005},
  number    = {7058},
  pages     = {529--533},
  volume    = {437},
  doi       = {10.1038/nature03972},
  publisher = {Springer Science and Business Media {LLC}},
}

@Article{Teskey2014,
  author    = {R. Teskey and T. Wertin and I. Bauweraerts and M. Ameye and M. A. McGuire and K. Steppe},
  journal   = {Plant, Cell and Environment},
  title     = {Responses of tree species to heat waves and extreme heat events},
  year      = {2014},
  number    = {9},
  pages     = {1699--1712},
  volume    = {38},
  doi       = {10.1111/pce.12417},
  publisher = {Wiley},
}

@Article{Cole_2014,
  author    = {L. E. S. Cole and S. A. Bhagwat and K. J. Willis},
  journal   = {Nature Communications},
  title     = {Recovery and resilience of tropical forests after disturbance},
  year      = {2014},
  number    = {1},
  volume    = {5},
  doi       = {10.1038/ncomms4906},
  publisher = {Springer Science and Business Media {LLC}},
}

@Article{CuevasGonzalez2009,
  author    = {M. Cuevas-Gonzalez and F. Gerard and H. Balzter and D. Riano},
  journal   = {Global Change Biology},
  title     = {Analysing forest recovery after wildfire disturbance in boreal Siberia using remotely sensed vegetation indices},
  year      = {2009},
  number    = {3},
  pages     = {561--577},
  volume    = {15},
  doi       = {10.1111/j.1365-2486.2008.01784.x},
  publisher = {Wiley},
}

@Article{Goetz_2006,
  author    = {S. J. Goetz and G. J. Fiske and A. G. Bunn},
  journal   = {Remote Sensing of Environment},
  title     = {Using satellite time-series data sets to analyze fire disturbance and forest recovery across Canada},
  year      = {2006},
  number    = {3},
  pages     = {352--365},
  volume    = {101},
  doi       = {10.1016/j.rse.2006.01.011},
  publisher = {Elsevier {BV}},
}

@Article{Bartels_2016,
  author    = {S. F. Bartels and H. Y. H. Chen and M. A. Wulder and J. C. White},
  journal   = {Forest Ecology and Management},
  title     = {Trends in post-disturbance recovery rates of Canada's forests following wildfire and harvest},
  year      = {2016},
  pages     = {194--207},
  volume    = {361},
  doi       = {10.1016/j.foreco.2015.11.015},
  publisher = {Elsevier {BV}},
}

@Article{Dobor_2018,
  author    = {L. Dobor and T. Hl{\'{a}}sny and W. Rammer and I. Barka and J. Trombik and P. Pavlenda and V. {\v{S}}ebe{\v{n}} and P. {\v{S}}t{\v{e}}p{\'{a}}nek and R. Seidl},
  journal   = {Agricultural and Forest Meteorology},
  title     = {Post-disturbance recovery of forest carbon in a temperate forest landscape under climate change},
  year      = {2018},
  pages     = {308--322},
  volume    = {263},
  doi       = {10.1016/j.agrformet.2018.08.028},
  publisher = {Elsevier {BV}},
}

@Article{Eppley1972,
  author  = {R. W. Eppley},
  journal = {Fishery Bulletin},
  title   = {Temperature and phytoplankton growth in the sea},
  year    = {1972},
  number  = {4},
  pages   = {1063-1085},
  volume  = {70},
}

@Article{Bissinger_2008,
  author    = {J. E. Bissinger and D. J. S. Montagnes and J. Harples and D. Atkinson},
  journal   = {Limnology and Oceanography},
  title     = {Predicting marine phytoplankton maximum growth rates from temperature: Improving on the Eppley curve using quantile regression},
  year      = {2008},
  number    = {2},
  pages     = {487--493},
  volume    = {53},
  doi       = {10.4319/lo.2008.53.2.0487},
  publisher = {Wiley},
}

@Article{Sherman_2016,
  author    = {E. Sherman and J. K. Moore and F. Primeau and D. Tanouye},
  journal   = {Global Biogeochemical Cycles},
  title     = {Temperature influence on phytoplankton community growth rates},
  year      = {2016},
  number    = {4},
  pages     = {550--559},
  volume    = {30},
  doi       = {10.1002/2015gb005272},
  publisher = {American Geophysical Union ({AGU})},
}

@Article{Regaudie_de_Gioux_2012,
  author    = {Regaudie-de-Gioux, A. and Duarte, C. M.},
  journal   = {Global Biogeochemical Cycles},
  title     = {Temperature dependence of planktonic metabolism in the ocean},
  year      = {2012},
  number    = {1},
  pages     = {1--10},
  volume    = {26},
  doi       = {10.1029/2010gb003907},
  publisher = {American Geophysical Union ({AGU})},
}

@Article{Zeebe_2013,
  author  = {R. E. Zeebe and J. C. Zachos},
  journal = {Philosophical Transactions of the Royal Society A: Mathematical, Physical and Engineering Sciences},
  title   = {{L}ong-term legacy of massive carbon input to the {E}arth system: {A}nthropocene versus {E}ocene},
  year    = {2013},
  number  = {2001},
  pages   = {20120006},
  volume  = {371},
  doi     = {10.1098/rsta.2012.0006},
}

@Article{Boers_2022,
  author  = {Boers, N. and Ghil, M. and Stocker, T. F.},
  journal = {Environmental Research Letters},
  title   = {{Theoretical and paleoclimatic evidence for abrupt transitions in the Earth system}},
  year    = {2022},
  number  = {9},
  pages   = {093006},
  volume  = {17},
  doi     = {10.1088/1748-9326/ac8944},
}

@Article{Kirtland_Turner_2017,
  author    = {S. Kirtland Turner and P. M. Hull and L. R. Kump and A. Ridgwell},
  journal   = {Nature Communications},
  title     = {A probabilistic assessment of the rapidity of {PETM} onset},
  year      = {2017},
  number    = {1},
  volume    = {8},
  doi       = {10.1038/s41467-017-00292-2},
  publisher = {Springer Science and Business Media {LLC}},
}

@Article{Gutjahr_2017,
  author    = {M. Gutjahr and A. Ridgwell and P. F. Sexton and E. Anagnostou and P. N. Pearson and H. Pälike and R. D. Norris and E. Thomas and G. L. Foster},
  journal   = {Nature},
  title     = {Very large release of mostly volcanic carbon during the Palaeocene{\textendash}Eocene Thermal Maximum},
  year      = {2017},
  number    = {7669},
  pages     = {573--577},
  volume    = {548},
  doi       = {10.1038/nature23646},
  publisher = {Springer Science and Business Media {LLC}},
}

@Article{Potter_1993,
  author    = {C. S. Potter and J. T. Randerson and C. B. Field and P. A. Matson and P. M. Vitousek and H. A. Mooney and S. A. Klooster},
  journal   = {Global Biogeochemical Cycles},
  title     = {Terrestrial ecosystem production: A process model based on global satellite and surface data},
  year      = {1993},
  number    = {4},
  pages     = {811--841},
  volume    = {7},
  doi       = {10.1029/93gb02725},
  publisher = {American Geophysical Union ({AGU})},
}

@Article{Lade_2018,
  author    = {S. J. Lade and J. F. Donges and I. Fetzer and J. M. Anderies and C. Beer and S. E. Cornell and T. Gasser and J. Norberg and K. Richardson and J. Rockström and W. Steffen},
  journal   = {Earth System Dynamics},
  title     = {Analytically tractable climate{\textendash}carbon cycle feedbacks under 21st century anthropogenic forcing},
  year      = {2018},
  number    = {2},
  pages     = {507--523},
  volume    = {9},
  doi       = {10.5194/esd-9-507-2018},
  publisher = {Copernicus {GmbH}},
}

@Article{Goodwin2007,
  author    = {P. Goodwin and R. G. Williams and M. J. Follows and S. Dutkiewicz},
  journal   = {Global Biogeochemical Cycles},
  title     = {Ocean-atmosphere partitioning of anthropogenic carbon dioxide on centennial timescales},
  year      = {2007},
  number    = {1},
  volume    = {21},
  doi       = {10.1029/2006gb002810},
  publisher = {American Geophysical Union ({AGU})},
}

@Article{Takahashi_1993,
  author    = {T. Takahashi and J. Olafsson and J. G. Goddard and D. W. Chipman and S. C. Sutherland},
  journal   = {Global Biogeochemical Cycles},
  title     = {Seasonal variation of CO2 and nutrients in the high-latitude surface oceans: A comparative study},
  year      = {1993},
  number    = {4},
  pages     = {843--878},
  volume    = {7},
  doi       = {10.1029/93gb02263},
  publisher = {American Geophysical Union ({AGU})},
}

@Article{Lenton2000,
  author  = {T. M. Lenton},
  journal = {Tellus B: Chemichal and Physical Meteorology},
  title   = {Land and ocean carbon cycle feedback effects on global warming in a simple Earth system model},
  year    = {2000},
  number  = {5},
  pages   = {1159--1188},
  volume  = {52},
}

@Article{Anderies2013,
  author    = {J. M. Anderies and S. R. Carpenter and W. Steffen and J. Rockström},
  journal   = {Environmental Research Letters},
  title     = {The topology of non-linear global carbon dynamics: from tipping points to planetary boundaries},
  year      = {2013},
  number    = {4},
  pages     = {044048},
  volume    = {8},
  doi       = {10.1088/1748-9326/8/4/044048},
  publisher = {{IOP} Publishing},
}

@Book{Zeebe2001,
  author    = {R. E. Zeebe and D. Wolf-Gladrow},
  publisher = {Elsevier},
  title     = {{CO2 in Seawater: {E}quilibrium, Kinetics, Isotopes}},
  year      = {2001},
}

@Article{Henson2011,
  author    = {S. A. Henson and R. Sanders and E. Madsen and P. J. Morris and F. Le Moigne and G. D. Quartly},
  journal   = {Geophysical Research Letters},
  title     = {A reduced estimate of the strength of the ocean's biological carbon pump},
  year      = {2011},
  number    = {4},
  pages     = {1--5},
  volume    = {38},
  doi       = {10.1029/2011gl046735},
  publisher = {American Geophysical Union ({AGU})},
}

@Article{Dunne2005,
  author    = {J. P. Dunne and R. A. Armstrong and A. Gnanadesikan and J. L. Sarmiento},
  journal   = {Global Biogeochemical Cycles},
  title     = {Empirical and mechanistic models for the particle export ratio},
  year      = {2005},
  number    = {4},
  pages     = {1--16},
  volume    = {19},
  doi       = {10.1029/2004gb002390},
  publisher = {American Geophysical Union ({AGU})},
}

@Article{Laws2000,
  author  = {E. A. Laws and P. G. Falkowski and W. O. Smith and H. Ducklow and J. J. McCarthy},
  journal = {Global Biochemical Cycles},
  title   = {Temperature effects on export production in the open ocean},
  year    = {2000},
  number  = {23},
  pages   = {1231--1246},
  volume  = {14},
}

@Article{Hammond_2022,
  author    = {William M. Hammond and A. Park Williams and John T. Abatzoglou and Henry D. Adams and Tamir Klein and Rosana L{\'{o}}pez and Cuauht{\'{e}}moc S{\'{a}}enz-Romero and Henrik Hartmann and David D. Breshears and Craig D. Allen},
  journal   = {Nature Communications},
  title     = {Global field observations of tree die-off reveal hotter-drought fingerprint for {E}arth's forests},
  year      = {2022},
  number    = {1},
  volume    = {13},
  doi       = {10.1038/s41467-022-29289-2},
  publisher = {Springer Science and Business Media {LLC}},
}

@Article{Mahecha_2010,
  author    = {M. D. Mahecha and M. Reichstein and N. Carvalhais and G. Lasslop and H. Lange and S. I. Seneviratne and R. Vargas and C. Ammann and M. A. Arain and A. Cescatti and I. A. Janssens and M. Migliavacca and L. Montagnani and A. D. Richardson},
  journal   = {Science},
  title     = {Global convergence in the temperature sensitivity of respiration at ecosystem level},
  year      = {2010},
  number    = {5993},
  pages     = {838--840},
  volume    = {329},
  doi       = {10.1126/science.1189587},
  publisher = {American Association for the Advancement of Science ({AAAS})},
}

@Article{Rising_2022,
  author    = {J. Rising and M. Tedesco and F. Piontek and D. A. Stainforth},
  journal   = {Nature},
  title     = {The missing risks of climate change},
  year      = {2022},
  number    = {7933},
  pages     = {643--651},
  volume    = {610},
  doi       = {10.1038/s41586-022-05243-6},
  publisher = {Springer Science and Business Media {LLC}},
}

@Article{Wu_2022,
  author    = {X. Wu and D. Jiang},
  journal   = {Environmental Research Letters},
  title     = {Probabilistic impacts of compound dry and hot events on global gross primary production},
  year      = {2022},
  number    = {3},
  pages     = {034049},
  volume    = {17},
  doi       = {10.1088/1748-9326/ac4c5b},
  publisher = {{IOP} Publishing},
}

@Book{IPCC.2021,
  author    = {IPCC},
  publisher = {Cambridge University Press},
  title     = {{Climate Change 2021: The Physical Science Basis, Working Group I Contribution to the Sixth Assessment Report of the Intergovernmental Panel on Climate Change, V. Masson-Delmotte, P. Zhai et al., eds.}},
  year      = {2021},
  month     = jun,
  doi       = {10.1017/9781009157896},
}

@Article{Noel2021,
  author    = {B. Noël and L. van Kampenhout and J. T. M. Lenaerts and W. J. van de Berg and M. R. van den Broeke},
  journal   = {Geophysical Research Letters},
  title     = {A 21st Century Warming Threshold for Sustained Greenland Ice Sheet Mass Loss},
  year      = {2021},
  number    = {5},
  volume    = {48},
  doi       = {10.1029/2020gl090471},
  publisher = {American Geophysical Union ({AGU})},
}

@Article{Boers2021,
  author    = {N. Boers and M. Rypdal},
  journal   = {Proceedings of the National Academy of Sciences},
  title     = {Critical slowing down suggests that the western Greenland Ice Sheet is close to a tipping point},
  year      = {2021},
  number    = {21},
  volume    = {118},
  doi       = {10.1073/pnas.2024192118},
  publisher = {Proceedings of the National Academy of Sciences},
}

@Article{Tedesco2016,
  author    = {M. Tedesco and S. Doherty and X. Fettweis and P. Alexander and J. Jeyaratnam and J. Stroeve},
  journal   = {The Cryosphere},
  title     = {The darkening of the Greenland ice sheet: trends, drivers, and projections (1981{\textendash}2100)},
  year      = {2016},
  number    = {2},
  pages     = {477--496},
  volume    = {10},
  doi       = {10.5194/tc-10-477-2016},
  publisher = {Copernicus {GmbH}},
}

@Article{Au_2023,
  author    = {J. Au and A. A. Bloom and N. C. Parazoo and R. M. Deans and C. Y. S. Wong and B. Z. Houlton and T. S. Magney},
  journal   = {Global Change Biology},
  title     = {Forest productivity recovery or collapse? Model-data integration insights on drought-induced tipping points},
  year      = {2023},
  number    = {19},
  pages     = {5652--5665},
  volume    = {29},
  doi       = {10.1111/gcb.16867},
  publisher = {Wiley},
}

@Article{Xu_2020,
  author    = {H. Xu and J. Xiao and Z. Zhang},
  journal   = {Environmental Research Letters},
  title     = {Heatwave effects on gross primary production of northern mid-latitude ecosystems},
  year      = {2020},
  number    = {7},
  pages     = {074027},
  volume    = {15},
  doi       = {10.1088/1748-9326/ab8760},
  publisher = {{IOP} Publishing},
}

@Article{Henson2012,
  author  = {S. A. Henson and R. Sanders and E. Madsen},
  journal = {Global Biogeochemical Cycles},
  title   = {Global patterns in efficiency of particulate organic carbon export and transfer to the deep ocean},
  year    = {2012},
  pages   = {1--14},
  volume  = {26},
}

@Article{Cook2017,
  author    = {J. M. Cook and A. J. Hodson and A. S. Gardner and M. Flanner and A. J. Tedstone and C. Williamson and T. D. L. Irvine-Fynn and J. Nilsson and R. Bryant and M. Tranter},
  journal   = {The Cryosphere},
  title     = {Quantifying bioalbedo: a new physically based model and discussion of empirical methods for characterising biological influence on ice and snow albedo},
  year      = {2017},
  number    = {6},
  pages     = {2611--2632},
  volume    = {11},
  doi       = {10.5194/tc-11-2611-2017},
  publisher = {Copernicus {GmbH}},
}

@Article{Williamson2019,
  author    = {C. J. Williamson and K. A. Cameron and J. M. Cook and J. D. Zarsky and M. Stibal and A. Edwards},
  journal   = {Frontiers in Microbiology},
  title     = {{G}lacier algae: {A} dark past and a darker future},
  year      = {2019},
  volume    = {10},
  doi       = {10.3389/fmicb.2019.00524},
  publisher = {Frontiers Media {SA}},
}

@Article{Stibal2017,
  author    = {M. Stibal and J. E. Box and K. A. Cameron and P. L. Langen and M. L. Yallop and R. H. Mottram and A. L. Khan and N. P. Molotch and N. A. M. Chrismas and F. C. Quaglia and D. Remias and C. J. P. P. Smeets and M. R. van den Broeke and J. C. Ryan and A. Hubbard and M. Tranter and D. van As and A. P. Ahlstr{\o}m},
  journal   = {Geophysical Research Letters},
  title     = {{A}lgae drive enhanced darkening of bare ice on the {G}reenland {I}ce {S}heet},
  year      = {2017},
  number    = {22},
  volume    = {44},
  doi       = {10.1002/2017gl075958},
  publisher = {American Geophysical Union ({AGU})},
}

@Article{Williamson2020,
  author  = {C. J. Williamson and J. Cook and A. Tedstone and M. Yallop and J. McCutcheon and E. Poniecka and D. Campbell and T. Irvine-Fynn and J. McQuaid and M. Tranter and R.Perkins and A. Anesio},
  journal = {Proceedings of the National Academy of Sciences},
  title   = {{A}lgal photophysiology drives darkening and melt of the {G}reenland {I}ce {S}heet},
  year    = {2020},
  number  = {11},
  pages   = {5694--5705},
  volume  = {117},
  doi     = {10.1073/pnas.1918412117},
}

@Article{Bochow2023,
  author    = {N. Bochow and A. Poltronieri and A. Robinson and M. Montoya and M. Rypdal and N. Boers},
  journal   = {Nature},
  title     = {{O}vershooting the critical threshold for the {G}reenland ice sheet},
  year      = {2023},
  number    = {7983},
  pages     = {528--536},
  volume    = {622},
  doi       = {10.1038/s41586-023-06503-9},
  publisher = {Springer Science and Business Media {LLC}},
}

@Article{Sabine_2004,
  author  = {C. L. Sabine and R. A. Feely and N. Gruber and R. M. Key and K. Lee and J. L. Bullister and R. Wanninkhof and C. S. Wong and D. W. R. Wallace and B. Tilbrook and F. J. Millero and T.-H. Peng and A. Kozyr and T. Ono and A. F. Rios},
  journal = {Science},
  title   = {{T}he {O}ceanic {S}ink for {A}nthropogenic {CO2}},
  year    = {2004},
  number  = {5682},
  pages   = {367--371},
  volume  = {305},
  doi     = {10.1126/science.1097403},
}

@Electronic{NOAA2023,
  address      = {1401 {C}onstitution {A}venue {NW}, {W}ashington, {DC} 20230, {U.S.}},
  author       = {NOAA},
  howpublished = {Online},
  month        = jul,
  organization = {National Oceanic and Atmospheric Administration},
  title        = {{Climate at a Glance Global Time Series}},
  url          = {https://www.ncei.noaa.gov/access/monitoring/climate-at-a-glance/global/time-series/globe/land_ocean/ytd/12/1850-2022},
  year         = {2023},
}

@Article{Sinclair.1994,
  author    = {Sinclair, T. R.},
  journal   = {Physiology and determination of crop yield},
  title     = {Limits to crop yield?},
  year      = {1994},
  pages     = {509--532},
  publisher = {Wiley Online Library},
}

@Article{Parisi.2023,
  author  = {Parisi, G.},
  journal = {Reviews of Modern Physics},
  title   = {{N}obel {L}ecture: {M}ultiple {E}quilibria},
  year    = {2023},
  pages   = {1--17},
  volume  = {95},
}

@Article{Sarmiento2002,
  author  = {Sarmiento, J. L. and Gruber, N.},
  journal = {Physics Today},
  title   = {Sinks for anthropogenic carbon},
  year    = {2002},
  number  = {8},
  pages   = {30--36},
  volume  = {55},
}

@Article{Lancaster2020,
  author    = {Lancaster, L. T. and Humphreys, A. M.},
  journal   = {Proceedings of the National Academy of Sciences},
  title     = {Global variation in the thermal tolerances of plants},
  year      = {2020},
  issn      = {1091-6490},
  number    = {24},
  pages     = {13580--13587},
  volume    = {117},
  doi       = {10.1073/pnas.1918162117},
  publisher = {Proceedings of the National Academy of Sciences},
}

@Article{Pistone2014,
  author    = {Pistone, K. and Eisenman, I. and Ramanathan, V.},
  journal   = {Proceedings of the National Academy of Sciences},
  title     = {{O}bservational determination of albedo decrease caused by vanishing {A}rctic sea ice},
  year      = {2014},
  issn      = {1091-6490},
  number    = {9},
  pages     = {3322--3326},
  volume    = {111},
  doi       = {10.1073/pnas.1318201111},
  publisher = {Proceedings of the National Academy of Sciences},
}

@Electronic{NSIDC2023,
  address      = {University of Colorado Boulder},
  author       = {NSIDC},
  howpublished = {Online},
  organization = {National Snow and Ice Data Centre},
  title        = {{S}ea {I}ce},
  url          = {https://nsidc.org/learn/parts-cryosphere/sea-ice/science-sea-ice},
  year         = {2023},
}

@Article{Myneni1995,
  author    = {Myneni, R. B. and Los, S. O. and Asrar, G.},
  journal   = {Geophysical Research Letters},
  title     = {Potential gross primary productivity of terrestrial vegetation from 1982–1990},
  year      = {1995},
  issn      = {1944-8007},
  number    = {19},
  pages     = {2617--2620},
  volume    = {22},
  doi       = {10.1029/95gl02562},
  publisher = {American Geophysical Union (AGU)},
}

@Article{Haverd2020,
  author    = {Haverd, V. and Smith, B. and Canadell, Josep. G. and Cuntz, M. and Mikaloff‐Fletcher, S. and Farquhar, G. and Woodgate, W. and Briggs, P. R. and Trudinger, C. M.},
  journal   = {Global Change Biology},
  title     = {Higher than expected {CO2} fertilization inferred from leaf to global observations},
  year      = {2020},
  issn      = {1365-2486},
  number    = {4},
  pages     = {2390--2402},
  volume    = {26},
  doi       = {10.1111/gcb.14950},
  publisher = {Wiley},
}

@Article{Jia2022,
  author    = {Jia, A. and Wang, D. and Liang, S. and Peng, J. and Yu, Y.},
  journal   = {Journal of Geophysical Research: Atmospheres},
  title     = {{Global Daily Actual and Snow‐Free Blue‐Sky Land Surface Albedo Climatology From 20‐Year MODIS Products}},
  year      = {2022},
  issn      = {2169-8996},
  number    = {8},
  volume    = {127},
  doi       = {10.1029/2021jd035987},
  publisher = {American Geophysical Union (AGU)},
}

@Article{Yallop2012,
  author    = {Yallop, M. L. and Anesio, A. M. and Perkins, R. G. and Cook, J. and Telling, J. and Fagan, D. and MacFarlane, J. and Stibal, M. and Barker, G. and Bellas, C. and Hodson, A. and Tranter, M. and Wadham, J. and Roberts, N. W.},
  journal   = {The ISME Journal},
  title     = {{Photophysiology and albedo-changing potential of the ice algal community on the surface of the Greenland ice sheet}},
  year      = {2012},
  issn      = {1751-7370},
  number    = {12},
  pages     = {2302--2313},
  volume    = {6},
  doi       = {10.1038/ismej.2012.107},
  publisher = {Springer Science and Business Media LLC},
}

@InBook{Lenton2016,
  author    = {Lenton, T. M. and Livina, V. N. and {Detecting and anticipating climate tipping points} and .},
  chapter   = {{Detecting and anticipating climate tipping points}},
  editor    = {Chavez, M. and Ghil, M. and Urrutia-Fucugauchi, J.},
  pages     = {51--62},
  publisher = {AGU - John Wiley \& Sons},
  title     = {{Extreme Events: Observations, Modelling, and Economics}},
  year      = {2016},
  isbn      = {978-1-119-15703-8},
  number    = {214},
  booktitle = {{Extreme Events: Observations, Modelling, and Economics}},
  doi       = {https://doi.org/10.1002/9781119157052.ch5},
}

@Article{Hartmann_2018,
  author    = {Hartmann, H. and Moura, C. F. and Anderegg, W. R. L. and Ruehr, N. K. and Salmon, Y. and Allen, C. D. and Arndt, S. K. and Breshears, D. D. and Davi, H. and Galbraith, D. and Ruthrof, K. X. and Wunder, J. and Adams, H. D. and Bloemen, J. and Cailleret, M. and Cobb, R. and Gessler, A. and Grams, T. E. E. and Jansen, S. and Kautz, M. and Lloret, F. and O’Brien, M.},
  journal   = {New Phytologist},
  title     = {Research frontiers for improving our understanding of drought‐induced tree and forest mortality},
  year      = {2018},
  issn      = {1469-8137},
  number    = {1},
  pages     = {15--28},
  volume    = {218},
  doi       = {10.1111/nph.15048},
  publisher = {Wiley},
}

@Article{RuizBenito2013,
  author    = {Ruiz-Benito, P. and Lines, E. R. and Gómez-Aparicio, L. and Zavala, M. A. and Coomes, D. A.},
  journal   = {PLoS ONE},
  title     = {{Patterns and Drivers of Tree Mortality in Iberian Forests: Climatic Effects Are Modified by Competition}},
  year      = {2013},
  issn      = {1932-6203},
  number    = {2},
  pages     = {e56843},
  volume    = {8},
  doi       = {10.1371/journal.pone.0056843},
  editor    = {Hector, Andrew},
  publisher = {Public Library of Science (PLoS)},
}

@Article{Lobell2011,
  author    = {Lobell, D. B. and Bänziger, M. and Magorokosho, C. and Vivek, B.},
  journal   = {Nature Climate Change},
  title     = {{Nonlinear heat effects on African maize as evidenced by historical yield trials}},
  year      = {2011},
  issn      = {1758-6798},
  number    = {1},
  pages     = {42--45},
  volume    = {1},
  doi       = {10.1038/nclimate1043},
  publisher = {Springer Science and Business Media LLC},
}

@InBook{Sage1999,
  author    = {R. F. Sage and D. A. Wedin and M. Li},
  chapter   = {{The Biogeography of C4 Photosynthesis: Patterns andControlling Factors}},
  editor    = {Rowan F. Sage and Russell K. Monson},
  pages     = {313--356},
  publisher = {Academic Press},
  title     = {{C4 Plant Biology}},
  year      = {1999},
  address   = {525 B Street, Suite 1900, San Diego, California 92101-4495, USA},
  isbn      = {0-12-614440-0},
}

@Article{Still2003,
  author    = {Still, C. J. and Berry, J. A. and Collatz, G. J. and DeFries, R. S.},
  journal   = {Global Biogeochemical Cycles},
  title     = {{Global distribution of C3 and C4 vegetation: Carbon cycle implications}},
  year      = {2003},
  issn      = {1944-9224},
  number    = {1},
  volume    = {17},
  doi       = {10.1029/2001gb001807},
  publisher = {American Geophysical Union (AGU)},
}

@Article{Nolan2018,
  author    = {Nolan, C. and Overpeck, J. T. and Allen, J. R. M. and Anderson, P. M. and Betancourt, J. L. and Binney, H. A. and Brewer, S. and Bush, M. B. and Chase, B. M. and Cheddadi, . and Djamali, M. and Dodson, J. and Edwards, M. E. and Gosling, W. D. and Haberle, S. and Hotchkiss, S. C. and Huntley, B. and Ivory, S. J. and Kershaw, A. P. and Kim, S.-H. and Latorre, C. and Leydet, M. and Lézine, A.-M. and Liu, K.-B. and Liu, Y. and Lozhkin, A. V. and McGlone, M. S. and Marchant, R. A. and Momohara, A. and Moreno, P. I. and Müller, S. and Otto-Bliesner, B. L. and Shen, C. and Stevenson, J. and Takahara, H. and Tarasov, P. E. and Tipton, J. and Vincens, A. and Weng, Ch. and Xu, Q. and Zheng, Z. and Jackson, S. T.},
  journal   = {Science},
  title     = {Past and future global transformation of terrestrial ecosystems under climate change},
  year      = {2018},
  issn      = {1095-9203},
  number    = {6405},
  pages     = {920--923},
  volume    = {361},
  doi       = {10.1126/science.aan5360},
  publisher = {American Association for the Advancement of Science (AAAS)},
}

@Article{Ferreira2011,
  author    = {Ferreira, D. and Marshall, J. and Rose, B.},
  journal   = {Journal of Climate},
  title     = {{Climate Determinism Revisited: Multiple Equilibria in a Complex Climate Model}},
  year      = {2011},
  issn      = {0894-8755},
  number    = {4},
  pages     = {992--1012},
  volume    = {24},
  doi       = {10.1175/2010jcli3580.1},
  publisher = {American Meteorological Society},
}

@Article{Fick.1855,
  author  = {Fick, A.},
  journal = {Annalen der Physik},
  title   = {{Ueber Diffusion}},
  year    = {1855},
  number  = {1},
  pages   = {59--86},
  volume  = {170},
  doi     = {10.1002/andp.18551700105},
}

@Book{CushmanRoisin.2011,
  author    = {Cushman-Roisin, B. and Beckers, J.-M.},
  publisher = {Academic Press},
  title     = {{Introduction to Geophysical Fluid Dynamics: Physical and Numerical Aspects (2nd ed.)}},
  year      = {2011},
}

@Article{Hawkins2017,
  author    = {Hawkins, E. and Ortega, P. and Suckling, E. and Schurer, A. and Hegerl, G. and Jones, P. and Joshi, M. and Osborn, T. J. and Masson-Delmotte, V. and Mignot, J. and Thorne, P. and van Oldenborgh, G. J.},
  journal   = {Bulletin of the American Meteorological Society},
  title     = {{Estimating Changes in Global Temperature since the Preindustrial Period}},
  year      = {2017},
  issn      = {1520-0477},
  number    = {9},
  pages     = {1841--1856},
  volume    = {98},
  doi       = {10.1175/bams-d-16-0007.1},
  publisher = {American Meteorological Society},
}

@Article{Taylor2012,
  author  = {Taylor, K. E. and Stouffer, R. J. and Meehl, G. A.},
  journal = {Bulletin of the American Meteorological Society},
  title   = {{An Overview of CMIP5 and the Experiment Design}},
  year    = {2012},
  issn    = {1520-0477},
  number  = {4},
  pages   = {485--498},
  volume  = {93},
  doi     = {10.1175/bams-d-11-00094.1},
}

@Article{Bland+A.1999,
  author  = {Bland, J. M. and Altman, D. G},
  journal = {Statistical Methods in Medical Research},
  title   = {Measuring agreement in method comparison studies},
  year    = {1999},
  number  = {2},
  pages   = {135--160},
  volume  = {8},
}

@TechReport{Boone2010,
  author      = {A. Boone},
  institution = {National Centre for Meteorological Research},
  title       = {{Description of the snow framework of the ISBA-ES model}},
  year        = {2010},
  address     = {Centre National de Recherches M´et´eorologiques, Meteo-France 42, avenue G. Coriolis, 31057 Toulouse Cedex France},
  url         = {10.1002/2015GL066275http://www.umr-cnrm.fr/IMG/pdf/snowdoc_v2.pdf},
}

@Article{Feingold2017,
  author    = {Feingold, G. and Balsells, J. and Glassmeier, F. and Yamaguchi, T. and Kazil, J. and McComiskey, A.},
  journal   = {Journal of Geophysical Research: Atmospheres},
  title     = {Analysis of albedo versus cloud fraction relationships in liquid water clouds using heuristic models and large eddy simulation},
  year      = {2017},
  issn      = {2169-8996},
  number    = {13},
  pages     = {7086--7102},
  volume    = {122},
  doi       = {10.1002/2017jd026467},
  publisher = {American Geophysical Union (AGU)},
}

@Article{Dutta2020,
  author    = {Dutta, S. and Di Girolamo, L. and Dey, S. and Zhan, Y. and Moroney, C. M. and Zhao, G.},
  journal   = {Geophysical Research Letters},
  title     = {{The Reduction in Near‐Global Cloud Cover After Correcting for Biases Caused by Finite Resolution Measurements}},
  year      = {2020},
  issn      = {1944-8007},
  number    = {23},
  volume    = {47},
  doi       = {10.1029/2020gl090313},
  publisher = {American Geophysical Union (AGU)},
}

@Article{Vrese2021,
  author  = {Vrese, P. and Stacke, T. and Caves, J.and Goodman, J. and Brovkin, V.},
  journal = {Communications Earth and Environment},
  title   = {{Snowfall-albedo feedbacks could have led to deglaciation of snowball Earth starting from mid-latitudes}},
  year    = {2021},
  issn    = {2662-4435},
  number  = {1},
  volume  = {2},
  doi     = {10.1038/s43247-021-00160-4},
}

@Article{Boone2001,
  author    = {Boone, A. and Etchevers, P.},
  journal   = {Journal of Hydrometeorology},
  title     = {{An Intercomparison of Three Snow Schemes of Varying Complexity Coupled to the Same Land Surface Model: Local-Scale Evaluation at an Alpine Site}},
  year      = {2001},
  issn      = {1525-7541},
  number    = {4},
  pages     = {374--394},
  volume    = {2},
  doi       = {10.1175/1525-7541(2001)002<0374:aiotss>2.0.co;2},
  publisher = {American Meteorological Society},
}

@Article{Pugh2020,
  author    = {Pugh, T. A. M. and Rademacher, T. and Shafer, S. L. and Steinkamp, J. and Barichivich, J. and Beckage, B. and Haverd, V. and Harper, A. and Heinke, J. and Nishina, K. and Rammig, A. and Sato, H. and Arneth, A. and Hantson, S. and Hickler, T. and Kautz, M. and Quesada, B. and Smith, B. and Thonicke, K.},
  journal   = {Biogeosciences},
  title     = {Understanding the uncertainty in global forest carbon turnover},
  year      = {2020},
  issn      = {1726-4189},
  number    = {15},
  pages     = {3961--3989},
  volume    = {17},
  doi       = {10.5194/bg-17-3961-2020},
  publisher = {Copernicus GmbH},
}

@Article{Pierrehumbert2011,
  author    = {Pierrehumbert, R. T. and Abbot, D. S. and Voigt, A. and Koll, D.},
  journal   = {Annual Review of Earth and Planetary Sciences},
  title     = {{Climate of the Neoproterozoic}},
  year      = {2011},
  issn      = {1545-4495},
  number    = {1},
  pages     = {417--460},
  volume    = {39},
  doi       = {10.1146/annurev-earth-040809-152447},
  publisher = {Annual Reviews},
}

@Article{RuizBenito2014,
  author    = {Ruiz-Benito, P. and Madrigal-González, J. and Ratcliffe, S. and Coomes, D. A. and Kändler, G. and Lehtonen, A. and Wirth, C. and Zavala, M. A.},
  journal   = {Ecosystems},
  title     = {{Stand Structure and Recent Climate Change Constrain Stand Basal Area Change in European Forests: A Comparison Across Boreal, Temperate, and Mediterranean Biomes}},
  year      = {2014},
  issn      = {1435-0629},
  number    = {8},
  pages     = {1439--1454},
  volume    = {17},
  doi       = {10.1007/s10021-014-9806-0},
  publisher = {Springer Science and Business Media LLC},
}

@Article{Kopp2023,
  author    = {Kopp, G.},
  journal   = {Solar Energy},
  title     = {Daily solar flux as a function of latitude and time},
  year      = {2023},
  pages     = {250--254},
  volume    = {249},
  doi       = {10.1016/j.solener.2022.11.022},
  publisher = {Elsevier BV},
}

@Article{Millar2024,
  author    = {Millar, J. L. and Broadwell, E. L. M. and Lewis, M. and Bowles, A. M. C. and Tedstone, A. J. and Williamson, C. J.},
  journal   = {Frontiers in Microbiology},
  title     = {Alpine glacier algal bloom during a record melt year},
  year      = {2024},
  issn      = {1664-302X},
  volume    = {15},
  doi       = {10.3389/fmicb.2024.1356376},
  publisher = {Frontiers Media SA},
}

@Article{Hoffman1998,
  author  = {Hoffman, P. F. and Kaufman, A. J. and Halverson, G. P. and Schrag, D. P.},
  journal = {Science},
  title   = {{A Neoproterozoic Snowball Earth}},
  year    = {1998},
  pages   = {1342--1346},
  volume  = {281},
}

@Article{Hoffman2002,
  author  = {Hoffman, {P. F.} and Schrag, {D. P.}},
  journal = {Terra Nova},
  title   = {The snowball {E}arth hypothesis: testing the limits of global change},
  year    = {2002},
  number  = {3},
  pages   = {129--155},
  volume  = {14},
}

@Article{Flanner2011,
  author    = {Flanner, M. G. and Shell, K. M. and Barlage, M. and Perovich, D. K. and Tschudi, M. A.},
  journal   = {Nature Geoscience},
  title     = {{Radiative forcing and albedo feedback from the Northern Hemisphere cryosphere between 1979 and 2008}},
  year      = {2011},
  issn      = {1752-0908},
  month     = jan,
  number    = {3},
  pages     = {151--155},
  volume    = {4},
  doi       = {10.1038/ngeo1062},
  publisher = {Springer Science and Business Media LLC},
}

@Article{Box2012,
  author    = {Box, J. E. and Fettweis, X. and Stroeve, J. C. and Tedesco, M. and Hall, D. K. and Steffen, K.},
  journal   = {The Cryosphere},
  title     = {Greenland ice sheet albedo feedback: thermodynamics and atmospheric drivers},
  year      = {2012},
  issn      = {1994-0424},
  number    = {4},
  pages     = {821--839},
  volume    = {6},
  doi       = {10.5194/tc-6-821-2012},
  publisher = {Copernicus GmbH},
}

@Article{Garbe2023,
  author    = {Garbe, Julius and Zeitz, Maria and Krebs-Kanzow, Uta and Winkelmann, Ricarda},
  journal   = {The Cryosphere},
  title     = {{The evolution of future Antarctic surface melt using PISM-dEBM-simple}},
  year      = {2023},
  issn      = {1994-0424},
  number    = {11},
  pages     = {4571--4599},
  volume    = {17},
  doi       = {10.5194/tc-17-4571-2023},
  publisher = {Copernicus GmbH},
}

@Article{Garbe2020,
  author    = {Garbe, Julius and Albrecht, Torsten and Levermann, Anders and Donges, Jonathan F. and Winkelmann, Ricarda},
  journal   = {Nature},
  title     = {{The hysteresis of the Antarctic Ice Sheet}},
  year      = {2020},
  issn      = {1476-4687},
  number    = {7826},
  pages     = {538--544},
  volume    = {585},
  doi       = {10.1038/s41586-020-2727-5},
  publisher = {Springer Science and Business Media LLC},
}

@TechReport{Danielson2011,
  author      = {{J. J.} Danielson and {D. B.} Gesch},
  institution = {U.S. Geological Survey},
  title       = {{Global multi-resolution terrain elevation data 2010 (GMTED2010)}},
  year        = {2011},
  number      = {2011-1073},
  type        = {Open-File Report},
}

@Article{Eisenman_2024,
  author    = {Eisenman, Ian and Armour, Kyle C.},
  journal   = {Nature Communications},
  title     = {{The radiative feedback continuum from Snowball Earth to an ice-free hothouse}},
  year      = {2024},
  issn      = {2041-1723},
  number    = {1},
  volume    = {15},
  doi       = {10.1038/s41467-024-50406-w},
  publisher = {Springer Science and Business Media LLC},
}

@Article{Hartin2015,
  author    = {Hartin, C. A. and Patel, P. and Schwarber, A. and Link, R. P. and Bond-Lamberty, B. P.},
  journal   = {Geoscientific Model Development},
  title     = {{A simple object-oriented and open-source model for scientific and policy analyses of the global climate system – Hector v1.0}},
  year      = {2015},
  issn      = {1991-9603},
  number    = {4},
  pages     = {939--955},
  volume    = {8},
  doi       = {10.5194/gmd-8-939-2015},
  publisher = {Copernicus GmbH},
}

@Article{Heydt2021,
  author    = {von der Heydt, Anna S. and Ashwin, Peter and Camp, Charles D. and Crucifix, Michel and Dijkstra, Henk A. and Ditlevsen, Peter and Lenton, Timothy M.},
  journal   = {Global and Planetary Change},
  title     = {Quantification and interpretation of the climate variability record},
  year      = {2021},
  issn      = {0921-8181},
  pages     = {103399},
  volume    = {197},
  doi       = {10.1016/j.gloplacha.2020.103399},
  publisher = {Elsevier BV},
}

@Article{Sherwood2020,
  author    = {Sherwood, S. C. and Webb, M. J. and Annan, J. D. and Armour, K. C. and Forster, P. M. and Hargreaves, J. C. and Hegerl, G. and Klein, S. A. and Marvel, K. D. and Rohling, E. J. and Watanabe, M. and Andrews, T. and Braconnot, P. and Bretherton, C. S. and Foster, G. L. and Hausfather, Z. and von der Heydt, A. S. and Knutti, R. and Mauritsen, T. and Norris, J. R. and Proistosescu, C. and Rugenstein, M. and Schmidt, G. A. and Tokarska, K. B. and Zelinka, M. D.},
  journal   = {Reviews of Geophysics},
  title     = {{An assessment of Earth’s climate sensitivity using multiple lines of evidence}},
  year      = {2020},
  issn      = {1944-9208},
  number    = {4},
  volume    = {58},
  doi       = {10.1029/2019rg000678},
  publisher = {American Geophysical Union (AGU)},
}

@Article{Winguth2015,
  author    = {Winguth, Arne M. E. and Shields, Christine A. and Winguth, Cornelia},
  journal   = {Palaeogeography, Palaeoclimatology, Palaeoecology},
  title     = {{Transition into a Hothouse world at the Permian–Triassic boundary—A model study}},
  year      = {2015},
  issn      = {0031-0182},
  month     = dec,
  pages     = {316--327},
  volume    = {440},
  doi       = {10.1016/j.palaeo.2015.09.008},
  publisher = {Elsevier BV},
}

@Article{Sun2012,
  author    = {Sun, Yadong and Joachimski, Michael M. and Wignall, Paul B. and Yan, Chunbo and Chen, Yanlong and Jiang, Haishui and Wang, Lina and Lai, Xulong},
  journal   = {Science},
  title     = {{Lethally hot temperatures during the early Triassic greenhouse}},
  year      = {2012},
  issn      = {1095-9203},
  month     = oct,
  number    = {6105},
  pages     = {366--370},
  volume    = {338},
  doi       = {10.1126/science.1224126},
  publisher = {American Association for the Advancement of Science (AAAS)},
}

@Article{Hall.ea.2019,
  author    = {Hall, Alex and Cox, Peter and Huntingford, Chris and Klein, Stephen},
  journal   = {Nature Climate Change},
  title     = {Progressing emergent constraints on future climate change},
  year      = {2019},
  number    = {4},
  pages     = {269--278},
  volume    = {9},
  publisher = {Nature Publishing Group UK London},
}

@Article{Zhu2019,
  author    = {Zhu, Jiang and Poulsen, Christopher J. and Tierney, Jessica E.},
  journal   = {Science Advances},
  title     = {{Simulation of Eocene extreme warmth and high climate sensitivity through cloud feedbacks}},
  year      = {2019},
  issn      = {2375-2548},
  number    = {9},
  volume    = {5},
  doi       = {10.1126/sciadv.aax1874},
  publisher = {American Association for the Advancement of Science (AAAS)},
}

@InBook{Forster2021,
  author    = {P. Forster and T. Storelvmo and K. Armour and W. Collins and J. L. Dufresne and D. Frame and D. J. Hunt and T. Mauritsen and M. D. Palmer and W. Watanabe and M. Wild and H. Zhang},
  chapter   = {7},
  editor    = {V. P. Masson-Delmotte and P. Zhai and A. Pirani and S. L. Connors and C. Pean and S. Berger and N. Cuad and Y. Chen and L. Goldfarb and M. I. Gomis and M. Huang and K. Leitzell and E. Lonnoy and J. B. R. Matthews and T. K. Mayckock and T. Waterfield and O. Yelekci and R. Yu and B. Zhou},
  pages     = {923--1054},
  publisher = {Cambridge University Press},
  title     = {{The Earth's Emergy Budget, Climate Feedbacks, and Climate Sensitivity. In Climate Change 2021: The Physical Science Basis. Contribution of Working Group I to the Sixth Assessment Report of the Intergovernmental Panel on Climate Change}},
  year      = {2021},
  address   = {Cambridge, United Kingdom and New Yorkd, NY, USA},
}

@Article{Caballero2013,
  author  = {R. Caballero and M. Huber},
  journal = {Proceedings of the National Academy of Science},
  title   = {State-dependent climate sensitivity in past warm climates and its implications for future climate projections},
  year    = {2013},
  number  = {35},
  pages   = {14162--14167},
  volume  = {110},
}

@Article{Aswhin2020,
  author  = {P. Aswhin and A. S. von der Heydt},
  journal = {Journal of Statistical Physics},
  title   = {Extreme sensitivity and climate tipping points},
  year    = {2020},
  number  = {5},
  pages   = {1531--1552},
  volume  = {179},
}

@Article{Wunderling_2024,
  author    = {Wunderling, Nico and von der Heydt, Anna S. and Aksenov, Yevgeny and Barker, Stephen and Bastiaansen, Robbin and Brovkin, Victor and Brunetti, Maura and Couplet, Victor and Kleinen, Thomas and Lear, Caroline H. and Lohmann, Johannes and Roman-Cuesta, Rosa Maria and Sinet, Sacha and Swingedouw, Didier and Winkelmann, Ricarda and Anand, Pallavi and Barichivich, Jonathan and Bathiany, Sebastian and Baudena, Mara and Bruun, John T. and Chiessi, Cristiano M. and Coxall, Helen K. and Docquier, David and Donges, Jonathan F. and Falkena, Swinda K. J. and Klose, Ann Kristin and Obura, David and Rocha, Juan and Rynders, Stefanie and Steinert, Norman Julius and Willeit, Matteo},
  journal   = {Earth System Dynamics},
  title     = {Climate tipping point interactions and cascades: a review},
  year      = {2024},
  issn      = {2190-4987},
  month     = jan,
  number    = {1},
  pages     = {41--74},
  volume    = {15},
  doi       = {10.5194/esd-15-41-2024},
  publisher = {Copernicus GmbH},
}

@Article{Winkelmann2011,
  author    = {Winkelmann, R. and Martin, M. A. and Haseloff, M. and Albrecht, T. and Bueler, E. and Khroulev, C. and Levermann, A.},
  journal   = {The Cryosphere},
  title     = {{The Potsdam Parallel Ice Sheet Model (PISM-PIK) – Part 1: Model description}},
  year      = {2011},
  issn      = {1994-0424},
  number    = {3},
  pages     = {715--726},
  volume    = {5},
  doi       = {10.5194/tc-5-715-2011},
  publisher = {Copernicus GmbH},
}

@Article{Madsen2022,
  author    = {Madsen, M. S. and Yang, S. and Aðalgeirsdóttir, G. and Svendsen, S. H. and Rodehacke, C. B. and Ringgaard, I. M.},
  journal   = {Climate Dynamics},
  title     = {{The role of an interactive Greenland ice sheet in the coupled climate-ice sheet model EC-Earth-PISM}},
  year      = {2022},
  issn      = {1432-0894},
  number    = {3–4},
  pages     = {1189--1211},
  volume    = {59},
  doi       = {10.1007/s00382-022-06184-6},
  publisher = {Springer Science and Business Media LLC},
}

@Article{Smith_2018,
  author    = {Smith, Christopher J. and Forster, Piers M. and Allen, Myles and Leach, Nicholas and Millar, Richard J. and Passerello, Giovanni A. and Regayre, Leighton A.},
  journal   = {Geoscientific Model Development},
  title     = {{FAIR v1.3: a simple emissions-based impulse response and carbon cycle model}},
  year      = {2018},
  issn      = {1991-9603},
  number    = {6},
  pages     = {2273--2297},
  volume    = {11},
  doi       = {10.5194/gmd-11-2273-2018},
  publisher = {Copernicus GmbH},
}

@Article{Jiang_2019,
  author    = {Jiang, Li-Qing and Carter, Brendan R. and Feely, Richard A. and Lauvset, Siv K. and Olsen, Are},
  journal   = {Scientific Reports},
  title     = {{Surface ocean pH and buffer capacity: past, present and future}},
  year      = {2019},
  issn      = {2045-2322},
  number    = {1},
  volume    = {9},
  doi       = {10.1038/s41598-019-55039-4},
  publisher = {Springer Science and Business Media LLC},
}

@Article{Kim2022,
  author    = {Kim, Soong-Ki and Kim, Hyo-Jeong and Dijkstra, Henk A. and An, Soon-Il},
  journal   = {Climate and Atmospheric Science},
  title     = {{Slow and soft passage through tipping point of the Atlantic Meridional Overturning Circulation in a changing climate}},
  year      = {2022},
  issn      = {2397-3722},
  month     = feb,
  number    = {1},
  volume    = {5},
  doi       = {10.1038/s41612-022-00236-8},
  publisher = {Springer Science and Business Media LLC},
}

@Article{Saltelli2010,
  author    = {Saltelli, Andrea and Annoni, Paola and Azzini, Ivano and Campolongo, Francesca and Ratto, Marco and Tarantola, Stefano},
  journal   = {Computer Physics Communications},
  title     = {Variance based sensitivity analysis of model output. Design and estimator for the total sensitivity index},
  year      = {2010},
  issn      = {0010-4655},
  number    = {2},
  pages     = {259--270},
  volume    = {181},
  doi       = {10.1016/j.cpc.2009.09.018},
  publisher = {Elsevier BV},
}

@Book{Saltelli2007,
  author    = {Saltelli, Andrea and Ratto, Marco and Andres, Terry and Campolongo, Francesca and Cariboni, Jessica and Gatelli, Debora and Saisana, Michaela and Tarantola, Stefano},
  publisher = {Wiley},
  title     = {{Global Sensitivity Analysis. The Primer}},
  year      = {2007},
  isbn      = {9780470725184},
  doi       = {10.1002/9780470725184},
}

@Article{Jansen1999,
  author    = {Jansen, Michiel J. W.},
  journal   = {Computer Physics Communications},
  title     = {Analysis of variance designs for model output},
  year      = {1999},
  issn      = {0010-4655},
  number    = {1–2},
  pages     = {35--43},
  volume    = {117},
  doi       = {10.1016/s0010-4655(98)00154-4},
  publisher = {Elsevier BV},
}

@Article{Hourdin.ea.2017,
  author    = {Hourdin, Fr{\'e}d{\'e}ric and Mauritsen, Thorsten and Gettelman, Andrew and Golaz, Jean-Christophe and Balaji, Venkatramani and Duan, Qingyun and Folini, Doris and Ji, Duoying and Klocke, Daniel and Qian, Yun and others},
  journal   = {Bulletin of the American Meteorological Society},
  title     = {The art and science of climate model tuning},
  year      = {2017},
  number    = {3},
  pages     = {589--602},
  volume    = {98},
  publisher = {American Meteorological Society},
}

@Article{Sobol2001,
  author    = {I. M. Sobol},
  journal   = {Mathematics and Computers in Simulation},
  title     = {{Global sensitivity indices for nonlinear mathematical models and their Monte Carlo estimates}},
  year      = {2001},
  issn      = {0378-4754},
  month     = feb,
  number    = {1–3},
  pages     = {271--280},
  volume    = {55},
  doi       = {10.1016/s0378-4754(00)00270-6},
  publisher = {Elsevier BV},
}

\end{refsegment}

%\noindent {\mg Remind me, pls., to get the numbers \#[XYZ] and  \#[X'Y'Z'] when accepted.}

\section*{Acknowledgements}

It is a pleasure to thank William A. Brock for criticism of earlier versions of this manuscript. The authors are grateful to Stephanie Henson for sharing observational data on the fraction of phytoplankton export to the deep ocean, Lester Kwiatkowski for discussions on phytoplankton dynamics, Steven Lade for suggestions  on zero-dimensional feedback process modelling, Colin Prentice for comments on the dynamics of land vegetation, Ailsa Roell for editorial comments, Christopher J. Williamson for discussing glacial algal blooms, Philipp de Vrese for sharing MPI-ESM snowball Earth output simulation datasets, Raymond Pierrehumbert for discussing with us plausible values of snowball Earth parameters, Paloma Ruiz Benito for discussing the use and interpretation of empirical vegetation death parameters, and Julius Garbe for discussing the dynamics of surface melting on the Antarctic ice sheet. This is TiPES contribution \#280 and ClimTip  contribution \#[X'Y'Z']; the Tipping Points in the Earth System (TiPES) and the Quantifying climate tipping points and their impacts (ClimTip) projects have received funding from the European Union's Horizon research and innovation programme under grant agreements No. 820970 and No. 101137601, respectively. We are grateful to Franklin Allen and the Brevan Howard Centre for Financial Analysis of Imperial College London for support provided. We are grateful also to the Singapore Green Finance Centre for support provided. M.G. also received support from the French Agence Nationale pour la Recherche (ANR) project TeMPlex under grant award ANR-23-CE56-0002.  \\

\clearpage

\section*{Supplementary Material}

%\newpage

\subsection*{Model parameters}
%
%\subsubsection*{Temperature equation parameters}

\begin{table}[H]
\centering \caption{{\bf Temperature equation parameters}}
\begin{tabular}{>{\centering}p{2cm}>{\raggedright}p{3.5cm}>{\raggedright}p{2.5cm}>{\raggedright}p{2.5cm}>{\raggedright}p{2.5cm}>{\raggedright}p{1.5cm}}
	\hline 
	Symbol  & Name  & Reference value  & Value used  & Source  & Equation\tabularnewline
	\hline 
	$c$  & Heat capacity  & 14 $\pm$ 6 W$\cdot$yr$\cdot$m$^{-2}\cdot$K$^{-1}$  & 10 W$\cdot$yr$\cdot$m$^{-2}\cdot$K$^{-1}$  & Schwartz \parencite{Schwartz2007}  & Eq. \eqref{eq:TempEq}\tabularnewline
	$Q_{0}$  & Incoming radiation  & 342.5 W$\cdot$m$^{-2}$  & 342.5 W$\cdot$m$^{-2}$  & Källén et al. \parencite{Kallen1979}  & Eq. \eqref{eq:TempEq}\tabularnewline
	$p$  & Fraction of land on the planet  & 0.29  & 0.29  & Källén et al. \parencite{Kallen1979}  & Eq. \eqref{eq:TempEq}\tabularnewline
	$\alpha_{{\rm min}}^{{\rm L}}$  & TOA land albedo lower bound & 0.36  & 0.36  & Boone et al. \parencite{Boone2010}; Yallop et al. \parencite{Yallop2012}  & Eq. \eqref{eq:alb_cases-1}\tabularnewline
	$\alpha_{{\rm max}}^{{\rm L}}$  & TOA land albedo upper bound  & 0.4  & 0.4  & Boone et al. \parencite{Boone2010}  & Eq. \eqref{eq:alb_cases-1}\tabularnewline
	$T_{\alpha_{L},\ell}$  & Land albedo lower bound temperature  & 290.5 K  & 290.5 K  & Robinson et al \parencite{Robinson_2012}   & Eq. \eqref{eq:alb_cases-1}\tabularnewline
	$T_{\alpha_{{L}},u}$  & Land albedo upper bound temperature  & 294 K  & 294 K  & Garbe et al \parencite{Garbe2020}, Box et al \parencite{Box2012}   & Eq. \eqref{eq:alb_cases-1}\tabularnewline
	$T_{\alpha_{O},\ell}$  & Ocean albedo lower bound temperature  & 200 K & 200 K  & Pierrehumbert et al. \parencite{Pierrehumbert2011}  & Eq. \eqref{eq:alb_cases-1}\tabularnewline
	$T_{\alpha_{O},u}$  & Ocean albedo upper bound temperature  & 299 K  & 299 K  & Källén et al. \parencite{Kallen1979}  & Eq. \eqref{eq:alb_cases-1}\tabularnewline
	$\alpha_{{\rm min}}^{{\rm O}}$  & TOA ocean albedo lower bound  & 0.25  & 0.26  & Källén et al. \parencite{Kallen1979}  & Eq. \eqref{eq:alb_cases-1}\tabularnewline
	$\alpha_{{\rm max}}^{{\rm O}}$  & TOA ocean albedo upper bound  & 0.55 -- 0.7  & 0.57  & De Vrese et al \parencite{Vrese2021}  & Eq. \eqref{eq:alb_cases-1}\tabularnewline
	$\kappa$  & Temperature linearization parameter  & 1.74  & 1.74  & Källén et al. \parencite{Kallen1979}  & Eq. \eqref{eq:TempEq}\tabularnewline
	$T_{\kappa}$  & Outgoing radiation linearization temperature  & 154 K  & 154 K  & Källén et al. \parencite{Kallen1979}  & Eq. \eqref{eq:TempEq}\tabularnewline
	$a$  & CO$_{2}$ radiative forcing  & 5.35 W$\cdot$m$^{-2}$  & 5.35 W$\cdot$m$^{-2}$  & Myhre et al. \parencite{Myhre1998}  & Eq. \eqref{eq:TempEq}\tabularnewline
	$C_{{\rm 0}}$  & Reference atmospheric carbon content  & 590 GtC  & 610 GtC  & Sarmiento and Gruber \parencite{Sarmiento2002}  & Eq. \eqref{eq:TempEq}\tabularnewline
	\hline 
	$T_{0}$  & Average equilibrium temperature with no anthropogenic carbon emissions  & 286.58 K & 286.5 K  & Hawkins et al \parencite{Hawkins2017} & Eq. \eqref{eq:TempEq}\tabularnewline
	\hline 
\end{tabular}

\label{tab:param-temp} 
\end{table}

%{\mg Pls. change all citations in this \& the following tables by replacing the date inserted by hand with a blank space between the name or the `et al.' \& the} \verb|\parencite|, {\mg which suffices for identification purposes.}

\clearpage

%\subsubsection*{Carbon equation parameters}

\begin{table}[H]
\centering \caption{{\bf Land carbon sink parameters}}
\begin{tabular}{>{\centering}p{2cm}>{\raggedright}p{3.5cm}>{\raggedright}p{2.5cm}>{\raggedright}p{2.5cm}>{\raggedright}p{2.5cm}>{\raggedright}p{1.5cm}}
	\hline 
	Symbol  & Name  & Reference value  & Value used  & Source  & Equation\tabularnewline
	\hline 
	$P_{{\rm m}}$  & Potential maximum annual GPP  & 179.9 GtC/yr  & 179 GtC/yr  & Myneni and Los \parencite{Myneni1995}  & Eq. \eqref{Eq.  Chavez-Ghil-Rombouts Cveg}\tabularnewline
	$C_{\rm h}$  & Half saturation carbon atmospheric content  & 1044 GtC [964--1,131]  & 1015 GtC  & Haverd et al \parencite{Haverd2020} & Eq. \eqref{Eq.  gc(C)}\tabularnewline
	$R_{L,0}$  & Reference land respiration rate  & 60 GtC/yr  & 60 GtC/yr  & Sarmiento and Gruber \parencite{Sarmiento2002}  & Eq. \eqref{Eq.  Global-Rs-Mahecha}\tabularnewline
	$Q_{10, \rm r}$  & Q$_{10}$ factor of the global land respiration  & 1.5  & 1.5  & Bond-Lamberty and Thomson \parencite{BondLamberty2010}  & Eq. \eqref{Eq.  Global-Rs-Mahecha}\tabularnewline
	$T_{v,\ell}^{-}$  & Lower vegetation viable temperature  & 257.75 $\pm$ 17.4 K & 257 K  & Lancaster and Humphreys \parencite{Lancaster2020}   & Eq. \eqref{eq:minimum}\tabularnewline
	$T_{v,u}$  & Upper vegetation viable temperature  & 324.45 $\pm$ 5.8 K & 324 K  & Lancaster and Humphreys \parencite{Lancaster2020}   & Eq. \eqref{eq:minimum}\tabularnewline
	$T_{o,\ell}^{+}$  & Upper minimum optimum vegetation temperature  & 298.15 K  & 298 K  & Still et al \parencite{Still2003}  & Eq. \eqref{eq:optimum}\tabularnewline
	$T_{o,\ell}^{-}$  & Lower minimum optimum vegetation temperature  & 290.15 K  & 290.15 K  & Huang et al. \parencite{Huang2019}  & Eq. \eqref{eq:optimum}\tabularnewline
	$T_{v,\ell}^{+}$  & Upper minimum viable vegetation temperature  & 288.15 K  & 289 K  & Sage et al \parencite{Sage1999}  & Eq. \eqref{eq:optimum}\tabularnewline
	$T_{o,u}$  & Lower maximum optimum vegetation temperature  & 302.15 K  & 302.15 K  & Huang et al. \parencite{Huang2019}  & Eq. \eqref{eq:optimum}\tabularnewline
	$e$  & Anthropogenic CO$_{2}$ emissions  & --  & -- GtC  & IIASA  \parencite{IIASA2009} & Eq. \eqref{eq:CarbAt}\tabularnewline
	%$K_{\rm l}$  & Environmental lower bound  & 0  & 0  & --  & Eq. \eqref{eq:bio}\tabularnewline
	%$K_{\rm u}$  & Environmental degradation upper bound  & 2.5  & 2.5  & --  & Eq. \eqref{eq:bio}\tabularnewline
	$a_{\rm d}$  & Linear mortality temperature response  & [-0.5, 0.9] K$^-1$  &  0.006K$^-1$ & Ruiz-Benito et al \parencite{RuizBenito2013}  & Eq. \eqref{eq:K-cumulative stress}\tabularnewline
	$b_{\rm d}$  & Nonlinear quadratic temperature response  &  [-0.6, 0.11] K$^{-2}$  & 0.0015 K$^{-2}$    & Ruiz-Benito et al \parencite{RuizBenito2013}   & Eq. \eqref{eq:K-cumulative stress}\tabularnewline
	$\mu$  & Land vegetation regeneration time  & 30 yrs  & 35 yrs  & Dobor et al. \parencite{Dobor_2018}  & Eq. \eqref{eq:EnvDegr}\tabularnewline
	\hline 
\end{tabular}

\label{tab:param-terrestrial-carbon} 
\end{table}

\clearpage

%\subsubsection*{Carbon equation parameters}

\begin{table}[H]
\centering \caption{{\bf Ocean carbon sink parameters}}
\begin{tabular}{>{\centering}p{2cm}>{\raggedright}p{3.5cm}>{\raggedright}p{2.5cm}>{\raggedright}p{2.5cm}>{\raggedright}p{2.5cm}>{\raggedright}p{1.5cm}}
	\hline 
	Symbol  & Name  & Reference value  & Value used  & Source  & Equation\tabularnewline
	\hline 
	$D_{{\rm T}}$  & Solubility temperature effect  & 0.0423  K$^{-1}$  & 0.0423 K$^{-1}$  & Takahashi et al. \parencite{Takahashi_1993}  & Eq. \eqref{eq:Ocean-carbon-partial-pressure-Lade-ea2018}\tabularnewline
	$r$  & Revelle buffer factor  & 12.5  & 12.5  &  Lade et al.  \parencite{Lade_2018} & Eq. \eqref{eq:Ocean-carbon-partial-pressure-Lade-ea2018}\tabularnewline
	$D$  & Atmosphere-ocean CO$_{2}$ equilibration time & 1 yr.$^{-1}$  & 1 yr.$^{-1}$  &  Lade et al. \parencite{Lade_2018} & Eq. \eqref{eq:Ocean-carbon-partial-pressure-Lade-ea2018}\tabularnewline
	$Q_{10,\rm phyto}$  & Q$_{10}$ value of phytoplankton carbon assimilation  & 1.5  & 1.5  &  Sherman et al. \parencite{Sherman_2016} & Eq. \eqref{eq:Biocarbon-flux}\tabularnewline
	$p_{0}$  & Reference phytoplankton GPP  & 50 GtC/yr  & 50 GtC/yr  &  Sarmiento and Gruber \parencite{Sarmiento2002} & Eq. \eqref{eq:Biocarbon-flux}\tabularnewline
	$w_{0}$  & Dissolved inorganic carbon flux rate from the mixed layer to the deep ocean  & 0.1 yr$^{-1}$  & 0.1 yr$^{-1}$  &  Lade et al. \parencite{Lade_2018} & Eq. \eqref{eq:Solubility-pump-Lade-ea2018}\tabularnewline
	$w_{T}$  & Weakening of the overturning circulation with temperature  & 0.1 K$^{-1}$  & 0.1 K$^{-1}$  &  Lade et al. \parencite{Lade_2018} & Eq. \eqref{eq:Solubility-pump-Lade-ea2018}\tabularnewline
	$a_{\rm exp}$  & Fraction of phytoplankton GPP export below 1,000 meters  & 0.0196  & 0.0196   &  Henson et al. \parencite{Henson2012} & Eq. \eqref{eq:Biocarbon-flux}\tabularnewline
	$b_{\rm exp}$  & Fraction of phytoplankton GPP export below 1,000 meters per degree increase  & 0.0526 K$^{-1}$ & 0.0526 K$^{-1}$ &  Henson et al. \parencite{Henson2012} & Eq. \eqref{eq:Biocarbon-flux}\tabularnewline
	
	\hline 
	$C_{\rm M,0}$  & Reference carbon stock in the mixed layer  &  900 GtC  & 890 GtC  &  Sarmiento and Gruber \parencite{Sarmiento2002} & Eq. \eqref{eq:Ocean-carbon-partial-pressure-Lade-ea2018} \tabularnewline
	\hline 
\end{tabular}

\label{tab:param-ocean-carbon} 
\end{table}

%\cm{Erik, Better to write “GtC/yr” than use the power; in any case, “yr” is a unit \& does not take either a period or an 's' after. The \emph{Sarmiento1998} ref. is also missing, as is the equation of Friedlingstein.}

\clearpage

\subsection*{Evolution of temperatures that affect land vegetation productivity}

\newpage

\subsection*{TCV model evaluation}

\subsubsection*{TCV temperatures vs. high-end models}

We rely here on the Bland-Altman methodology \parencite{Bland+A.1999} to compare the reliability of temperature estimates using our TCV model versus instrumental observations, on the one hand, and the ensemble mean of 42 GCMs used in CMIP5 \parencite{Taylor2012}, on the other. This methodology was devised in the biomedical context and it appears to be new in the climate sciences; hence we indicate briefly the essence thereof.  Unlike the RMSE statistic that  provides a single aggregate error metric, this test characterises biases and how differences vary across the whole range of values.

Given two sets of measurements, $S_1 = \{s_k^{(1)} : k = 1, 2, \cdots n\}$ and $S_2 = \{s_k^{(2)} : k = 1, 2, \cdots n\}$ of the same $n$ terms of a sequence $S_0 = \{s_k^{(0)} : k = 1, 2, \cdots n\}$, we plot for each index $k$ the mean $x_k = (s_k^{(1)} + s_k^{(2)})/2$ of the two separate measurements and their difference $y_k = s_k^{(1)} - s_k^{(2)}$ in a cartesian $(x, y)$ plot. In the clinical context, the two measurements are typically the results of two different instruments or two distinct measurement techniques. Here, they are estimates of a given climate variable, like global temperature, by the TCV model and by one of the 42 GCMs. 

\begin{figure}[H]

%\centering \subfigure[TCV simulation vs. instrumental temperatures]{ \label{fig:BA_obs}
% label of upper left panel    
%\centering \includegraphics[scale=0.3]{./FigSI9a}
%} 

\centering

\subfigure[TCV simulation vs. instrumental temperatures]{ \label{fig:BA_obs}
 %label of upper left panel    
\centering \includegraphics[scale=0.3]{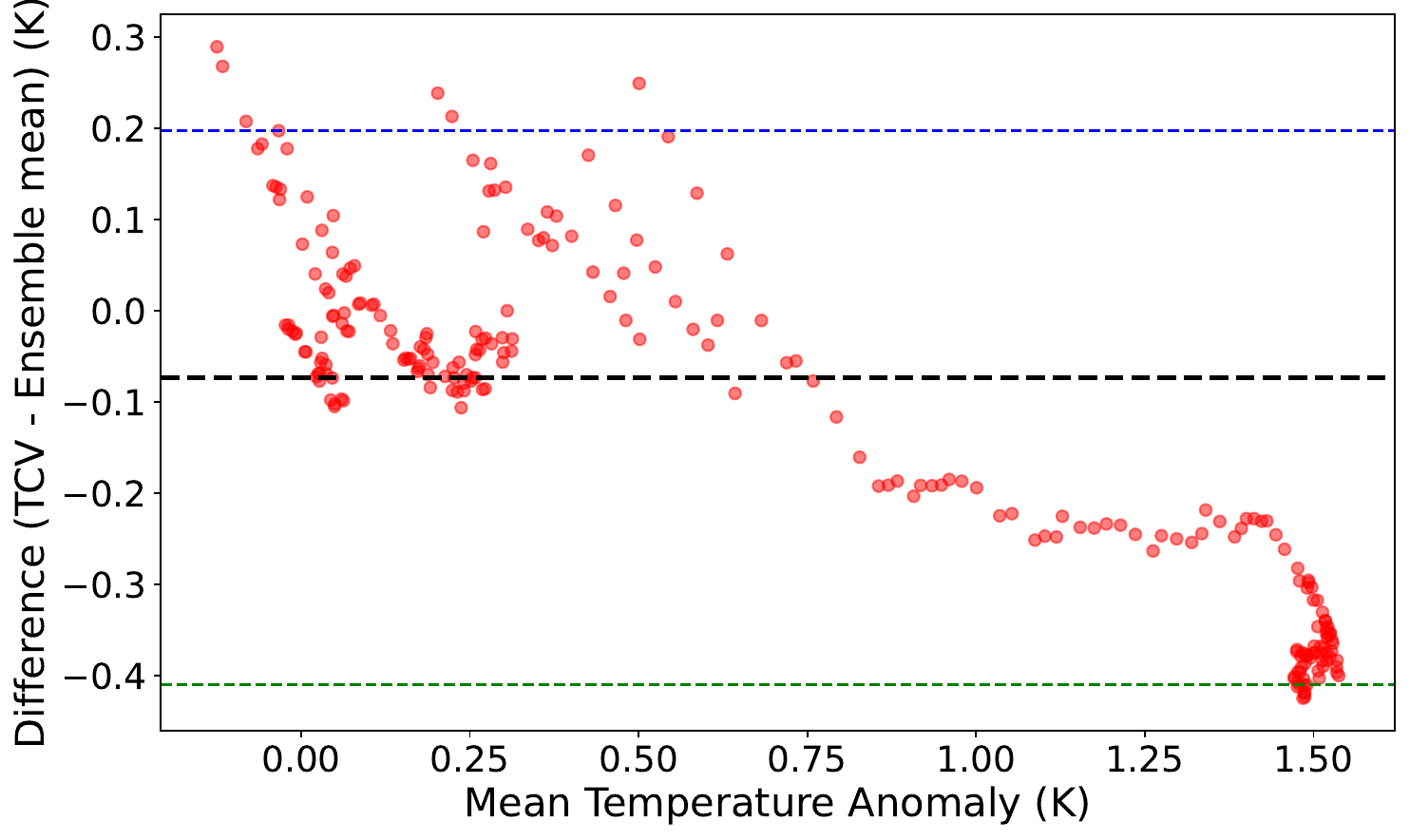}
} \subfigure[TCV vs GCMs average - RCP 4.5]{ \label{fig:BA_rcp45}
 %label of upper right panel    
\centering \includegraphics[scale=0.3]{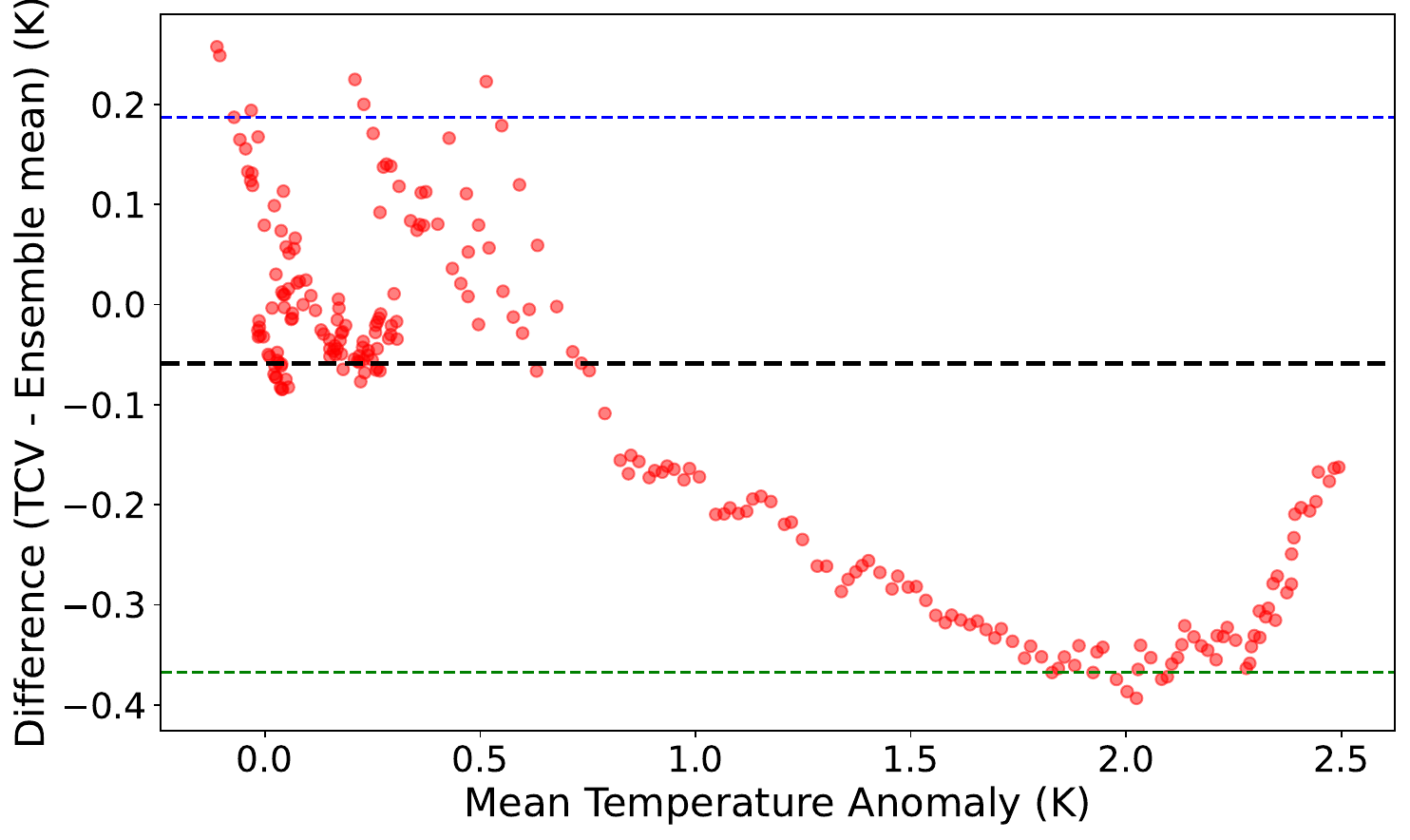}
}

%%%%%%%%%%%%%%%%%%%%%%%%%%%%%
%%%%%    lower panels     %%%%%%%%%%%%%%%%
%\subfigure[TCV vs GCMs average - RCP 6.0]{ \label{fig:BA_rcp60}
% label of upper left panel    
%\centering \includegraphics[scale=0.3]{./FigSI9d}
%} \subfigure[TCV vs GCMs average - RCP 8.5 (before global tipping)]{ \label{fig: BA_rcp85}
% label of upper right panel    
%\centering \includegraphics[scale=0.3]{./FigSI9e}
%}

\caption{{\footnotesize{\bf Bland-Altman non-parametric test plots comparing TCV temperature simulations to observations and to GCM ensemble median.} The red dots display the differences between TCV model and ensemble mean. The heavy dashed black line displays the bias of the TCV model with respect to the median of observations in (a) and to the GCM ensemble median in (b). The light blue dashed lines mark the 97.5$^{ \mathrm{th}}$ percentile of differences and the dashed green line mark the 2.5$^{ \mathrm{th}}$ percentile of differences --  the space between these two dashed lines describes the range within which the central 95 \% of the measured differences lie and allow to visualize the difference between the TCV model and the benchmark of comparison across temperature anomaly values.}}
\label{Fig:Bland-Altman} % label of the full figure
\end{figure}

One defines the mean $\bar s$ of the entire sample, i.e., the sum of the $x_k$'s, represented in each panel of Fig.~\ref{Fig:Bland-Altman} by the dashed blue line, and studies the scatter of the differences  $y_k$. The dotted red lines in the panels indicate, respectively, the values $\bar s \pm 1.96 \sigma$, which bracket the 95 \% confidence interval for a Gaussian scatter. We conclude that the TCV model's estimates are statistically indistinguishable from those of the set of the 42 IPCC-class GCMs .

Visually, the test allows to compare the differences between TCV model and ensemble mean. We analyse firstly the bias of the TCV model with respect to the median of observations and also to the GCM ensemble median. The test allows allows to visualize the 97.5$^{ \mathrm{th}}$ percentile of differences and the the 2.5$^{ \mathrm{th}}$ percentile of differences in two horizontal lines --  the space between these two lines describes the range within which the central 95 \% of the measured differences lie and allow to visualize the difference between the TCV model and the benchmark of comparison across temperature anomaly values. Test results show that TCV model estimates are consistently positioned between these two lines across all tested RCP scenarios (not shown here).

%\newpage

\subsubsection*{Carbon dynamics and validation}

\begin{figure}[H]
	
\centering

\subfigure[Ocean carbon flux -- GCP models  \& TCV]{ \label{fig:Ocean_carbon_marked}
% label of upper left panel    
\centering \includegraphics[scale=0.3]{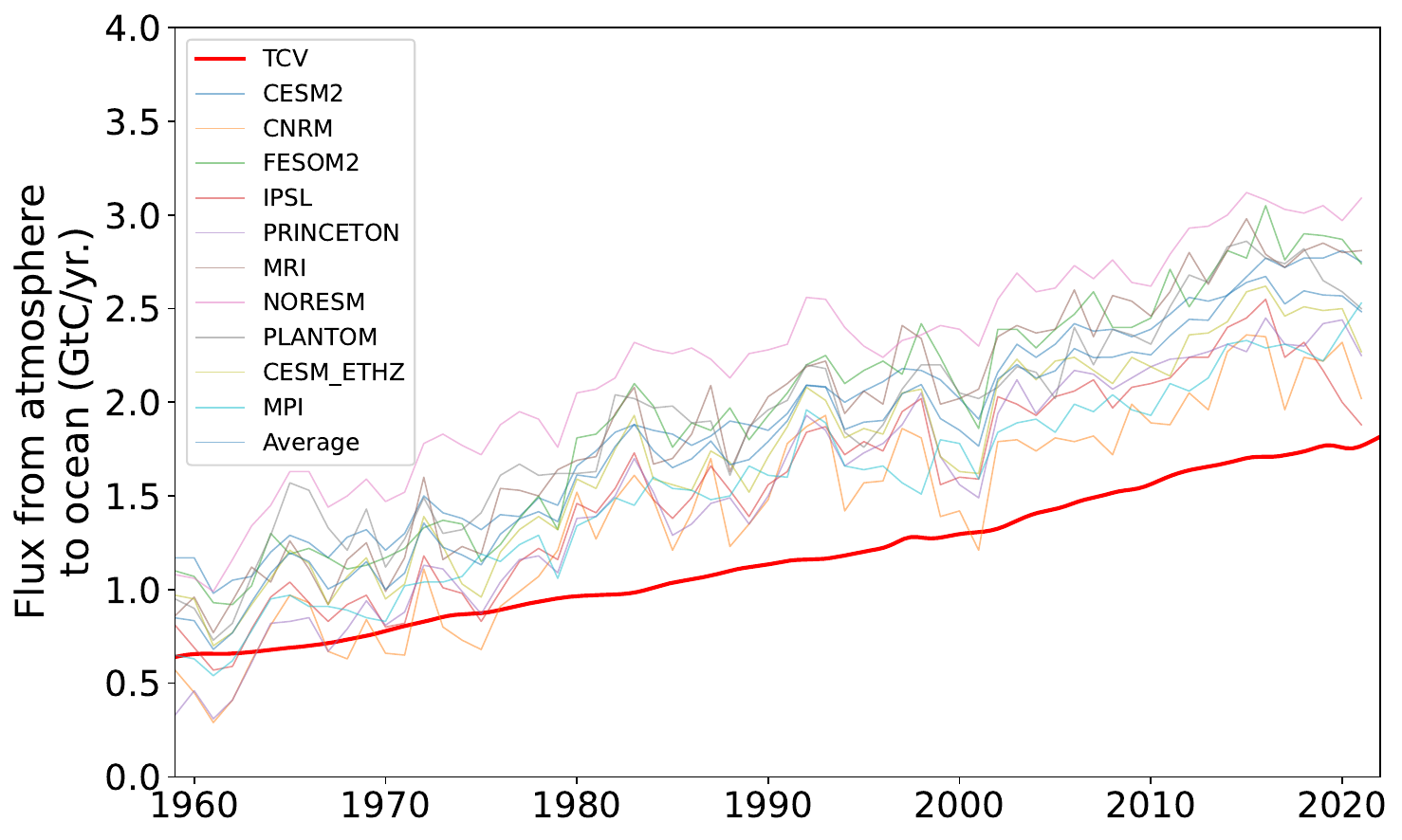}
} \subfigure[Land carbon flux -- GCP models  \& TCV]{ \label{fig: Land_carbon_marked}
% label of upper right panel    
\centering \includegraphics[scale=0.3]{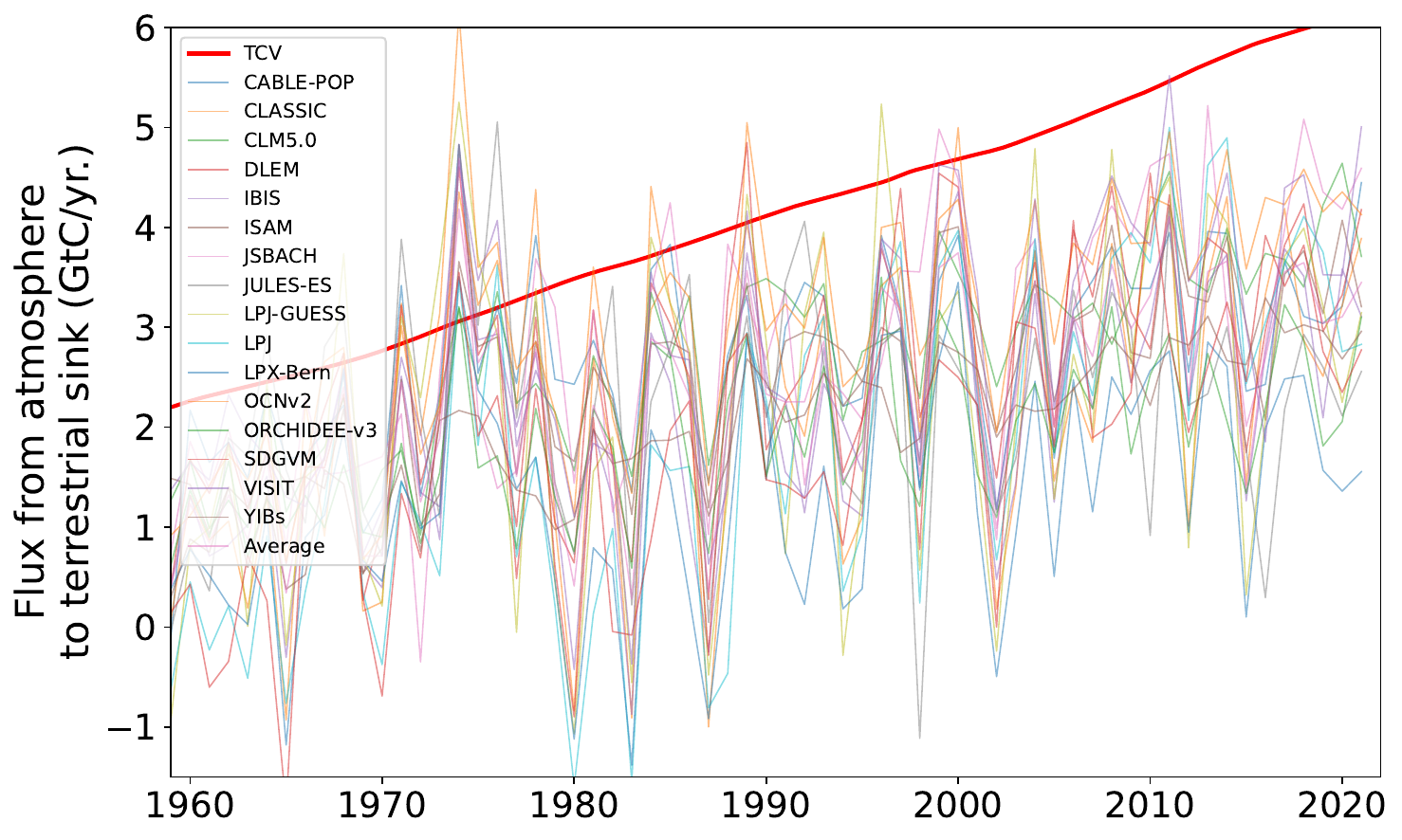}
}

%%%%%%%%%%%%%%%%%%%%%%%%%%%%%
%%%%%   upper panels     %%%%%%%%%%%%%%%%
\subfigure[Atmosphere to ocean GCP carbon fluxes range]{ \label{fig:OceanC_Range}
% label of upper left panel    
\centering \includegraphics[scale=0.3]{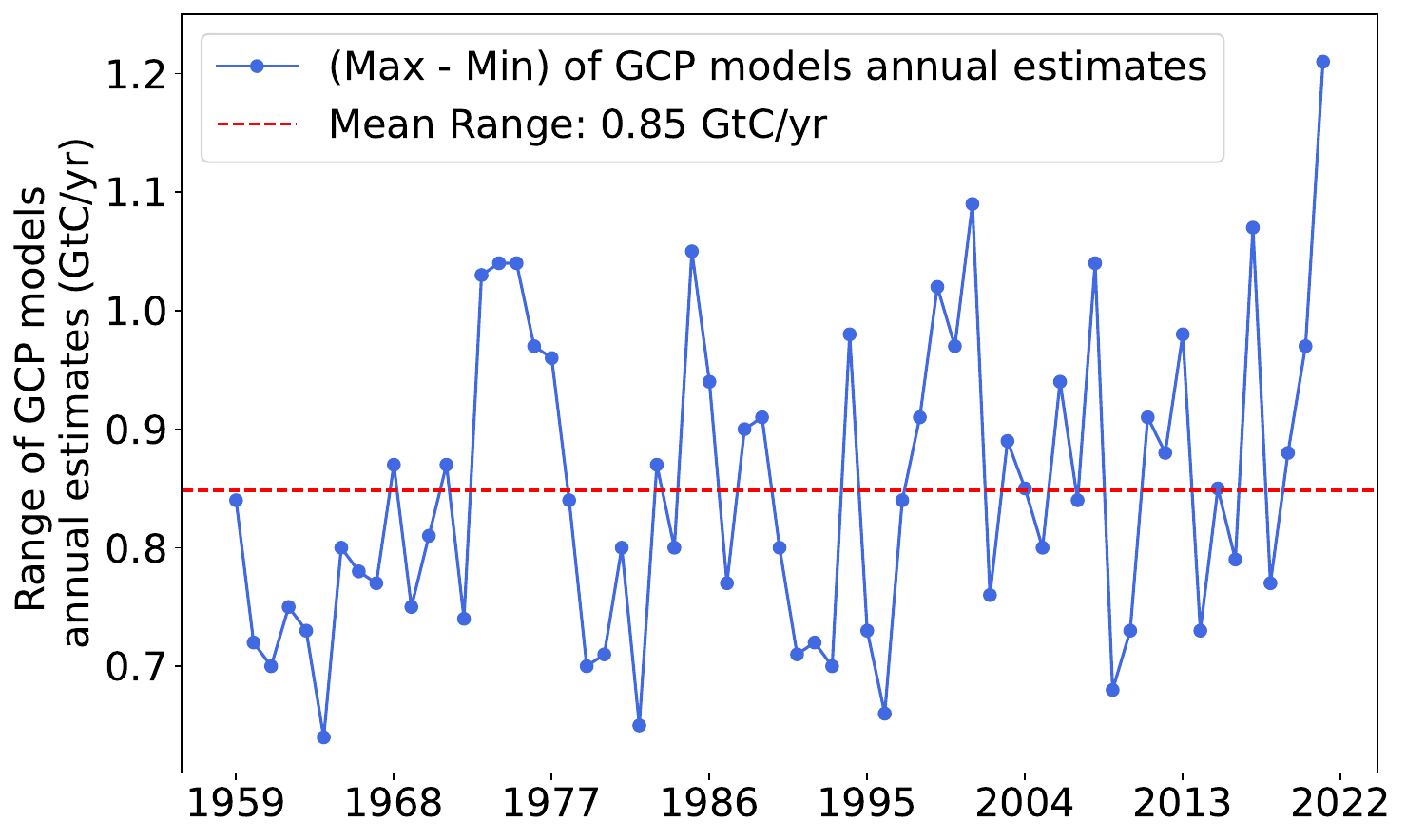}
} \subfigure[Atmosphere to land GCP carbon fluxes]{ \label{fig: LandC_Range}
% label of upper right panel    
\centering \includegraphics[scale=0.3]{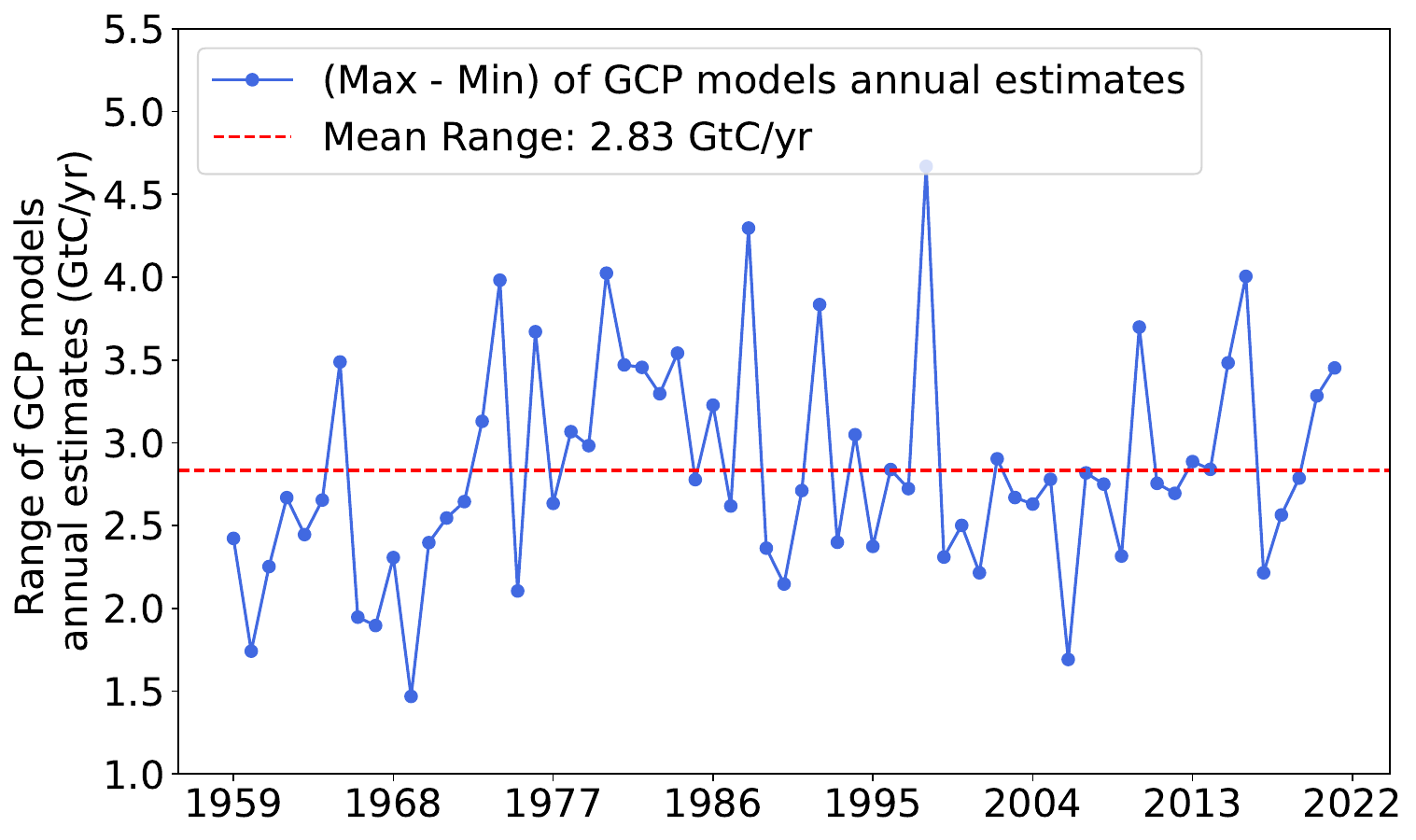}
}

%%%%%%%%%%%%%%%%%%%%%%%%%%%%%
%%%%%   lower panels     %%%%%%%%%%%%%%%%
\subfigure[RMSE of TCV vs. GCP ocean flux models]{ \label{fig:OceanC_RMSE}
% label of upper left panel    
\centering \includegraphics[scale=0.3]{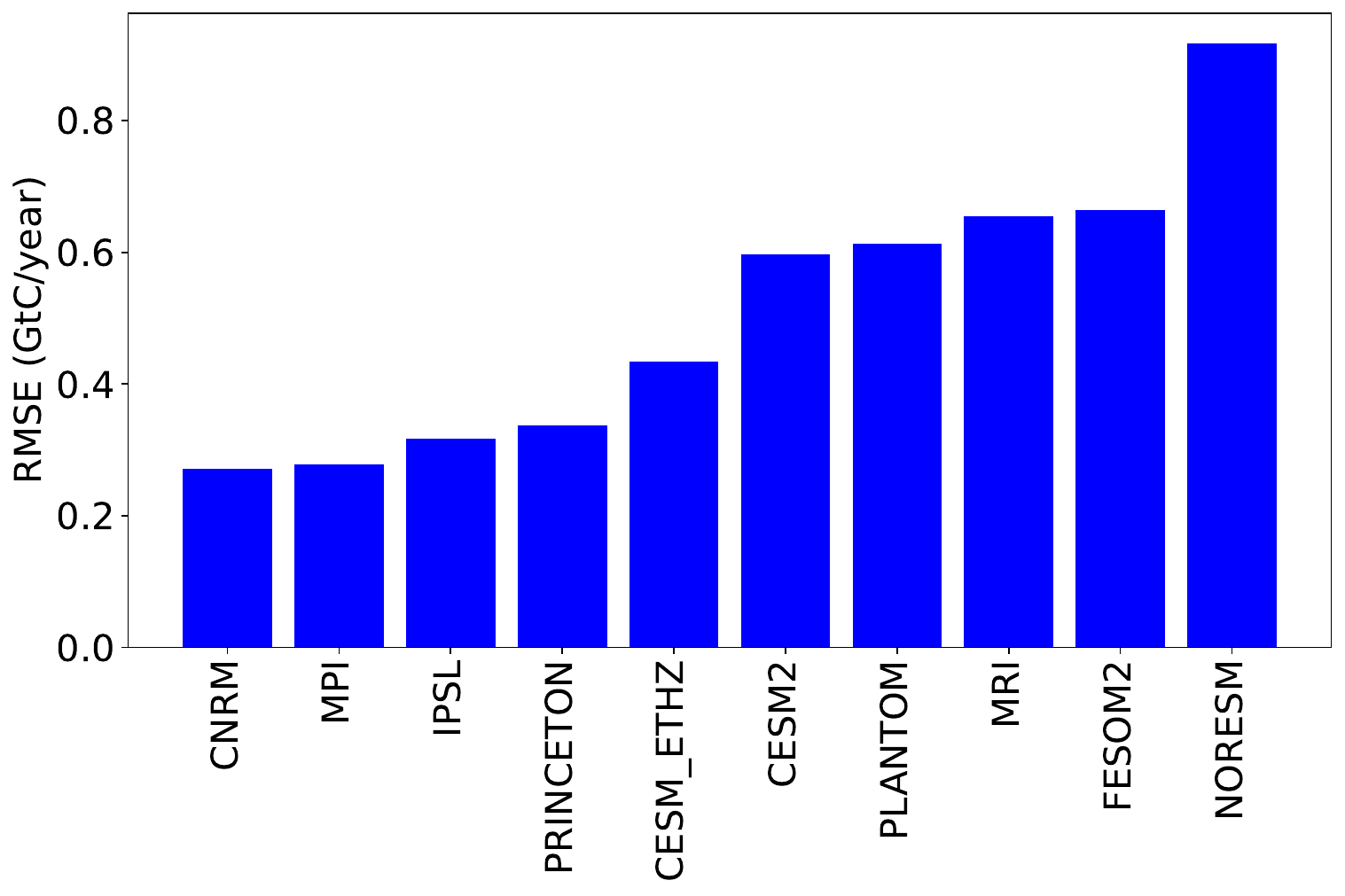}
} \subfigure[RMSE of TCV vs. GCP land flux models]{ \label{fig: TerrestrialC_RMSE}
% label of upper right panel    
\centering \includegraphics[scale=0.3]{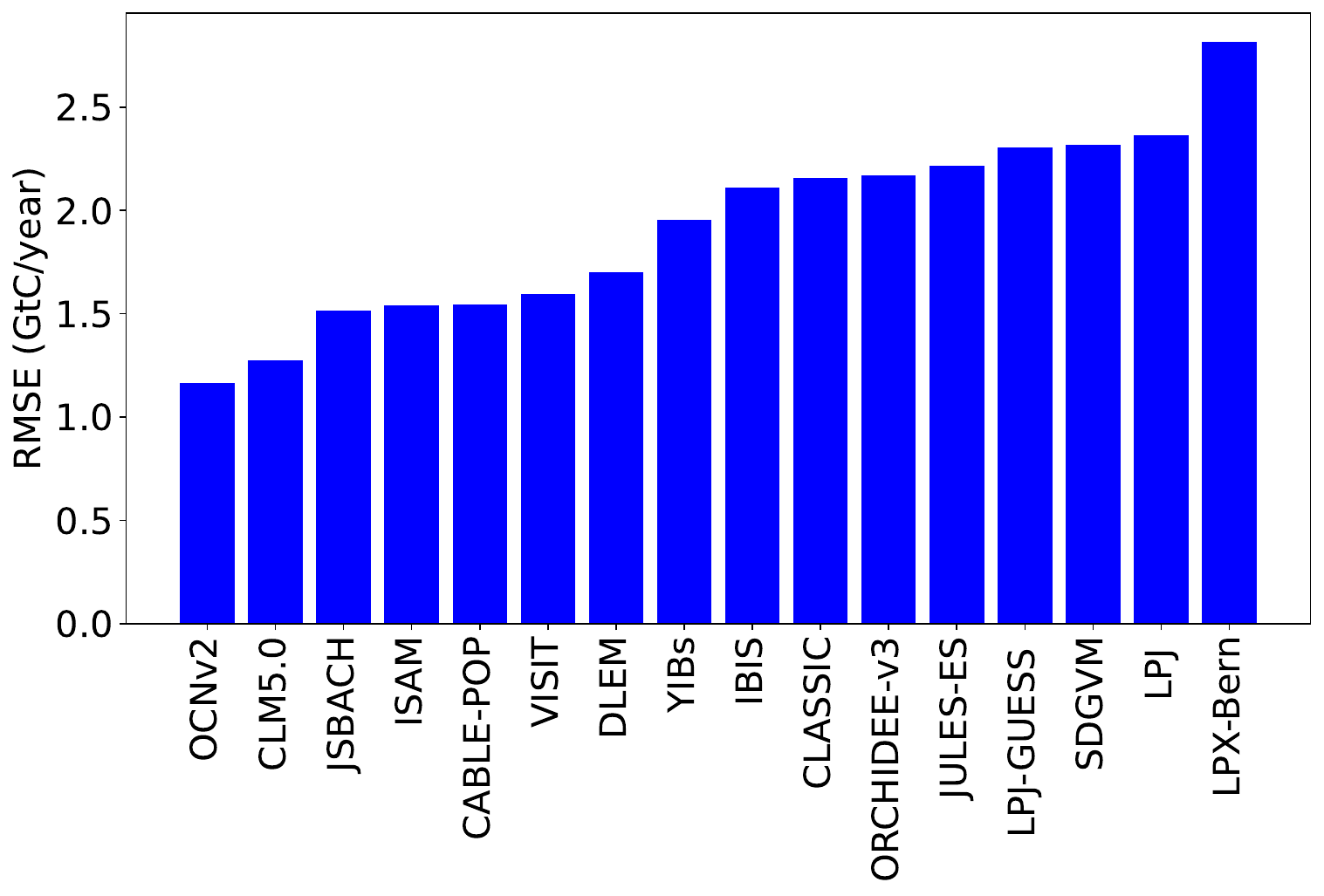}
}

\caption{{\footnotesize{\bf Variability of the GCP carbon fluxes and comparison with the TCV model fluxes.} (a,b) The annually estimated carbon fluxes from all GCP models \parencite{Friedlingstein_2022} (colors) for (a) atmosphere-to-ocean fluxes (eleven models), and (b) for atmosphere-to-land fluxes (16 models), over the time interval of 1959--2021, and the corresponding TCV model fluxes (solid red). (c,d) The differences each year between the maximum GCP flux vs. the corresponding minimum (solid blue) and the mean over time of the blue line (dashed red); and (e,f) the individual RMSEs between the TCV model and each GCP model.			
	%			compared with the RMSE over theentire time series for each GCP model with respect to the TCV model: in (c) for the flux to the ocean mixed layer,and in (d) for the flux to the land vegetation.
	}}
	\label{Fig:Carbon_robustness_test} % label of the full figure
\end{figure}

\newpage

\subsubsection*{Model Sensitivity Analysis}
\begin{figure}[H]
	%\centering  %\subfigure[
	%Hysteresis - Saddle Node bifurcation diagram
	\centering \includegraphics[width=1\textwidth]{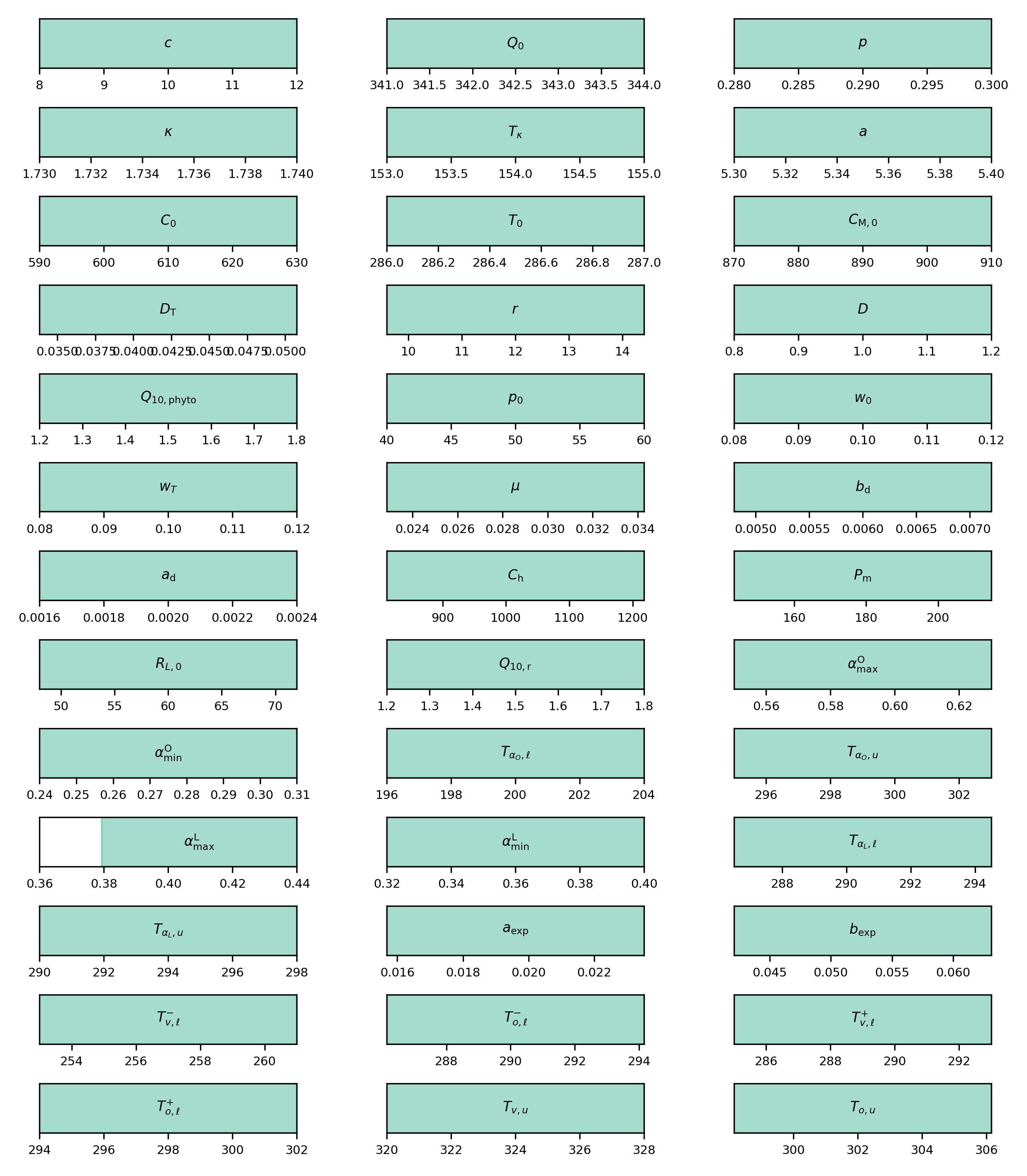}
	\caption{{\footnotesize{}{}{} {\bf Parameter ranges that yield bistability in the ``one-at-a-time" (OAT) sweep.} Each subplot shows one of the 39 parameters with the $x$‑axis spanning the range of values explored. Filled boxes indicate intervals within which the model exhibits more than one steady state, while the blank intervals correspond to a single steady state for $V = 0.$}}
	\label{fig:oat} % label of the full figure
\end{figure}

%{\mg
We study the sensitivity of the TCV model's bistability across the range of anthropogenic emissions $e$ and the ecosystem quality degradation levels $V$ to parameter values. With $\mathbf{x}$ a vector of parameter values, we define the binary response as follows:
\begin{equation}
Y\bigl(\mathbf{x},V\bigr)
\;=\;
\begin{cases}
	1, & \text{if the model admits $>1$ steady states at $(\mathbf{x},V)$},\\
	0, & \text{otherwise}.
\end{cases}
\end{equation}
and its trajectory‐wise probability \(p_{\text{bist}} = \mathbb{E}[Y]\).  The one‑at‑a‑time (OAT) sensitivity analysis plot displayed in Fig.~\ref{fig:oat} is carried out as follows for a given $V$ value. For every parameter $\{x_i : i=1,\dots,k\},$ we sweep \(x_i\) across a plausible interval \([x_i^{\min},x_i^{\max}]\), while the remaining \((k-1)\) parameters are fixed at their reference values $\mathbf{x}^{(0)}$ listed in Tables~\ref{tab:param-temp}, \ref{tab:param-terrestrial-carbon}, and \ref{tab:param-ocean-carbon}. We record the envelope of values:
\begin{equation}
\mathcal{R}_{i}
= \bigl\{\,Y(\mathbf{x}^{(0)}+\delta\,\mathbf{e}_i) \quad \mbox{for} \quad
\delta\in[x_i^{\min}-x_i^{(0)},\,x_i^{\max}-x_i^{(0)}]\bigr\}.
\end{equation}

Because \(Y\) is binary, \(\mathcal{R}_{i}\subseteq\{0,1\}\) and the extent of the colored segment on the plot is exactly \(\max\mathcal{R}_{i,j}-\min\mathcal{R}_{i,j}\). A fully shaded bar across the entire parameter range signals that changing \(x_i\) can shift the model between single to bi-stable state within the emission levels tested of $-15 \le e \le 70$~GtC/yr. All parameter intervals \([x_i^{\min},x_i^{\max}]\) are chosen to cover a broad set of possible values. All temperatures span a 4~K range, ocean minimum and maximum albedos span the ranges from 0.25 to 0.32, and 0.55 to 0.62, respectively, as reported in the literature  \cite{Budyko.1969, Sellers.1969,Vrese2021},
% \cm{Sentence not clear at all here!}
while terrestrial albedo bounds vary in a similarly broad range centered on the reference values. Incoming solar radiation varies within a $\pm$1.5 Wm$^2$K$^{-1}$ range of the estimated preindustrial value, 
% preindustrial temperature and 
while oceanic and atmospheric carbon contents vary within a $\pm$20~GtC range with respect to reference values. All other parameters are allowed to vary within a $\pm$20 \% range.

The output plotted in Fig.~\ref{fig:oat} provides a first filter to discriminate between parameters that never alter \(Y\) across the entire \((x_i)\) grid vs. those that may do so.  
% \cm{From the Fig., it seems that only $\alpha_{max}^L$ does so. Why pick 4, \& state in the Main Text that it's 2 out of 39? Pls. clarify!}
It turns out that, in fact, it is only $\alpha_{{\rm max}}^L$ that clearly does affect the TCV model's possesing either one or two stable steady states. For completeness, we consider the four related parameters $\mathcal{S} = \{T_{\alpha_L, \ell}, T_{\alpha_L, u}, \alpha_{{\rm max}}^L, \alpha_{{\rm min}}^L\}$ herewith.

\begin{figure}[H]
		\label{fig.sobol} % label of upper left panel    
		\centering 
		\includegraphics[width=0.7\textwidth]{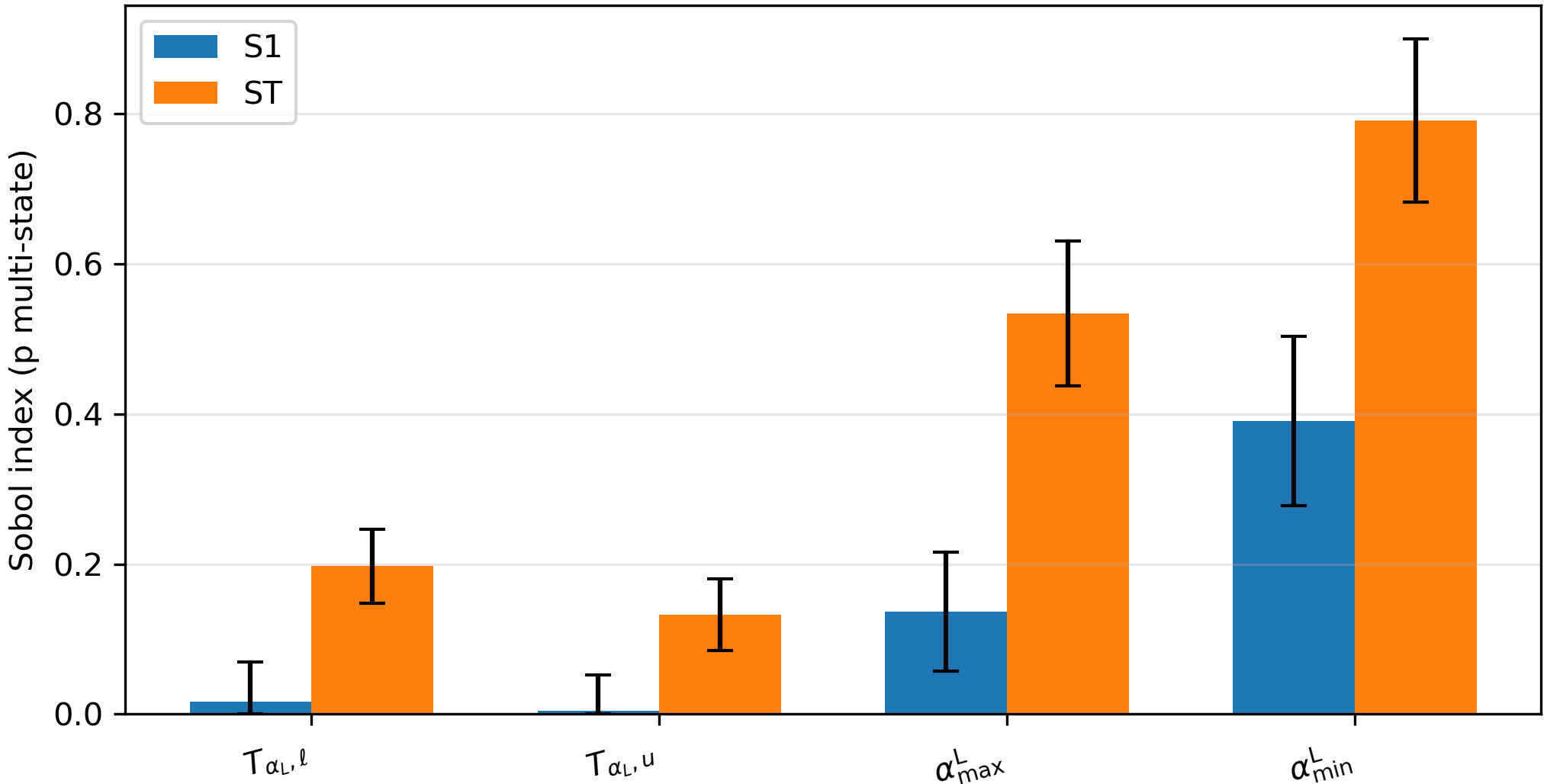}
	\caption{{\footnotesize{}{}{}{\bf Variance-based Sobol sensitivity analysis of the TCV model to selected parameters.} First–order ($S_{1}$, blue bar) and total–effect ($S_{T}$, orange bar) Sobol sensitivity indices of the probability of bistable behaviour ($p_{\text{bist}}$) at $V = 0$. Bars show point estimates, and error bars denote analytic 95\% confidence intervals. The parameter \(\alpha_{\mathrm{max}}^{\mathrm{L}}\) 
%			\cm{inconsistent with parameters in figure} and \(\alpha_{\mathrm{min}}^{\mathrm{L}}\) 
			dominates both individually ($S_{1}$) and including higher-order interactions ($S_{T}$), whereas the remaining parameters contribute only weakly to the variance of $p_{\text{bist}}$.}}
	\label{fig:sobol} % label of the full figure
\end{figure}

For the reduced set of inputs \(\mathcal{S}\), we estimate the two Sobol indices of main-effect $S_{1,i}$ and total-effect $T_{i}$ \cite{Sobol2001,Saltelli2007} defined by
\begin{equation}
	S_{1,i} =
	\frac{\operatorname{Var}_{x_i}\!\bigl[
		\mathbb{E}_{\mathbf{x}_{\sim i}}(Y\mid x_i)\bigr]}
	{\operatorname{Var}(Y)}, 
	\qquad
	T_{i} =
	1 -
	\frac{\operatorname{Var}_{\mathbf{x}_{\sim i}}\!\bigl[
		\mathbb{E}_{x_i}(Y\mid\mathbf{x}_{\sim i})\bigr]}
	{\operatorname{Var}(Y)}.
\end{equation}
The main effect \(S_{1,i}\) is displayed in blue bars in Fig.~\ref{fig:sobol} and captures the fraction of output variance that can be removed if \(x_i\) were fixed to be a constant. The total effect \(T_i\) displayed in orange bars captures the variance share attributable to \(x_i\) when including interactions with the other parameters. Because \(T_i \ge S_{1,i}\) by construction, the difference \(T_i-S_{1,i}\) measures the strength of all the interaction effects that involve \(x_i\).

Both indices are estimated here using the Saltelli \cite{Saltelli2007} “$AB$–$BA$’’ Monte‑Carlo scheme, which uses two independent \(N\times k'\) base matrices \(A\) and \(B\). For \(S_{1,i}\) the unbiased Jansen estimator \cite{Jansen1999} is: %\cm{There is a Jansen (1999) problem similar to the earlier Parisi one. I've gotten over the Saltelli (2007) one by citing it first in the Main Text, but don't think that Jansen belongs there.}}
\begin{equation}
\widehat{S}_{1,i} \;=\;
\frac{\dfrac1N\sum_{n=1}^{N}
	f(B)_n\bigl[f(A^{(i)})_n-f(A)_n\bigr]}
{\operatorname{Var}_{\!N}(Y)};
\end{equation}

its sampling variance, under mild regularity conditions, is
\begin{equation}
	\operatorname{Var}\!\bigl[\widehat{S}_{1,i}\bigr]
	\simeq
	\frac{1}{N}\Bigl(
	\mu_4 - \mu_2^2
	\Bigr),
\end{equation}
where \(\mu_2\) and \(\mu_4\) are second‑ and fourth‑order mixed moments that can be estimated from the same model runs \cite{Jansen1999}. Analogous expressions hold for \(\widehat{T}_{i}\) \cite{Saltelli2010}. Assuming asymptotic normality, the 95 \% analytic confidence intervals are computed as follows:
\begin{equation}
	\widehat{S}\;\pm\;1.96\,\sqrt{\operatorname{Var}[\widehat{S}]},\qquad
	\widehat{T}\;\pm\;1.96\,\sqrt{\operatorname{Var}[\widehat{T}]}.
\end{equation}

%{\mg
The narrow intervals on the three parameters ${\mathcal S} \setminus \{\alpha_{{\rm min}}^{{\rm L}}\}$ confirm that $\alpha_{{\rm min}}^{{\rm L}}$ alone dominates the output variance, while the other factors contribute only marginally. These results are entirely consistent with the qualitative OAT screening stage illustrated in Fig.~\ref{fig:oat}. In order to identify which parameters interaction dominate the total variance, for any two inputs $x_i,x_j\;(i<j)$ of subset  ${\mathcal S}$ we also report the second-order Sobol index defined as follows:

\begin{equation}
	S_{2,ij} \;=\;
	\frac{\operatorname{Var}_{x_i,x_j}\!\bigl[
		\mathbb{E}_{\mathbf{x}_{\sim\{i,j\}}}(Y\mid x_i,x_j)\bigr]
		- V_i - V_j}
	{\operatorname{Var}(Y)},
\end{equation}

where $V_i=\operatorname{Var}_{x_i}\!\bigl[\mathbb{E}(Y\mid x_i)\bigr]$. $S_{2,ij}$ measures the fraction of variance that can only be explained by the joint variation of $x_i$ and $x_j$ and cannot be attributed to either parameter alone. Using the extended Saltelli ``AB–BA'' sampling scheme, the unbiased estimator \cite{Saltelli2010} is defined by:

\begin{equation}
\widehat{S}_{2,ij}
\;=\;
\frac{\dfrac1N\sum_{n=1}^{N}
	\bigl[f(B^{(i)})_n - f(A)_n\bigr]
	\bigl[f(A^{(j)})_n - f(A)_n\bigr]}
{\widehat{\operatorname{Var}}_N(Y)}.
\end{equation}

\begin{figure}[H]
	%\centering  %\subfigure[
	%Hysteresis - Saddle Node bifurcation diagram
	\centering \includegraphics[width=1\textwidth]{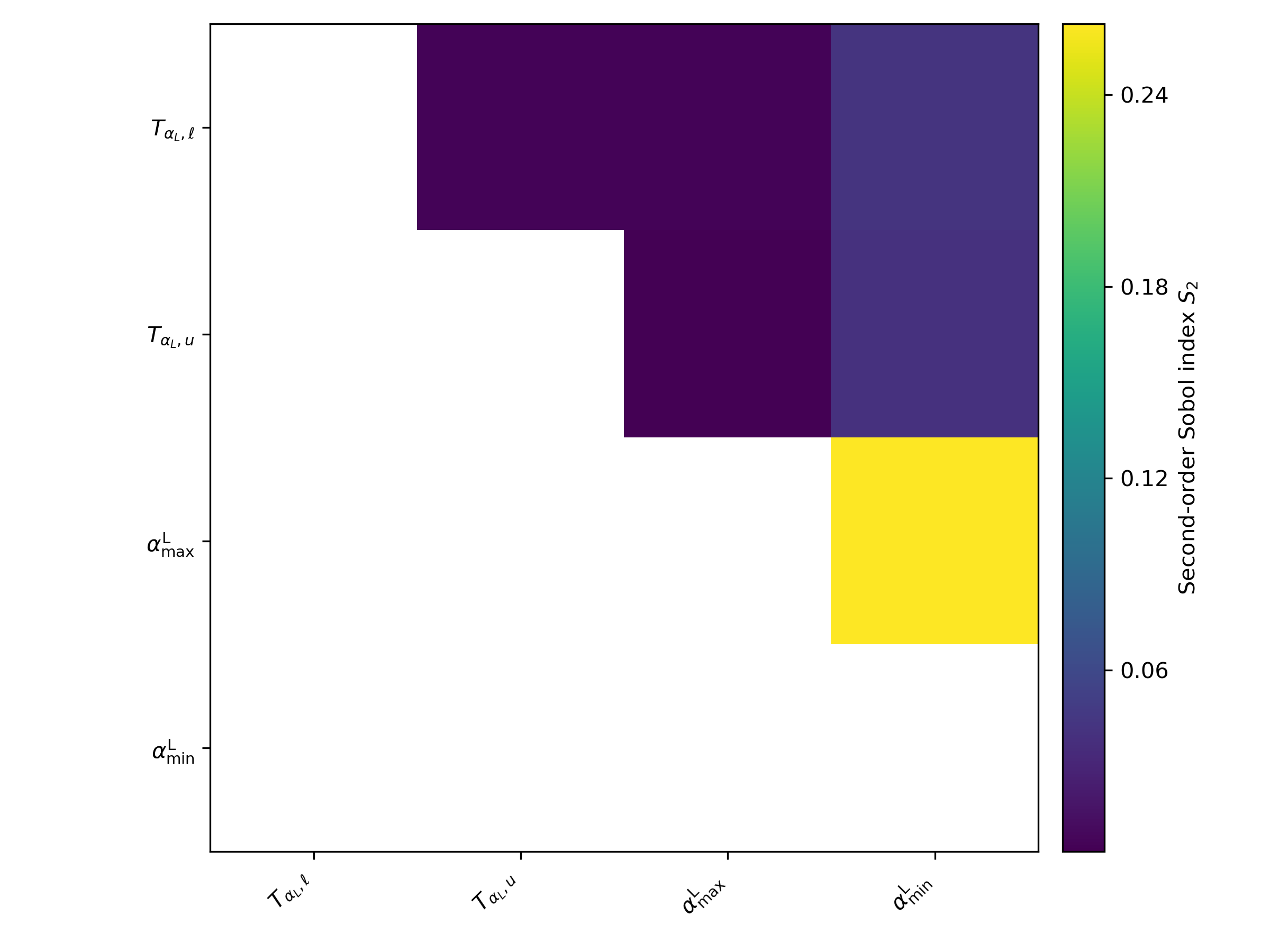}
	\caption{{\footnotesize{}{}{} {\bf Pairwise TCV terrestrial albedo parameters interaction Sobol indices ($S_{2}$).} Second-order Sobol indices of pairwise interactions for the four parameters set ${\mathcal S}$.  Colour scale indicates the proportion of $ST$ output variance that can be attributed only to the interaction of each parameter pair. The interaction $\alpha_{\min}^{\mathrm L}\!\times\!\alpha_{\max}^{\mathrm L}$ of terrestrial albedo bounds dominate and accounts for $\approx25 \%$ of the variance, while interactions involving the threshold temperatures ($T_{\alpha_L,\ell}$, $T_{\alpha_L,u}$) account for less than $\approx5 \%$.}}
	\label{fig:sobolint} % label of the full figure
\end{figure}

Figure~\ref{fig:sobolint} shows that the only significant pairwise interaction is $\alpha_{\min}^{\mathrm L}\!\times\!\alpha_{\max}^{\mathrm L}$ between the two terrestrial albedo bounds with $S_{2,\alpha_{\min}^{\mathrm L}\times\alpha_{\max}^{\mathrm L}}\approx 0.25$ indicating that approximately 25 \% of the output (bistability) variance arises from the non-additive effect of modifying the two terrestrial albedo bounds jointly. All other pairwise parameter second order Sobol indices have $|S_{2,ij}|<0.05$ indicating negligible interaction. Combined with the large $S_{1}$ and $ST$ gap for $\alpha_{\min}^{\mathrm L}$ (Fig.~\ref{fig:sobol}), this result confirms that the bistability of the TCV model is primarily sensitive to the contrast of terrestrial albedo bounds rather than by the threshold temperatures $T_{\alpha_L,\ell}$ and $T_{\alpha_L,u}$.

%} \\
%\cm{Pls. turn all multi into bi, and replace `multi' in the caption above.}

\subsection*{Global average land albedo latitudinal weights}

%\newpage

\subsection*{Visualization of the tipping to the hothouse}

The following three videos are attached to the present submission. The first two videos are made using the DAE model governed by the two ODEs~\eqref{eq:TempEq3} and \eqref{eq:TotalCarb}, with $V$ fixed. In the third video, $V = V(t)$ is allowed to change according to the ODE~\eqref{eq:EnvDegr3} of the DAE model. In all three cases, the emissions $e = e(t)$ evolve according to the RCP 8.5 scenario that corresponds to Fig.~\ref{Fig:bdry}(c).

In all three videos, the left panel shows the changes in the model's nullclines 
$F_1(T,C)=0$ and $F_2(T,C)=0$, along with the trajectory of the actual solution $(T(t), C(t))$ in the corresponding model's phase plane $(T, C).$ The right panel illustrates the evolution of the solution's temperature $T(t)$ in time. The initial point of the trajectory being computed is taken close to, but not identical to the climate for preindustrial conditions  in 1750 and computations are carried out till 2200. Note that, at the end of the simulation, $e(t) = $~28.8 GtC/yr and the corresponding temperature of $P_2$ is roughly 14~K higher than at $e = 0$. A rapid shift of 10~K from the present-climate--like steady state to a hothouse Earth steady state starts approximately in 2085 and it is completed over a 20-year time interval.  

%\subsubsection*{Tipping for variable vegetation degradation $V$}

The short video labeled \verb|TCV_video-V=V(t)| displays the results of a simulation that is carried out with 
the complete DAE model governed by equations~\eqref{eq:TC-3box}, i.e., the
ecosystem degradation $V = V(t)$ varies in time following equation~\eqref{eq:EnvDegr3} as $e(t)$ follows the anthropogenic RCP 8.5 emission scenario. 
The video documents the merger of the stable and unstable fixed points $P_1$ and $P_3$ that leads to the sudden tipping to the hothouse Earth, as well as the overshoot of  the globally averaged temperature $T(t)$ 
before reaching its new equilibrium at roughly 12~K above the preindustrial level. Notice the good fit to the observed global temperatures (black line) over the instrumental interval 1880--2022.

\section*{Author contributions}

E.C., M.G., and J.R. designed the study, E.C. and J.R. carried out the computations, and all three authors participated in the analysis of the results and in the writing.

\section*{Competing interests} The authors declare no competing interests.

\end{document}